\documentclass[twocolumn]{article}
\usepackage[utf8]{inputenc}
\usepackage{amsmath}
\usepackage{bbm}
\usepackage{bm}
\usepackage[margin=0.9in]{geometry}
\usepackage{slashed}
\usepackage{graphicx}
\usepackage{xcolor}
\usepackage{authblk}
\usepackage{appendix}
\usepackage{soul}
\usepackage{ulem}
\usepackage{widetext}

\usepackage{amssymb}
\usepackage[backend=bibtex,natbib=true,style=phys,biblabel=brackets,maxnames=20]{biblatex}
\usepackage{sectsty}
\sectionfont{\Large}
\subsectionfont{\large}
\subsubsectionfont{\large}
\paragraphfont{\large}

\DeclareMathOperator*{\SumInt}{%
\mathchoice%
  {\ooalign{$\displaystyle\sum$\cr\hidewidth$\displaystyle\int$\hidewidth\cr}}
  {\ooalign{\raisebox{.14\height}{\scalebox{.7}{$\textstyle\sum$}}\cr\hidewidth$\textstyle\int$\hidewidth\cr}}
  {\ooalign{\raisebox{.2\height}{\scalebox{.6}{$\scriptstyle\sum$}}\cr$\scriptstyle\int$\cr}}
  {\ooalign{\raisebox{.2\height}{\scalebox{.6}{$\scriptstyle\sum$}}\cr$\scriptstyle\int$\cr}}
}

\definecolor{teal}{RGB}{106,117,179}

\newcommand{\ic}{{\rm i}}
\newcommand{\E}{{\rm E}}
\newcommand{\e}{{\rm e}}

\newcommand{\Dcal}{{\cal D}}
\newcommand{\mc}{{\mathfrak{m}}}

\title{\bf Absence of $CP$ violation in the strong interactions}
\author[1]{Wen-Yuan Ai\footnote{wenyuan.ai@uclouvain.be}}
\author[2]{Juan S. Cruz\footnote{juan.cruz@tum.de}}
\author[2]{Bj\"orn Garbrecht\footnote{garbrecht@tum.de}}
\author[2]{Carlos Tamarit\footnote{carlos.tamarit@tum.de}}

\affil[1]{\it Centre for Cosmology, Particle Physics and Phenomenology, \par
Universit\'e catholique de Louvain, Louvain-la-Neuve B-1348, Belgium}

\affil[2]{\it Physik-Department T70, James-Franck-Stra{\ss}e,\par Technische Universit\"at M\"unchen, 85748 Garching, Germany}

\date{}

\newif\ifarXiv

\arXivfalse

\newif\ifSupplementary
\Supplementarytrue
\Supplementaryfalse

\newcommand\acknos{\paragraph*{Acknowledgements}
We would like to thank C.~Bonati, D.J.H.~Chung, J.~de~Vries, Jean-Marc Gerard, Feng-Kun Guo, E.~Mereghetti, J.~Redondo, A.~Ringwald and A.~Shindler for discussions and comments.
WYA is supported by the FSR Postdoc Incoming Fellowship of UC Louvain. This work has also been supported in part by SFB~1258 of the Deutsche Forschungsgemeinschaft. BG is thankful to D.J.H.~Chung and the physics department at UW-Madison for hospitality and support during initial stages of this work.}

\begin{document}

\ifSupplementary{}
\else
{

\onecolumn

\begin{flushleft}
{\tt CP3-20-02}\hspace{12.4cm} {\tt TUM-HEP-1249/20}
\end{flushleft}

{\let\newpage\relax\maketitle}

\begin{abstract}
We derive correlation functions for massive fermions with a complex mass in the presence of a general vacuum angle. For this purpose, we first build the Green's functions in the one-instanton background and then sum over the configurations of background instantons. The quantization of topological sectors follows for saddle points of finite Euclidean action in an infinite spacetime volume and the fluctuations about these. For the resulting correlation functions, we therefore take the infinite-volume limit before summing over topological sectors.  In contrast to the opposite order of limits, the chiral phases from the mass terms and from the instanton effects then are aligned so that, in absence of additional phases, these do not give rise to observables violating charge-parity symmetry. This result is confirmed when constraining the correlations at coincident points by using the index theorem instead of instanton calculus.
\end{abstract}

\twocolumn

\sectionfont{\large}

\ifarXiv{\section*{Digest}}
\else
{

\section{Introduction}
The theoretical formulation of the strong interactions in general allows for a Lagrangian term
\begin{align}
\label{topologcical:term}
1/(16\pi^2) \theta\, {\rm tr}\,F \widetilde F
\end{align}
that is odd (i.e. it changes sign) under charge-parity ($CP$) conjugation. Here,  $F$ is the gauge field strength tensor and $\widetilde{F}$ is its Hodge dual, with electric and magnetic components being interchanged. One may expect in general that this term also leads to phenomena that violate $CP$.

Conceivable in particular is a permanent electric dipole moment of the neutron~\cite{Baluni:1978rf,Crewther:1979pi}, which, together with other potential indications of strong $CP$-violation, has not been observed to date. Since in first place, there is no reason to prefer $\theta=0$ (or an integer multiple of $\pi$), it is therefore argued that the absence of such signals constitutes a shortcoming of the theory, referred to as the strong $CP$ problem, and that it requires an extension of the Standard Model of particle physics. Theoretical research in this direction is extensive, and there is a number of experiments hunting for a proposed particle, the axion, that arises in many of these extensions~\cite{10.1093/ptep/ptaa104}.

From the Lagrangian, the action follows by integration over the spacetime. Since the $CP$-odd term~(\ref{topologcical:term}) turns out to be a total derivative, the corresponding contribution to the action is determined by the boundary conditions on the gauge fields. Taking these to be vanishing physical fields, i.e. pure gauge configurations, at the boundary of spacetime, the integrals over the $CP$-odd term yield $\theta$ times integer values $\Delta n$---{to be referred to as winding number or topological charge}---corresponding to so-called homotopy classes that categorize maps of a three-dimensional sphere onto itself, where maps in different classes cannot be continuously transformed into one another~\cite{Callan:1976je,Jackiw:1976pf}.

This topological quantization is of central relevance when evaluating the effects from the term~(\ref{topologcical:term}). One implication is, for example, that if the predictions of the theory depend on $\theta$, they must be periodic in this parameter. This is because in the quantized theory, the action enters the path integral as a phase. The theory is therefore invariant under replacements $\theta\to\theta+2\pi n$, where $n\in \mathbbm Z$. Therefore, $\theta$ is sometimes referred to as the vacuum angle. Further, topological quantization implies that observables are to be calculated from an interference of amplitudes from different topological sectors, i.e. from path integrals for a given $\Delta n$ or homotopy class, in the infinite spacetime.

To state a principle leading to vanishing physical boundary conditions and therefore to topological quantization, we note that the nonvanishing contributions to the Euclidean path integral arise from saddle points of finite action and fluctuations around these. Saddle points correspond to solutions to the Euclidean equations of motion, and for these to exist in the infinite spacetime volume, the physical boundary conditions must vanish. As a consequence, the path integrals for the different topological sectors must then be evaluated in infinite spacetime volumes first. Otherwise, there would be no reason to assume topological quantization. In a second step, amplitudes from the different topological sectors are then to be interfered.

On the other hand, for boundary conditions imposed on finite spacetime volumes, saddle points and solutions to the equations of motion exist for nonvanishing physical fields at the boundaries as well. Moreover, the ground state configuration, that should determine the boundary conditions on finite spacetime volumes, is neither a field eigenstate nor a pure gauge configuration, i.e. it does not correspond to vanishing physical fields. In contrast, the Euclidean path integral in infinite volumes automatically projects the pure gauge field eigenstates on the corresponding accessible ground states. Nonetheless, if there were a principle that would lead to topological quantization for boundary conditions imposed on some finite surface, one could interfere the topological sectors prior to taking the spacetime volume to infinity.

Here, we show that the material consequence of the order of the limits is as follows: When taking the spacetime volume to infinity before interfering the topological sectors, $CP$-violating phenomena are absent in the strong interactions without extending the theory or setting the $CP$-odd term to zero. On the other hand, interfering the topological sectors before taking the spacetime volume to infinity, one concludes that correlation functions exhibit $CP$-violation that cannot be removed by field redefinitions~\cite{tHooft:1986ooh}.

The question of whether there is $CP$ violation in general in the strong interactions of massive quarks should not be a matter of choice but be a prediction of the theory. Appended to this letter is therefore extensive supplementary material that addresses many aspects of the limiting procedure as well as pertaining matters such as the principle of cluster decomposition.

 Technically, we arrive at our conclusions by computing the correlation functions for massive fermions, where we keep $\theta$ as well as the phase of the determinant of the matrix of quark masses general. As one of the methods, we use the leading approximation to a dilute gas of instantons so that the spacetime-dependence of the correlations can be recovered. As an alternative route, using arguments based on factorization properties of path integrals and the Atiyah-Singer index theorem \cite{Atiyah:1963zz}, we confirm that the coincident limit of the fermion correlations does not exhibit $CP$ violation,  provided the interference of the topological sectors takes place among infinite spacetime volumes. Hence, the main results of this work hold beyond the perturbative expansion about instanton configurations. They crucially rely on how topological quantization emerges in spacetimes of infinite volume and the order in which the pertaining limits are carried out.

\section{Topological charge, massive quarks, and charge-parity violation}

In electrodynamics, the topological term~(\ref{topologcical:term}) is immaterial because its volume integral can be traded for a surface integral over the boundaries of spacetime where it can be shown that finite action configurations have fields decaying fast enough such that the integral vanishes. This is not true for the strong interactions, where, due to the self-interactions, extended field configurations with finite action, so-called instantons, exist while the surface term no longer vanishes~\cite{Belavin:1975fg}. For this reason, it has been proposed that values of $\theta\not= \pi m$ ($m\in \mathbbm Z$) may imply $CP$-violation~\cite{Callan:1976je,Jackiw:1976pf,Baluni:1978rf,Crewther:1979pi}.

While the topological term is local in the first place, and while in singular gauges the topological flux can be constrained to infinitesimal surfaces about the centres of the instantons~\cite{Jackiw:1976dw}, Eq.~\eqref{topologcical:term} is nonetheless equivalent to a surface term at the boundary of the spacetime at infinite distance. It is therefore an essential point  whether it affects local observables in quantum field theory. The standard view is that this is the case because of a change in the local vacuum structure imposed by the boundary term. On the other hand, as illustrated in Figure~\ref{fig:locality}, one can approximate observables by including the fluctuations in a subvolume of the spacetime with all possible boundary conditions on its surface. One may expect---and it is possible to show this---that the theory in the subvolume is then independent of the boundary conditions in the infinite distance so that these have no material impact.

\begin{figure}[h!]
\centering
\includegraphics[width=.24\textwidth]{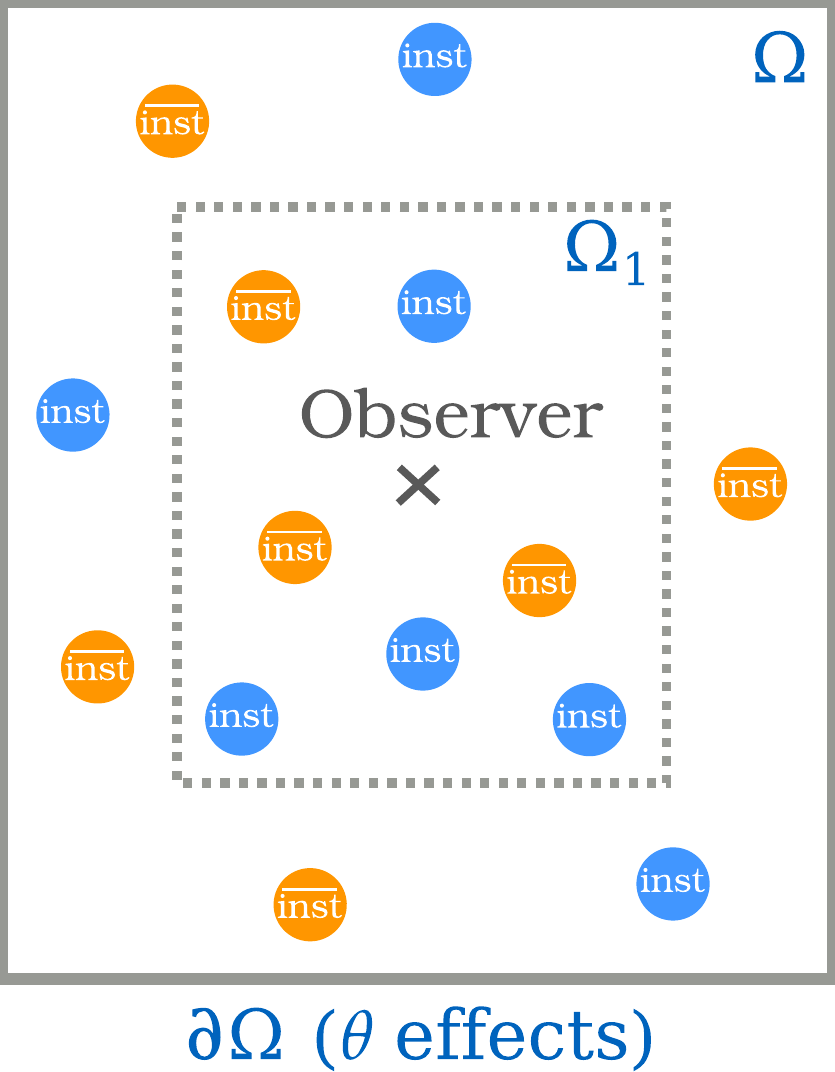}
 \caption{\label{fig:locality}\footnotesize
 For local quantum field theory, an observer is expected to be only sensitive to fluctuations in a local subvolume $\Omega_1\subset\Omega$ in the limit of an infinite volume of the spacetime $\Omega$. The $\theta$-parameter { influences} the conditions at the boundary $\partial\Omega$. It can be shown that these do not affect the fluctuations in the subvolume. Fluctuations
 corresponding to instantons and anti-instantons
 are depicted as  blue and orange circles, respectively.}
 \end{figure}

Intricately related with the topological term are $CP$-odd contributions to quark masses that can be expressed through $\bar\psi_j m_j {\rm e}^{{\rm i}\alpha_j\gamma^5}\psi_j$ where $j=1,\ldots,N_f$ and $N_f$ is the number of quark flavours. The quark fields are denoted by the spinors $\psi_j$, $\gamma^5$ is a matrix in spinor space and the phases $\alpha_j$ are $CP$-odd. The phases $\alpha_j$ can in principle be removed by redefinitions of the quark fields. However, since the so-called chiral symmetry of the quark fields is anomalous~\cite{Bell:1969ts,Adler:1969gk}, the quark phases are tied to the vacuum-angle $\theta$. In particular, $\bar\theta=\theta+\bar\alpha$, where $\bar\alpha=\sum_{j=1}^{N_f}\alpha_j$, is a phase that remains invariant under field redefinitions.

In order to calculate the most important $CP$-violating effects from the topological term, one derives effective fermion interactions caused by the instantons as Lagrangian terms of the form~\cite{tHooft:1976rip,tHooft:1976snw,tHooft:1986ooh}
\begin{align}
\label{digest:eff:operator}
-\Gamma_{N_f} {\rm e}^{{\rm i}\xi} \prod_{j=1}^{N_f}(\bar\psi_j P_{\rm L}\psi_j)-\Gamma_{N_f}{\rm e}^{-{\rm i}\xi}\prod_{j=1}^{N_f}(\bar\psi_j P_{\rm R}\psi_j)\,,
\end{align}
where $\Gamma_{N_f}$ is a coefficient and the left and right chiral projectors are $P_{\rm L,R}=(1\mp\gamma^5)/2$.

The interaction~(\ref{digest:eff:operator}) implies that there is no chiral symmetry with an overall ${\rm U}(1)$ phase. In the effective chiral Lagrangian for low energies, where quantum chromodynamics (QCD) confines, there thus is the corresponding term
\begin{align}\label{digest:eq:LC}
|\lambda| {\rm e}^{-\ic\xi} f_\pi^4\,{\rm det}\, U+|\lambda| {\rm e}^{\ic\xi} f_\pi^4\,{\rm det}\, U^\dagger\,,
\end{align}
where $f_\pi$ is the pion decay constant, $U$ is a field of the form of a unitary matrix describing the mesons and $\lambda$ is a coefficient within the effective theory.

The aforementioned invariance of $\bar\theta$ under field redefinitions leaves two possibilities for the phase $\xi$ compatible with the chiral anomaly (assuming that $\xi$ is a function of $\alpha$ and $\theta$, and that the effective action is periodic in these parameters):
\begin{itemize}
\item
$\xi=\theta$, i.e. in general misaligned with mass terms such that there is $CP$ violation,
\item
$\xi=-\bar\alpha$, i.e. aligned with mass terms such that there is no $CP$ violation.
\end{itemize}
The restriction to the above choices can be understood in terms of a spurious chiral symmetry under which $\theta$ transforms or simply by demanding that the relative phases between the interactions of Eq.~\eqref{digest:eff:operator} and the tree-level mass terms remain invariant under field redefinitions. Based on the topological quantization of the path integral and the ensuing order of limits, we derive here the effective operator~(\ref{digest:eff:operator}) and show that the second possibility, $\xi=-\bar\alpha$, is the one that is realized what implies that there is no $CP$ violation in the strong interactions.

When relating these remarks to the literature, we note that the possibility $\xi=\theta$ is implied in most of the papers without dismissing $\xi=-\bar \alpha$. The early papers, as well as literature following these, on phenomenological $CP$ violation in the strong interactions make use of the freedom of chiral field redefinitions in order to set $\theta=0$ and attribute the $CP$-odd phases to the quark masses~\cite{Baluni:1978rf,Crewther:1979pi}. In the context of the present discussion, this corresponds to setting $\xi=\theta=0$ while $\bar\alpha\not=0$ in general. The case of $\xi=-\bar \alpha$ is apparently not pursued. Also more recent discussions of the coefficients of the operator~(\ref{digest:eq:LC}), e.g. Ref.~\cite{Srednicki:2007qs}, do not mull over this latter possibility.

Reference~\cite{tHooft:1986ooh} appears to contain the only direct calculation leading to $\xi=\theta$, making use of the dilute instanton gas approximation. As we point out in the present work, this conclusion relies on computing the interference among topological sectors in finite spacetime volumes and taking these to infinity afterwards. Reversing this order of limits, as it is indicated when topological quantization emerges from the requirement of finite saddles in the action in infinite spacetimes, we show in the present work that one is led to conclude that $\xi=-\bar \alpha$ instead.

The interactions~(\ref{digest:eq:LC}) are directly related to $CP$-violating observables such as a permanent electric dipole moment of the neutron or the decays $\eta^\prime \to 2\pi$ (Section~\ref{sec:chiral_Lagrangian}). For $\xi=\theta$, one recovers the standard results~\cite{Baluni:1978rf,Crewther:1979pi,Cheng:1987gp}, while for $\xi=-\bar\alpha$, these signals vanish.

\section{Fermion correlations in a dilute instanton gas}

In this section we show that $\xi=-\bar{\alpha}$ by computing the quark correlation function in the approximation of a dilute instanton gas.
In order to simplify notation, we set $N_f=1$ and drop the index for the quark flavour. One should keep in mind that for a single quark flavour, the instanton effects amount to an addition to the quark mass. However, the generalization to the cases with $N_f>1$ relevant for the potentially $CP$-violating phenomenology follows along the lines of the simplified analysis.

\begin{widetext}
To compute the correlation function, we use the following Green's function in the background of $n$ instantons {(with topological charge $+1$)} and $\bar n$ anti-instantons  {(with topological charge $-1$)} located at $x_{0,\nu}$, $x_{0,\bar \nu}$ respectively:
\begin{align}
\label{digest:eq:proptotal}
 {\rm i}S_{n,\bar n}(x,x^{\prime})\approx{\rm i} S_{\rm 0inst}(x,x^{\prime}){+}\sum_{\bar\nu=1}^{\bar n}\frac{\varphi_{0{\rm L}}(x-x_{0,\bar\nu}){\varphi^\dagger_{0{\rm L}}}(x^{\prime}-x_{0,\bar \nu})}{m {\rm e}^{-{\rm i}\alpha}}{+}\sum_{\nu=1}^n\frac{\varphi_{0{\rm R}}(x-x_{0,\nu}){\varphi^\dagger_{0{\rm R}}}(x^{\prime}-x_{0,\nu})}{m {\rm e}^{{\rm i}\alpha}}\,.
\end{align}
This approximation is valid for a dilute instanton gas and quark masses such that $m$ is small compared to $1/\varrho$, where $\varrho$ is the radius of the instantons (which is not fixed). The spinors $\varphi_{0{\rm L,R}}$ are the analytic continuation of the zero modes of the Euclidean Dirac operator in the (anti-)instanton background, that determines the equation of motion for the quark fields, in the massless limit, and
\begin{align}
\label{digest:S:0inst}
{\rm i}S_{0\rm inst}(x,x^{\prime})=(-\gamma^{\mu}\partial_\mu+{\rm i}m {
\rm e}^{-{\rm i}\alpha \gamma^5})\int\frac{{\rm d}^4p}{(2\pi)^4} {\rm e}^{-{\rm i}p(x-x^{\prime})}\frac{1}{p^2-m^2+{\rm i}\epsilon}
\end{align}
is the solution in a background without instantons and is approximately valid at large distances from the individual locations, i.e. in between the instantons and anti-instantons.
\end{widetext}

Further, we readily assume here Minkowski metric. The approximation~(\ref{digest:eq:proptotal}) has been used e.g. in Ref.~\cite{Diakonov:1985eg} for $\alpha=0$. The generalization to $\alpha\not=0$ may appear obvious but there are some complications when transforming the spectrum of the Dirac operator from Euclidean to Minkowski spacetime. Yet, these can be addressed in detail thus confirming the form of the propagator~(\ref{digest:eq:proptotal}) (Section~\ref{sec:Green's}).
Note that the Green's function~(\ref{digest:eq:proptotal}) is independent of $\theta$ because the topological term has not yet entered the derivation. However, it needs to be taken into account when summing configurations corresponding to different homotopy classes in the path-integral expression for the correlation function.

Here, we use the saddle point approximation to the path integral, where we sum over all instanton and anti-instanton numbers $n$ and $\bar n$ and  integrate over the locations of instantons and anti-instantons as well as over the remaining collective coordinates such as the radii $\varrho$ and gauge orientations (which are independent for each instanton and anti-instanton).

The question of whether $\xi=-\alpha$ or $\xi=\theta$ is decided by the treatment of the summation over $n$ and $\bar n$ in conjunction with how boundary conditions are imposed on the path integral. Let $\Omega$ denote the volume of spacetime and first consider Minkowski space such that $\Omega$ is infinite. The case of finite $\Omega$ is discussed below. Boundary conditions on the path integral are fixed by requiring that the physical gauge fields (as well as all other fields) vanish on the boundary of $\partial\Omega$, such that the action takes finite values at its saddle points~\cite{Coleman:1985rnk} (Section~\ref{sec:bc}). For the gauge field, this leaves open the possibility of pure gauge configurations.

These remarks apply to field configurations that are regular in $\Omega$. In calculations aiming for interactions beyond the dilute instanton gas~\cite{Callan:1977gz,Diakonov:1985eg}, it can be advantageous to use the singular gauge~\cite{Jackiw:1976dw} so that one avoids working with integrands that are not manifestly square integrable. The price to pay for this is that there are singularities at the centres of the instantons or their approximate deformations. While spacetime needs to be punctured at these singularities, there are no apparent problems in constructing saddle point approximations to the path integral. Since the singular contributions at the centres of the instantons are pure gauges, the topological flux through an infinitesimal ball around such a point is again quantized. Then, for infinite $\Omega$ the fields still must vanish on $\partial \Omega$ but there is no topological flux through $\partial \Omega$, in contrast to the regular gauge. In effect, topological quantization results from the requirement of finite saddle point configurations in infinite spacetime volumes also in the singular gauge. In contrast, when restricting spacetime to finite $\Omega$, there are finite saddle points for arbitrary nonsingular boundary conditions on $\partial \Omega$. Hence, some different principle would again be necessary to impose topological quantization for finite boundaries.

Both $\partial\Omega$ and ${\rm SU}(2)\subset{\rm SU(3)}$ (i.e. the subgroup of the group of gauge symmetries of the strong interactions) are homeomorphic to the three-dimensional sphere $S^3$ such that the gauge configurations fall into classes according to the third homotopy group. These characterize the number of times $\Delta n$ a three-dimensional hypersurface can be wrapped around $S^3$. In the context of strong interactions, the class of configurations with boundary conditions corresponding to a certain $\Delta n$ are sometimes referred to as a topological sector.

This property is of relevance for the present case because in the saddle point approximation $\Delta n=n-\bar n$.  Furthermore, it is possible to define { vacuum} states $|n_{\rm CS}\rangle$ with a certain { integer} Chern--Simons number $n_{\rm CS}$. Taking the matrix element characterized by $\langle m_{\rm CS}|$ and $|n_{\rm CS}\rangle$ corresponds to fixing the topological sector $\Delta n=m_{\rm CS}-n_{\rm CS}$. We also note that the states $|n_{\rm CS}\rangle$ are not gauge invariant as the Chern--Simons number (defined on a {spatial} hypersurface) can change by all possible integer values through so-called large gauge transformations that are not continuously connected to the identity component. Thus, the true vacuum state should be constructed as a superposition of all Chern--Simons numbers of equal weight, but there may be relative phases proportional to $n_{\rm CS}$. These phases are effectively equivalent with the topological term in the action when calculating expectation values using the path integral approach. Since different topological sectors are distinguished by the boundary conditions which are taken at infinity, contributions to the path integral within a fixed topological sector must be evaluated for infinite spacetime volumes $\Omega$. Note that this reasoning  also applies to spacetime manifolds with compact spatial hypersurfaces yet with an infinite time direction. The possibility of restricting the integration to finite subvolumes of spacetime is discussed below.

\begin{widetext}
The fermion correlator should therefore be evaluated as
\begin{align}
 \langle\psi(x)\bar\psi(x')\rangle
 =&\,\,\lim_{N\to\infty \atop N\in \mathbbm N} \lim_{\Omega\to\infty}\frac{1}{Z(N,\Omega)}\sum_{{m_{\rm CS},n_{\rm CS}}\atop{|m_{\rm CS}-n_{\rm CS}|\leq N}}\,{}\langle m_{\rm CS}|\psi(x)\bar\psi(x')|n_{\rm CS}\rangle
 \notag\\
 =&\!\lim_{N\to\infty \atop N\in \mathbbm N} \lim_{\Omega \to\infty}\frac{\sum_{\Delta n=-N}^{N}\sum_{n}{}\langle n_{\rm CS}+\Delta n|\psi(x)\bar\psi(x')| n_{\rm CS}\rangle}{\sum_{\Delta n=-N}^{N} Z_{\Delta n}(\Omega)}
 =\!\lim_{N\to\infty \atop N\in \mathbbm N} \lim_{\Omega\to\infty}\!\!\frac{\sum\limits_{\Delta n=-N}^{N}\langle \psi(x) \bar\psi(x^\prime)\rangle_{\!\Delta n}}{\sum_{\Delta n=-N}^{N} Z_{\Delta n}(\Omega)}
\label{digest:correlation:limits}
\,.
\end{align}
where $Z(N,\Omega)$ and $Z_{\Delta n}(\Omega)$ are the partition function summed for all sectors $|\Delta n|\leq N$ and that for a single topological sector, respectively. The dependence on $N$ and $\Omega$ needs to be kept before taking these parameters to infinity. The order of the two limits in the last expression determines whether one arrives at $\xi=-\alpha$ or $\xi=\theta$, as we discuss next.

Now we need to consider the fermion correlator in a fixed topological sector. { For a single flavour one has:}
\begin{align}
&\langle \psi(x) \bar\psi(x^\prime)\rangle_{\!\Delta n}
=\!\!\!\!\!\!
\sum\limits_{\bar n,n\geq 0 \atop n-\bar n=\Delta n}\!\!\!\!\!\!\frac{1}{\bar n! n!}
\Big[
{\,\bar h(x,x^\prime)}\left(\frac{\bar n}{m {\rm e}^{-{\rm i}\alpha}} P_{\rm L}+\frac{n}{ m {\rm e}^{{\rm i}\alpha} }P_{\rm R}\right) \Omega^{\bar n+n -1}
+{\rm i} S_{0{\rm inst}}(x,x^\prime) \Omega^{\bar n+n}
\Big]\times
(-{\rm i}\kappa)^{\bar n +n}
{\rm e}^{{\rm i}\Delta n(\alpha + \theta)}
\notag\\
=&\left[\!\left({\rm e}^{{\rm i}\alpha}\! I_{\Delta n+1}(2{\rm i}\kappa \Omega)P_{\rm L}+{\rm e}^{\!\!-{\rm i}\alpha}\! I_{\Delta n-1}(2{\rm i}\kappa \Omega) P_{\rm R}\right)\frac{{\rm i}\kappa}{m}\,\bar h(x,x^\prime)
+I_{\Delta n}(2{\rm i}\kappa \Omega){\rm i} S_{0{\rm inst}}(x,x^\prime)\!\right]\times
(-1)^{\Delta n}
{\rm e}^{{\rm i}\Delta n(\alpha + \theta)}\,.
\label{digest:correlation:fixed}
\end{align}
In this expression, $\bar h(x,x^\prime)$ is a spinor correlation that remains after the integration of the instanton and anti-instanton locations as well as the collective coordinates and $\kappa$ includes the exponential suppression of the instanton action---as these correspond to tunneling processes---as well as extra factors that appear when evaluating the path integral to one-loop accuracy (Section~\ref{sec:path:int:fixed:top}). Finally, $I_\alpha(x)$ is the modified Bessel function of order $\alpha$.
\end{widetext}
\begin{figure}
\centering
\includegraphics[width=.48\textwidth]{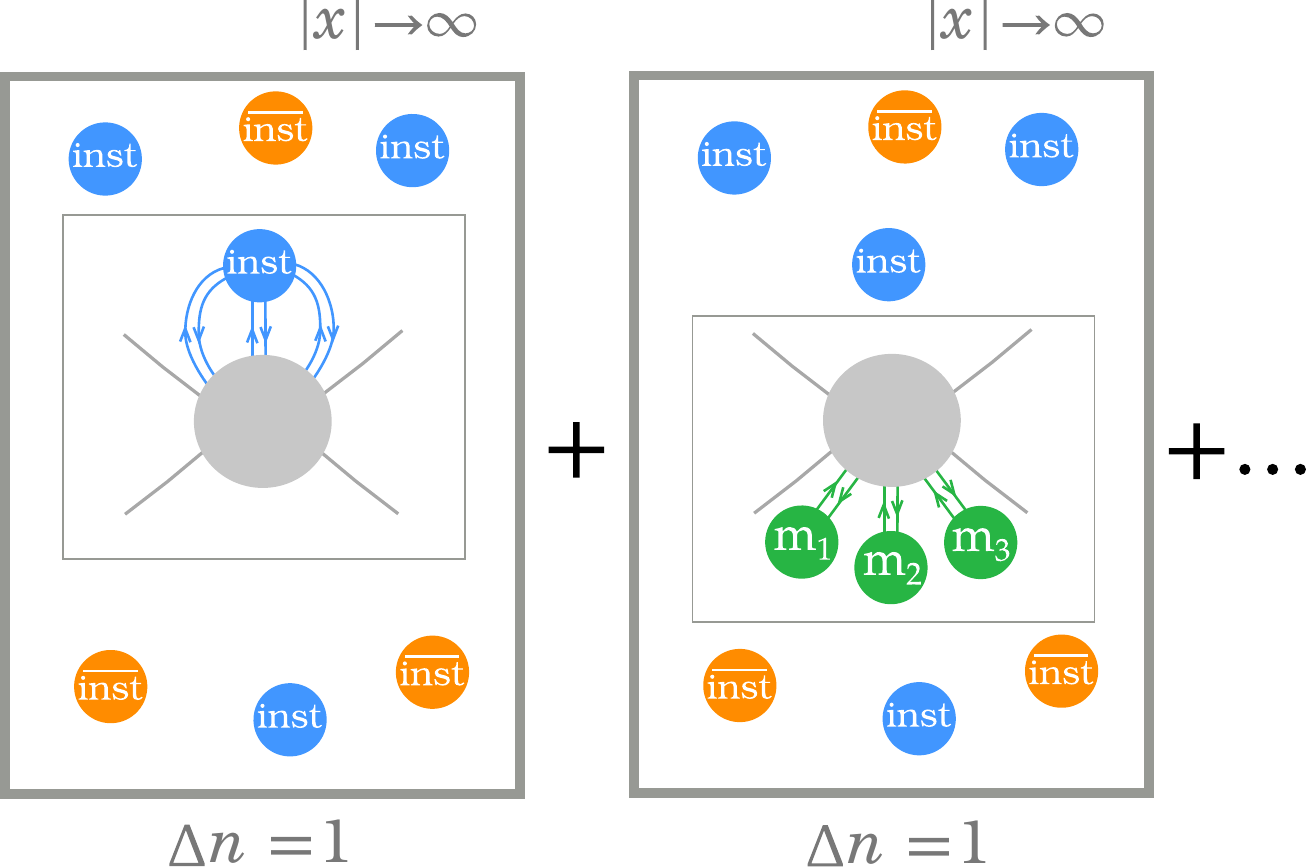}
\caption{\label{fig:phase:attribution}
\footnotesize
Schematically shown are two contributions to a four-point correlation function in some multi-instanton background. The shaded blobs represent some subdiagram. On the left, there is a piece induced by a six-point fermion Green's function in the background of an instanton (corresponding to $\bar h$ in the two-point case). On the right, an interaction of the same chiral structure is induced by the fermion mass terms $m_{1,2,3}$(corresponding to ${\rm i}S_{0{\rm inst}}$ in the two-point case).  When integrating over the subvolumes indicated by the thin grey boxes only, the left piece would acquire a relatively misaligned phase $\theta+\alpha$ ($\alpha$ being here the sum of the quark mass phases) compared to the right piece because the phases come from the topological term and the fermion determinants. When instead correctly computing the path integral over the full spacetime volume (represented by the thick grey boundaries), the phase for both pieces is aligned and given by $\Delta n (\theta+\alpha)$. For infinite spacetime volumes, the interferences between the different sectors $\Delta n$ moreover are immaterial.
}
\end{figure}

It is clear that the dilute instanton gas approximation does not apply directly to QCD. Rather, one could think of a nonabelian gauge theory whose particle content is made up such that the running coupling remains perturbative in the infrared and there is asymptotic freedom in the ultraviolet. In such a model, the scale invariance is broken radiatively such that there is no dilatational modulus and instead a preferred instanton size. That the symmetry properties with respect to $CP$ of such a theory in principle also apply to QCD should therefore be taken as a more or less plausible assumption. In Section~\ref{sec:digest:indexthrm}, we thus also present a derivation of the coincident fermion correlations that does not rely on the dilute instanton gas approximation.

The volume factors $\Omega$ in Eq.~(\ref{digest:correlation:fixed}) are resulting here from the integration of the instanton locations over the entire spacetime. These appear in the same form even when taking these volumes to be finite for a given topological sector before interfering between these~\cite{Callan:1977gz,Diakonov:1985eg,tHooft:1986ooh}. It is then understood that $\Omega$, which is taken to infinity after interfering the topological sectors, is much larger than other scales that appear in the dilute instanton gas. This includes the mean separation between instantons and anti-instantons as well the typical size of these. In fact, restricting $\Omega$ to small volumes given by some physical length scale so that these only contain few instantons would substantially alter the results of e.g. Refs.~\cite{Callan:1977gz,Diakonov:1985eg,tHooft:1986ooh} that do not impose such truncations on the path integral. The fact that the instanton locations are to be integrated over the entire spacetime is tied to translational invariance and mathematically derives from trading the translational moduli for collective coordinates~\cite{tHooft:1976snw,Bernard:1979qt}. It can also be seen in analogy with the calculation of the partition function for a classical ideal gas, where the individual positions of the particles are integrated over the entire configuration space. Beyond the dilute gas approximation, the spacetime integrations should be modified to account for the overlap of instantons and anti-instantons due to their finite size while yet, the individual locations are still to be integrated over infinite volumes~\cite{Diakonov:1985eg}. For a theory to which the dilute gas approximation applies, omitting such corrections only amounts to a controllable error.

From Eq.~(\ref{digest:correlation:fixed}), we see explicitly that in a fixed topological sector and large spacetime volumes $\Omega$, the modulus of the coefficients of the left and right chiral contributions tends to the same value. In particular, for $x\to\infty$ and $|\arg(x)|<\pi/2$, $I_\alpha(x)\sim\exp(x)/\sqrt{2\pi x}$, i.e. these functions become independent of their index. As a consequence, for $\Omega\to\infty$, all topological sectors contribute in precisely the same way.  Moreover, the chiral phases from the mass term {contained in $S_{0{\rm inst}}$ (see Eq.~\eqref{digest:S:0inst})} and those induced by instanton effects are aligned, as a consequence of these phases (that originate from the fermion determinants and the topological term) being fixed by the boundary conditions on the topological sector $\Delta n$ as we illustrate in Figure~\ref{fig:phase:attribution}. When normalizing by the partition function, the modified Bessel functions as well as the phase proportional to $\Delta n$ cancel and we obtain
\begin{align}
\label{digest:correlation}
 \langle\psi(x)\bar\psi(x')\rangle
={\rm i}S_{0\text{inst}}(x,x^\prime)
+\frac{{\rm i}\kappa}{m}
\bar h(x,x^\prime) {\rm e}^{-{\rm i}\alpha\gamma^5}\,,
\end{align}
such that the explicit phase can be identified with $\xi=-\alpha$. In contrast, if we were turning around the order of limits in Eq.~(\ref{digest:correlation:limits}), we would sum over two independent exponential series for $n$ and $\bar n$ and find $\theta$ rather than $-\alpha$ in Eq.~(\ref{digest:correlation}) so that $\xi=\theta$ (Section~\ref{subsec:summation}).

Taking the correct order of limits, i.e. $\Omega\to\infty$ before summing over topological sectors therefore explains the absence of $CP$ violation in the strong interactions.
This result can be generalized to an arbitrary number of fermion flavours (Section~\ref{sec:several:flavours}).

\section{Chiral correlations from the index theorem}
\label{sec:digest:indexthrm}
In this section we provide an alternative derivation of the previous results without using instantons. The starting point are the factorization properties of the path integration when the full spacetime volume $\Omega$ is divided into subvolumes $\Omega_1$ and $\Omega_2$.
Following standard textbook arguments used in the context of cluster decomposition \cite{Weinberg:book:vol2:1996}, the fact that the topological charge $\Delta n$ is a surface flux allows to write the partition function of the full spacetime volume $\Omega=\Omega_1\cup\Omega_2$ as
\begin{align}\label{eq:Zcluster}
 Z_{\Delta n}(\Omega)=\sum_{\Delta n_1=-\infty}^\infty Z_{{\Delta n}_1}(\Omega_1) Z_{{\Delta n}-{\Delta n}_1}(\Omega_2).
\end{align}
For convenience, in this section we work in Euclidean space, as this simplifies the tracking of the complex phases. First we can extract the $\theta$-dependent phase,  $Z_{\Delta n}(\Omega)\propto {\rm e}^{\ic \Delta n\theta}$. Any additional complex phases can only come from the integration over fermionic fluctuations. To leading order in a loop expansion around saddle points, these integrations have the form of determinants of the Dirac operator in each saddle-point background. Here we make no approximation of the saddle points in terms of a dilute instanton gas.

Parity transformations relate pairs of eigenfunctions of the massive  Dirac operator with mutually conjugate eigenvalues, except for those eigenfunctions that, being zero modes of the massless operator, have eigenvalues given by the complex fermion masses, resulting in opposite phases for right-handed and left-handed modes (Section~\ref{sec:cplx:mass:Eucl}). Hence the phase of the full determinant within a topological class characterized by $\Delta n$ is determined by the difference between the number of right and left-handed zero modes of the massless Dirac operator,  which according to the Atiyah-Singer index theorem coincides with $\Delta n$ \cite{Atiyah:1963zz}. This gives a phase of ${\rm e}^{\ic \Delta n\bar\alpha}$ for the product of all fermion determinants. As a consequence, we may write
\begin{align}\label{eq:Ztildeg}
 Z_{\Delta n}(\Omega) =  {\rm e}^{\ic \Delta n\bar\theta}\tilde g_{\Delta n}(\Omega)
\end{align}
with real $\tilde g_{\Delta n}(\Omega)$. Equation~\eqref{eq:Zcluster} gives then the relations
\begin{align}\label{eq:clustertilde0}
 \tilde g_{{\Delta n}}(\Omega_1+\Omega_2)=\!\!\sum_{{\Delta n}_1=-\infty}^\infty\tilde g_{{\Delta n}_1}(\Omega_1) \tilde g_{{\Delta n}-{\Delta n}_1}(\Omega_2).
\end{align}

Setting $\Omega_i=0$ above can be seen to imply that $\tilde g_{\Delta n}(0) = \delta_{\Delta n,0}$. We next note that parity transformations relate $\Delta n$ with $-\Delta n$. As the $\tilde g_{\Delta n}$ are real and not sensitive to parity-violating effects from the complex fermion masses, one has $\tilde g_{-\Delta n}(\Omega)=\tilde g_{\Delta n}(\Omega)$. The former results motivate the Ansatz
\begin{align}
\label{eq:Ansatz0}
 \tilde g_{\Delta n}(\Omega)=\Omega^{|{\Delta n}|} f_{|{\Delta n}|}(\Omega^2), \quad f_{|{\Delta n}|}(0)\neq0.
\end{align}
Remarkably, assuming analyticity in $\Omega$ (and as shown in Section~\ref{sec:CDC1}), there is a unique solution which, upon substitution in Eq.~\eqref{eq:Ztildeg}, gives
\begin{align}\label{eq:Zbeta0}
 Z_{\Delta n}(\Omega) =  I_{\Delta n}(2\beta\Omega)   \,{\rm e}^{\ic \bar\theta\Delta n}\,,
\end{align}
where $\beta$ depends on the parameters of the theory and is not determined at the present level of generality.
This has the same form as the result for the partition function in the dilute gas approximation (Section~\ref{sec:path:int:fixed:top}).

Finally we note that since all dependence on the complex fermion masses is included in $\bar\theta$, $\beta$ can only depend on the moduli of the complex fermion masses $\mathfrak{m}_j\equiv m_j {\rm e}^{\ic \alpha_j}$: $\beta=\beta(\mathfrak{m}_j\mathfrak{m}_j^*)$.  In order to obtain fermion correlators, it suffices to note that $\mathfrak{m}_j$ and  $\mathfrak{m}_j^*$ can be seen as sources for integrated two-point functions. Within a fixed topological sector $\Delta n$, the volume averages of the fermionic correlators can be obtained as
\begin{align}\begin{aligned}
 \frac{1}{
 \Omega}\int {\rm d}^4 x\,\langle \bar\psi_i P_{\rm R} \psi_i \rangle_{\Delta n}=&\,-\frac{1}{
 \Omega}\frac{\partial}{\partial \mc_i}Z_{\Delta n}, \\
 \frac{1}{
 \Omega}\int {\rm d}^4 x\,\langle \bar\psi_i P_{\rm L} \psi_i \rangle_{\Delta n}=&\,-\frac{1}{
 \Omega}\frac{\partial}{\partial \mc^*_i}Z_{\Delta n}.
\end{aligned}\end{align}
Using Eq.~\eqref{eq:Zbeta0} and summing over topological sectors after taking the limit $\Omega\rightarrow\infty$ as before gives
correlators whose phases are aligned with the tree-level masses, leading to no CP violation:
\begin{align}\begin{aligned}
  \frac{1}{\Omega}\int d^4 x\,\langle \bar\psi_i P_{\rm R} \psi_i \rangle = {-}2\mc^*_i\,\partial_{\mc_i \mc^*_i} \beta(\mc_k \mc^*_k),\\
  \frac{1}{\Omega}\int d^4 x\,\langle \bar\psi_i P_{L} \psi_i \rangle ={-}2\mc_i\,\partial_{\mc_i \mc^*_i} \beta(\mc_k \mc^*_k).
\end{aligned}\end{align}
By taking additional derivatives with respect to the masses $\mathfrak{m}_j$,$\mathfrak{m}_j^*$, the results can be extended to correlation functions involving more fermion fields.

\section{Finite subvolumes, periodic boundary conditions and fixed topological sectors}

To view this result from additional angles, we discuss what one would obtain for fixed topological sectors or for finite spacetime volumes. Taking the order of the limits as in Eq.~(\ref{digest:correlation:limits}), we have seen that the modified Bessel functions in Eq.~(\ref{digest:correlation:fixed}) tend to a common limit. This can be seen as a consequence of $\Delta n/\Omega\to 0$. Taking $\Omega\rightarrow \infty$ before summing over different topological sectors may therefore be viewed to be equivalent with setting $\Delta n=0$ from the outset. This explains why taking limits as in Eq.~(\ref{digest:correlation:limits}) leads to the alignment between the various chiral phases. {We note that a relevant example for finite $\Omega$ and fixed $\Delta n$ is given by boundary conditions that are periodic in all four dimensions. This setup is mostly chosen in lattice simulations, where $\Delta n$ freezes in the continuum limit.}

In the approximation of the dilute instanton gas, it can be shown that fixing $\Delta n$ in an infinite spacetime volume is compliant with the principle of cluster decomposition (Section~\ref{sec:clustering:inf:vol}). In finite spacetime volumes $\Omega$, corrections to the asymptotic form of correlators required by the cluster decomposition principle then vanish, provided $\Omega$ is chosen large enough to meet a given precision (Section~\ref{sec:cluster_decomposition_finite_VT}). This observation has also been made in Refs.~\cite{Brower:2003yx,Aoki:2007ka} through different calculational methods. We therefore conclude that it is possible to describe the strong interactions in a fixed sector with finite $\Delta n$, provided $\Omega$ is large enough or infinite, and that there are no $CP$-violating effects in this theory.

With the above observation and working in a single topological sector with fixed $\Delta n$, we can evaluate the path integral in a finite subvolume $\Omega_1\subset\Omega$ according to Figure~\ref{fig:locality}, no matter whether the full spacetime volume is finite or infinite.
For such a setup, we need to sum or integrate over boundary conditions of a certain winding number $\Delta n_1$ (which is not necessarily integer because instantons can be located at the boundary). The full winding number $\Delta n$ is however fixed by the boundary conditions on $\partial \Omega$. In particular, let $\Omega_2=\Omega\setminus\Omega_1$ and $\Delta n_2$ be the winding number within $\Omega_2$. Then, $\Delta n=\Delta n_1+\Delta n_2$ remains fixed such that the total phase proportional to $\Delta n$ separates just like in Eq.~(\ref{digest:correlation:fixed}) and cancels within observables. One can then obtain expectation values from a path integration restricted to $\Omega_1$ in which the $\theta$ dependence is absent, and once more the result~(\ref{digest:correlation}) is recovered (Sections~\ref{sec:clustering:inf:vol} and~\ref{sec:cluster_decomposition_finite_VT}).


We emphasize that the fermion correlations evaluated according to Eq.~\eqref{digest:correlation:limits} are compatible with the enhanced mass of the $\eta'$-meson compared to those mesons associated with spontaneously broken symmetries that are not anomalous (Section~\ref{sec:chiral_Lagrangian}). This can be explained in more detail when observing that the chiral susceptibility evaluated in finite subvolumes of spacetime agrees with known results from the dilute instanton gas approximation and moreover when noting that even within a fixed topological sector, there is an $\eta'$-meson with enhanced mass (Section~\ref{subsec:suceptibility}).  Then one can also show that under reasonable assumptions the mass of the $\eta'$ is proportional to the topological susceptibility of the pure gauge theory evaluated in finite subvolumes, which generalizes classic results derived for large numbers of colours in  Refs.~\cite{Witten:1979vv,Veneziano:1979ec}. Finally, we note that arguments linking the topological susceptibility with $CP$ violation~\cite{Shifman:1979if} rely on assuming analyticity in $\theta$ of the partition function for the full volume, which does not apply when the infinite volume limit is taken before summing over the topological sectors (Sections~\ref{sec:chiral_Lagrangian} and \ref{subsec:suceptibility}).

\section{Conclusions}

In this work, we have derived fermion correlations in instanton backgrounds, investigated the cases of finite and infinite spacetime volumes and checked the compliance with cluster decomposition. If there were a valid principle that would allow the limit of infinite spacetime volume to be taken after the summation over topological sectors, we would recover $CP$-violating correlations proportional to the rephasing-invariant parameter $\bar \theta$. However, based on the reasoning that the quantization of the topological sectors comes from the fact that the path integral receives its nonvanishing contributions from saddle points of finite action and fluctuations about these, boundary conditions in Euclidean space should be imposed at infinity before the summation over topological sectors. The conclusion then is that the theory of strong interactions with massive fermions does not predict $CP$-violating phenomena, irrespective of the value of $\bar\theta$.

}
\fi

\sectionfont{\Large}

\ifarXiv{}
\else
{
\acknos
}
\fi

}
\fi

\onecolumn

\ifarXiv
\else
\noindent{\huge\bf Supplementary material}

\renewcommand\thesection{S\arabic{section}}
\setcounter{equation}{0}
\setcounter{section}{0}
\renewcommand\theequation{S\arabic{equation}}

\section{Outline}

We present here the technical details that corroborate the statements made in the main text.
\fi

The anomalous violation of chiral fermion number through instanton and sphaleron transitions is a characteristic feature of the strong interactions, and for the weak interactions, it is likely to be of key importance for the generation of the baryon asymmetry of the Universe~\cite{Bell:1969ts,Adler:1969gk,Belavin:1975fg,tHooft:1976rip,tHooft:1976snw,Klinkhamer:1984di,Kuzmin:1985mm}. Upon the
discovery of the Belavin-Polyakov-Schwartz-Tyupkin (BPST) instanton~\cite{Belavin:1975fg},  it was soon realized by 't~Hooft that these instanton solutions can also solve the axial ${\rm U}(1)$ problem~\cite{Weinberg:1975ui}, which queries why
there is no pseudo-Goldstone boson associated with flavour-diagonal chiral rephasings---the
$\eta'$ is much heavier than the mesons in the octet. Although the Adler-Bell-Jackiw (ABJ) anomaly~\cite{Bell:1969ts,Adler:1969gk} implies that the axial ${\rm U}(1)$ current is not conserved, it was believed for a while that the anomalous term vanishes when integrated over the whole spacetime because it is a total derivative. However, for the BPST instanton, the anomaly turns out to be nonvanishing globally, thus providing extra breaking for the axial ${\rm U}(1)$ symmetry and giving rise to the splitting of $\eta'$ from the meson octet. The violation of chiral fermion number induced by instantons is typically suppressed by the tunneling exponent. At finite temperature, it is however possible to have thermal transitions instead of tunneling. These are described by the sphaleron, i.e. an unstable saddle point of the energy functional for the gauge fields~\cite{Klinkhamer:1984di}.

In the context of thermal field theory and since the instanton corresponds to a Euclidean saddle point solution, calculations are typically carried out using imaginary time. Nonetheless, some of the main phenomenological applications are within scattering theory or kinetic theory such that it is necessary to transfer the results to the real time of Minkowski space. This is generally possible through the analytic continuation of Green's functions. Nonetheless, it remains of interest to achieve a formulation directly in Minkowski spacetime because it would allow for a first-principle derivation of kinetic theory involving instantons, e.g. in the Schwinger-Keldysh formalism~\cite{Schwinger:1960qe,Keldysh:1964ud}, or a more systematic treatment of fermions that are not of the Dirac type, e.g. in chiral gauge theories. A real-time approach would also serve as a check for the correct interpretation of the analytically continued quantities. In view of this, we also
discuss in this paper some details on the correlation functions in Minkowski spacetime.

Real-time calculations are typically only feasible when expanding about a saddle point of the action. However, there is no saddle for the action in  Minkowski spacetime that would correspond to an instanton configuration. The saddle is recovered when extending the path integral over the degrees of freedom of the bosonic fields into the complex plane and deforming the integration contour. Convergent integration contours that go through the saddle of interest can be found using the Picard--Lefschetz theory~\cite{Witten:2010cx} which has led to a number of applications and further developments, for instance, in Refs.~\cite{Tanizaki:2014xba,Cherman:2014sba,Andreassen:2016cvx,Ai:2019fri,Mou:2019gyl,Mou:2019tck}. Effects from
the chiral anomaly for real background fields in Minkowksi space are calculated
e.g. in Refs.~\cite{Nielsen:1983rb,Domcke:2018gfr,Domcke:2019qmm}.

It is advantageous to derive the Green's function for fermions from a spectral sum, this way the contribution of modes that account for the chiral anomaly, i.e. the
zero modes in the massless limit, is readily isolated~\cite{tHooft:1976rip,tHooft:1976snw,Shifman:1979uw,tHooft:1986ooh}. Given the spectrum of the massless Dirac operator in the instanton
background, this construction is straightforward for the case of a real mass term in Euclidean space. Assuming the mass acts as a perturbation to the eigenspectrum, it is also obvious how to insert a complex mass into the zero-mode contribution to the Green's function. In Section~\ref{sec:continuation}, we therefore note this result along with some well-known generalities about analytic continuation of the problem. We focus for simplicity on setups with Dirac fermions in the fundamental representation of the gauge group, as in quantum chromodynamics (QCD). It is less clear how to construct the spectral sum in Euclidean space in the presence of a complex mass that cannot be treated as a small perturbation. This is because of the occurrence of $\gamma^5$; the complex mass term is not proportional to an identity matrix. In Section~\ref{sec:cplx:mass:Eucl}, we show that the spectral sum can be built in terms of the eigenfunctions of the massless Dirac operator after an additional orthogonal transformation among the pairs of modes with opposite eigenvalues. As for the eigenmodes, there is a complication in the analytic continuation because the improperly normalizable Euclidean continuum modes will in general not
be normalizable when evaluated in real time~\cite{Ai:2019fri}.
In Section~\ref{sec:cplx:mass:arbitrary}, we therefore discuss in detail
how the spectral decomposition of the Green's function can be continued from
Euclidean to Minkowski space by rotating the temporal coordinate axis by an angle $\vartheta$. This
requires a particular procedure for the continuation of the dual eigenvectors
that we refer to as $\vartheta$-conjugation, and in \ifarXiv\else{Section}\fi~\ref{app:freeprop}, we exemplify this on the Green's function for a Dirac fermion in the homogeneous and isotropic background spacetime.
As a result, in Section~\ref{sec:cplx:mass:Mink}, we then show how the spectral sum can be understood in terms of the eigenmodes of the Dirac operator directly in Minkowski spacetime, which requires discussion because this operator
is non-Hermitian since the analytically continued gauge field configuration of the instanton is complex.

Having reported the results for Green's functions of fermion with complex masses (i.e. nonzero chiral phase) in (anti-)instanton backgrounds, we proceed in Section~\ref{sec:fermi:corr} to derive correlation functions, starting with two-point functions in a theory with a single fermion. The correlation functions do not trivially coincide with the Green's functions because in the path integral, the sum over the number of individual instantons as well as the integral over their locations are yet to be carried out. We observe that for a given number of instantons with positive and negative winding numbers, chiral phases
from the fermion determinant as well as from the $\theta$-vacuum of the gauge theory multiply all structures---left and right chiral contributions as well as pieces corresponding to the homogeneous background between instantons---by the same factor (see Eq.~\eqref{digest:correlation:fixed}). The boundary conditions on the path integral must be chosen such that there are saddle points of finite action~\cite{Coleman:1985rnk}. Some  comments concerning this point in the context of the present work are made in Section~\ref{sec:bc}. Therefore, the physical fields must be vanishing at infinity, which allows pure gauge configurations of the gluon field. This implies the topological quantization of the winding number.
As a consequence, the integration over the infinite spacetime
volume must first be done for configurations with fixed total winding number, as we carry out in Section~\ref{sec:path:int:fixed:top}. The summation over the different winding numbers is then performed subsequently, as shown in Section~\ref{subsec:summation}.
After the summations and integrations, the chiral phase of the mass
term is aligned with the phase associated with the effects from the instantons breaking chiral symmetry. While for simplicity, the derivations are carried out in detail for the case of a single fermion flavour, we subsequently discuss the generalization to the realistic case of several flavours in Section~\ref{sec:several:flavours}. We also show in Section~\ref{sec:general:correlations} how to calculate higher-point correlation functions in theories with several flavours and complex mass terms and demonstrate that again, the $\theta$-angle drops out of the final result. A consequence of these findings is that there are no $CP$-violating effects in the strong interactions. To clarify this, in Section~\ref{sec:chiral_Lagrangian}, we eventually discuss how the chiral phases that we have computed for the fermion correlation function determine couplings in the effective Lagrangian that governs the strong interactions at low energies.

As an alternative to computing the correlation functions from the fluctuations about the ensemble of instantons and anti-instantons, in Section~\ref{sec:CDC1} we constrain the dependence of the partitions $Z_{\Delta n}$ on the spacetime volume and the fermion phases using cluster decomposition and the index theorem. Again, we verify the phase alignment between terms from topological effects and from fermion masses.

Evaluating the contributions from the single topological sectors in the infinite-volume
limit may be viewed as equivalent to fixing the winding number altogether. In
Section~\ref{sec:CDC}, we therefore verify that fixing the topological sector in large
spacetime volumes does not violate the principle of cluster decomposition. We do so by deriving the expectation values from a path integral restricted to a subvolume of the
full spacetime. In addition, we discuss observables such as the density of winding number
and the topological susceptibility for finite or infinite spacetimes with free and fixed
topological sectors.
\ifarXiv
\else{}
\fi

{While with Section~\ref{sec:Green's}, we devote a large part of this \ifarXiv\else{material}\fi{} to the discussion on the analytic continuation between Euclidean and Minkowskian Green's functions and fermionic functional determinants, we note that all of the main conclusions are equally reached when working entirely in Euclidean space. A reader not concerned with the analytic continuation may take the Green's function~\eqref{eq:proptotal} and the ratio of functional determinants~\eqref{eq:MinkowskiDet} as a starting point. Their Euclidean counterparts can be rederived straightforwardly.

\section{Green's function for fermions in a one-instanton background in
Min\-kow\-ski space}
\label{sec:Green's}

\subsection{Analytic continuation of the instanton solutions and fermion
fluctuations between Euclidean and Minkowski space}
\label{sec:continuation}

We discuss here some generalities of the continuation
of the instanton solution, the Dirac operator and its Green's function between Euclidean and Minkowski spacetime.  For definiteness, we consider Dirac fermions in the fundamental representation in the background of $\rm SU(2)$ BPST (anti-)instantons. We construct the fermion Green's function by regulating the divergence from the fermion zero-mode by a mass term with a nonzero chiral phase. While such a phase can straightforwardly be inserted into the well-known results for the Green's function e.g. from Ref.~\cite{Shifman:1979uw}, the explicit discussion of this matter serves us to introduce the general context as well as some notation.

The strong interactions are described by QCD,
where the Euclidean action reads (leaving aside the topological term for the moment)
\begin{align}
S_{\rm E}=\int{\rm d}^4 x\left[\frac{1}{4g^2}F^{\E a}_{mn}F^{\E a}_{mn}+\psi^{\E\dagger}_i\left(\gamma^{\E}_m D^{\E}_m+ M_{ij}\right)\psi^\E_j\right]\,.
\end{align}
The super- and subscripts ``E'' indicate that a quantity is defined in Euclidean space.  Our conventions for the  Euclidean coordinates are such that
\begin{align}\label{eq:Euc4v}
     x^{\E}_m=\{\vec{x},x_4\}\,,
\end{align}
and tensorial quantities are labelled with Latin indices $m,n\dots$, taking values between $1$ and $4$. Contractions of indices are carried out through the metric $\delta_{mn}$. Further, in Euclidean space one does not necessarily need to distinguish upper and lower indices, and we use lower indices throughout (except for $\gamma^5$).

The various quantities in the action are defined as follows. First, $F^{\E a}_{mn}\equiv\partial_m A^{\E a}_n-\partial_n A_{m}^{\E a}+ f^{abc}A^{\E b}_m A^{\E c}_n$ are the components of the field strength tensor and $\widetilde{F}^\E_{mn}\equiv \frac{1}{2}\varepsilon_{mnlq}F^\E_{lq}$ are their Hodge dual ($\varepsilon_{mnlq}$ is the totally anti-symmetric tensor). The Euclidean Dirac operator is $\slashed{D}^\E\equiv\gamma_m^\E D^\E_m$, where $\gamma^\E_m$ are the Dirac matrices in Euclidean space, satisfying $\{\gamma^\E_m,\gamma^\E_n\}=2\delta_{mn}$.
Since chiral fermions are central to the discussion of the ABJ anomaly~\cite{Bell:1969ts,Adler:1969gk} and the strong CP problem, it is convenient to employ the Weyl basis for the Dirac matrices:
\begin{align}
\gamma^\E_m=\begin{pmatrix}
0 & -\ic\sigma^\E_m \\
\ic\bar{\sigma}^\E_m & 0
\end{pmatrix},\quad
\gamma^5=\begin{pmatrix}
-\mathsf{1}_2 & 0\\
0 & \mathsf{1}_2
\end{pmatrix}\,,
\end{align}
where $\sigma^\E_m=\left(\vec{\tau},\ic\mathsf{1}_2\right)$ and $\bar{\sigma}^\E_m=(\vec{\tau},-\ic\mathsf{1}_2)$ with $\mathsf{1}_2$ being the unit $2\times 2$ matrix and $\vec{\tau}^i$ the Pauli matrices.
The covariant derivative takes the form
\begin{align}
D^\E_m\psi^\E_i=\left(\partial_m-\ic A_m^{\E a} T^a\right)\psi^\E_i
\end{align}
when $\psi^\E_i$ lives in the fundamental representation of the gauge group, and
\begin{align}
D^\E_m\psi^\E_i=\partial_m\psi^\E_i-\ic A_m^{\E a}[T^a,\psi^\E_i]
\end{align}
when $\psi^\E_i$ lives in the adjoint representation. Here $T^a\equiv \tau^a/2$ are the generators of the gauge group and satisfy $[T^a,T^b]= \ic f^{abc}T^c$ and ${\rm tr} T^a T^b=\delta^{ab}/2$. {When restricting to the subgroup ${\rm SU}(2)$, the structure constants are $f^{abc}\equiv \varepsilon^{abc}$.}  The subscript $i$ on the fermions is the flavor index, to be distinguished from the spatial vector index.

In four-dimensional Euclidean space the BPST instanton with the collective coordinates corresponding to the location set to zero and with winding number $\eta=+1$ is given in terms of the vector potential {
\begin{align}
\label{A:antiinstanton}
A^{\E a}_m(\vec{x},x_4)=2\eta_{amn}\frac{x_n}{(x^\E)^2+\varrho^2}\,,
\end{align}
}
where the 't Hooft symbols $\eta_{a mn}$ are defined as~\cite{tHooft:1976snw}
\begin{align}
\eta_{a m n}=\begin{cases}\varepsilon_{amn},\quad &m,n=1,2,3 \\ -\delta_{an}, &m=4 \\ \delta_{am}, &n=4 \\ 0, &m=n=4 \end{cases}\,.
\end{align}
The expression for the anti-instanton with $\eta=-1$, which is the parity conjugate of Eq.~\eqref{A:antiinstanton}, is obtained when replacing
$\eta_{amn}\to\bar\eta_{amn}$ where the $\bar{\eta}_{amn}$ differ from $\eta_{amn}$ by a change in the sign of $\delta$.

The continuation of Euclidean time to an arbitrarily rotated time contour is parametrized as  (cf. Ref.~\cite{Ai:2019fri})
\begin{align}
\label{Euclideanvector}
x_4\rightarrow {\rm e}^{-\ic(\vartheta-\frac{\pi}{2})}t\, ,
\end{align}
where $t$ is a real parameter. Then, for $\vartheta=\pi/2$, $t$ is just Euclidean time whereas for $\vartheta=0^+$, it corresponds to Minkowskian time. Here the infinitesimal $0^+$ that regulates  the continuation of the instanton configuration to Minkowski spacetime can be understood as a prescription to ensure that the path integration captures the transition amplitude from the true vacuum state onto itself \cite{Ai:2019fri}. We simply take $0^+$ to be zero whenever it does not play a role. For a fixed value of $\vartheta$ characterizing a choice of time contour, we label the real coordinates of the (time-rotated) spacetime as
\begin{align}
\label{Minkowskivector}
x^\mu=(x^0,\vec{x})=(t,\vec{x})\,,
\end{align}
where Greek indices run from $0$ to $3$.
With this parameterization, all equations of motion as well as their solutions do in general depend on $\vartheta$. The $\vartheta$-dependent instanton solutions for the gauge fields can be simply obtained by performing the substitution of Eq.~\eqref{Euclideanvector} into Eq.~\eqref{A:antiinstanton} or the corresponding Euclidean solution for $\eta=+1$. In particular, the solutions in Minkowski spacetime are obtained
when taking $\vartheta=0^+$. In the following, we clarify when necessary whether we are referring to quantities for general $\vartheta$ or for a particular choice. For the remainder of this section we consider the continuation from Euclidean into Minkowski spacetime, maintaining a superscript ``E'' for Euclidean quantities, and omitting labels for their Minkowskian counterparts.

First, one should note that when recasting expressions in terms of
Minkowskian metric tensors (e.g. $-\delta_{mn}\to \eta_{\mu\nu}\equiv{\rm diag}(1,-1,-1,-1)$) and Dirac matrices, it is natural to define the components $A_\mu$ of the Minkowski gauge field as:
\begin{align}
A_0(x^0,\vec{x})={\rm i}A^{\E}_4(\vec{x},x_4=\ic x^0)\;\,\text{and}\;\,A^{{}}_i(x^0,\vec{x})= A^{\E}_i(\vec{x},x_4=\ic x^0)\;\,\text{for}\;\,i=1,2,3\,.
\end{align}
When expressing
{ $A_\mu=(\tau^a/2)A^a_\mu$},  this implies however that the components $A^{{\rm }a}_\mu$ when evaluated for the $\eta=-1$ instanton solution (Eq.~\eqref{A:antiinstanton})  continued to
$\vartheta=0^+$ are in general complex. Since the physical fields $A^{{\rm }a}_\mu$ are however real, a deformation of the integration contour of the path integral is required in order to capture
the analytically continued solution, which constitutes then a complex saddle point from which  appropriate complex integration contours that lead to well-behaved integrands can be obtained by  means of steepest-descent flows~\cite{Witten:2010cx,Tanizaki:2014xba}. In Ref.~\cite{Ai:2019fri} it is derived how to evaluate the path integration of bosonic fluctuations on the deformed contours using Picard-Lefschetz theory, which would have
to be applied here in order to deal with the fluctuations of the gauge field. The saddle point for the fermion field is still given by the vanishing field configuration, and the path integral of the Gra\ss mannian fermion fluctuations can be carried out as usual.

In chiral representation the Dirac matrices for Minkowski spacetime are given by
\begin{align}
\gamma^{{\rm}0}=\gamma^{\E}_4\;\,\text{and}\;\,\gamma^{{\rm} i}={\rm i}\gamma^{\E}_i\;\,\text{for}\;\,i=1,2,3\,.
\end{align}
Note that the form of $\gamma^5$ is the same for Euclidean and Minkowski space, and it is defined as $\gamma^5={\rm i}\gamma^0\gamma^1\gamma^2\gamma^3=\gamma^\E_1\gamma^\E_2\gamma^\E_3\gamma^\E_4$.
The Minkowskian Dirac operator is then obtained from the Euclidean one by performing the analytic continuation of Eq.~\eqref{Euclideanvector} to $\vartheta=0^+$:
\begin{align}\begin{aligned}
\label{Dirac:continuation}
\slashed D^{\E}=(\slashed\partial^{\E}_m{\color{teal}-{\rm i}}\gamma^{\E}_m A^{\E}_m)\rightarrow&\,
\left(
-{\rm i}\frac{\partial}{\partial x^0}\gamma^{\E}_4
+\vec \gamma^{\E}\cdot\nabla-{\rm i}\gamma^{\E}_4 A^{\E}_4(\vec{x},x_4=\ic x^0) -{\rm i} \vec\gamma^{\E} \cdot \vec A^{\E}(\vec{x},x_4=\ic x^0)\right)\\
=&\,-{\rm i}\left(
\frac{\partial}{\partial x^0}\gamma^{0}
+\vec \gamma\cdot\nabla-{\rm i}\gamma^{0} A_0(x^0,\vec{x}) -{\rm i} \vec\gamma \cdot \vec A(x^0,\vec{x})\right)=-{\rm i}\slashed D
\,,
\end{aligned}\end{align}
where $\vec{\gamma}^{}\cdot\nabla\equiv \sum_i\gamma^{i}\partial_i$ and accordingly for $\vec{\gamma}\cdot\vec{A}$.
We can generalize this continuation such as to include a complex mass
$m {\rm e}^{{\rm i}\alpha}\equiv m_{\rm R}+{\rm i} m_{\rm I}$, resulting in
\begin{align}
\label{Dirac:massive}
\slashed D^{\E} +m_{\rm R}+{\rm i} \gamma^5 m_{\rm I}\rightarrow-\left({\rm i}\slashed D-m_{\rm R}-{\rm i}\gamma^5 m_{\rm I}\right)\,.
\end{align}
On the right-hand side, we recover the standard Dirac operator for a massive fermion in Minkowski spacetime. It is a non-Hermitian operator leading to a Lagrangian
term that is however Hermitian when sandwiched between $\bar\psi=\psi^\dagger\gamma^{0}$
and $\psi$ and when $A_\mu^{a}$ is real. As noted above, the latter condition
is not met for the complex saddle corresponding to the instanton.

When including a complex fermion mass,
the Euclidean Green's function $S^\E(x^\E,x^{\E\prime})$ satisfies
\begin{align}
\label{Euclidean:Green:massless}
(\slashed D^{\E}+m_{\rm R}+{\rm i}\gamma^5 m_{\rm I}) S^{\E}(x^{\E},x^{\E\prime})=\delta^4(x^{\E}-x^{\E\prime})\,.
\end{align}
The most straightforward way of constructing it is from
the spectral sum in the massless limit. It is constituted by the solutions to the eigenvalue problem
\begin{align}
\label{eveq:massless}
\slashed D^{\E}\hat\psi^{\E}_\lambda=\left(\slashed \partial^{\E} - {\rm i}\gamma^{\E}_m A^{\E}_m\right)\hat\psi^{\E}_\lambda=\lambda^{\E}\hat\psi^{\E}_\lambda\,,
\end{align}
as
\begin{align}
\label{Euclidean:Green:massless:sum}
S^{\E}(x^{\E},x^{\E\prime})
=\SumInt\limits_{\lambda^{\E}}
 \frac{\hat\psi^{\E}_\lambda(x^{\E})\hat\psi^{\E\dagger}_\lambda(x^{\E\prime})}{\lambda^{\E}}\,.
\end{align}
Since the Euclidean Dirac operator $\slashed D^{\E}$ is anti-Hermitian, its
eigenfunctions can readily be assumed to be orthonormal and Eq.~(\ref{Euclidean:Green:massless})
be immediately verified. Yet, Eq.~(\ref{Euclidean:Green:massless:sum})
is ill-defined because of the fermionic zero mode $\lambda^{\E}=0$ in the instanton background. The Euclidean index theorem relates the winding number to the difference between the number of right-handed and left-handed zero modes. This gives one left-handed zero-mode for a $\eta=-1$ background, and a right-handed zero mode for $\eta=1$. The former is given by
\begin{align}
\label{eq:zeromodeanti}
\psi^{\rm E}_{0{\rm L}}(x^{\rm E})=\left(\begin{array}{c}\chi^{\rm E}_0(x^{\rm E})\\[1mm]\left(\begin{array}{c}0\\0\end{array}\right)\end{array}\right)\,,\quad\text{where}\quad
\chi^{\rm E}_0(x^{\rm E})=\frac{\varrho  u}{\pi\left[\varrho^2+{(x^{\rm E})^2}\right]^{\frac32}}
\end{align}
and $u$ is a $2\times2$ antisymmetric matrix with a Weyl index $\alpha$ and an index $b$ labelling the fundamental representation of ${\rm SU}(2)$, i.e.  $u^{\alpha b}=\varepsilon^{\alpha b}$,  with $\varepsilon^{12}=1$.  As anticipated the mode is left chiral,
i.e. $P_{\rm L}\psi^{\rm E}_{0{\rm L}}(x^{\rm E})=\psi^{\rm E}_{0{\rm L}}(x^{\rm E})$, where $P_{\rm L,R}=\frac{1\mp\gamma^5}{2}$
are the chiral projectors. The solution $\psi^{\rm E}_{0{\rm R}}$ in the $\eta=+1$ instanton background can be obtained by switching the chiral block in Eq.~\eqref{eq:zeromodeanti}.

A small complex mass term can serve as a regulator of the zero-mode contribution to Eq.~\eqref{Euclidean:Green:massless:sum} because, for fermions in the fundamental representation
of the gauge group in the $\eta=-1$ instanton background, one obtains at first order in perturbation
theory~\cite{Shifman:1979uw}
\begin{align}
\label{eq:masspertE}
S^{\E}(x^{\E},x^{\E\prime})
=\frac{\hat\psi^{\E}_0(x^{\E})\hat\psi_0^{\E\dagger}(x^{\E\prime})}{m {\rm e}^{-{\rm i}\alpha}}+\SumInt\limits_{\lambda^{\E}\not=0}
\frac{\hat\psi^{\E}_\lambda(x^{\E})\hat\psi^{\E\dagger}_\lambda(x^{\E\prime})}{\lambda^{\E}}\,.
\end{align}
From Eq.~(\ref{Dirac:massive}), it then follows that we may analytically continue this solution as
\begin{align}
\label{eq:analyCon;Greens}
{\rm i}S^{}(x,x^{\prime})
=S^{\E}(x^{\E},x^{\E\prime})\big|_{x_4={\rm i} x^0,\, x'_4={\rm i} x^{\prime0}}\,,
\end{align}
where the dependence on $x^{(\prime)}$ is understood to refer to the components $x^{(\prime)0}$ and $\vec x^{(\prime)}$ of the corresponding four-vector $x^{(\prime)\mu}$ as in Eq.~\eqref{Minkowskivector}. This
Minkowski-space Green's function
approximately solves the equation
\begin{align}
\label{Dirac:massive:Minkowski}
\left({\rm i}\slashed D-m_{\rm R}-{\rm i}\gamma^5 m_{\rm I}\right){\rm i}S(x,x^{\prime})={\rm i}\delta^4(x-x^{\prime})\,.
\end{align}
The above equation can be obtained from an analytic continuation of Eq.~\eqref{Euclidean:Green:massless}, with the continuation of the  delta function giving $\delta^4(x^{\E}-x^{\E\prime})\rightarrow-{\rm i}\delta^4(x-x')$. (For example, one can start with the representation of $\delta(x)$ in terms of its Fourier-transform and analytically continue $x$ away from the real line.) On the other hand, taking Eqs.~\eqref{Euclidean:Green:massless} and~\eqref{Dirac:massive:Minkowski} as the definitions of the Euclidean and Minkowskian Green's functions, respectively, one can infer from the path integral the following correspondence between the Green's functions and the fermion propagators in the one-instanton background:
\begin{align}
    S^{\E}(x^{\E},x^{\E\prime})=&\langle\psi^{\E}(x^{\E})\psi^{\E\dagger}(x^{\E\prime})\rangle,&{\rm i} S^{}(x^{},x^{\prime}) =\langle \psi^{}(x^{}) \bar{\psi}^{}(x^{\prime})\rangle\,.
\end{align}
Recalling that the mapping between Euclidean and Minkowskian fermion fields goes as
$\psi^{\E}(\vec{x},x_4={\rm i}x^0)={\psi}(x^0,\vec{x})$, $\psi^{\E\dagger}(\vec{x},x_4={\rm i}x^0)=\bar{\psi}(x^0,\vec{x})$, one can confirm that the  Euclidean and Minkowskian Green's functions are indeed related by the analytic continuation of Eq.~\eqref{eq:analyCon;Greens}. (The present notation differs from that used in  Ref.~\cite{Vainshtein:1981wh} where $\psi^{\E\dagger}(\vec{x},x_4={\rm i}x^0)=\ic\bar{\psi}(x^0,\vec{x})$.)
Note however that, as it is elaborated upon in Section~\ref{sec:cplx:mass:Mink}, it is not straightforward to show that this analytic continuation has a well-defined spectral representation in terms of (im)properly normalizable eigenfunctions of the Dirac operator in Minkowski spacetime  \cite{Ai:2019fri}.

Equations \eqref{eq:masspertE} and \eqref{eq:analyCon;Greens} show that a mass term with a complex phase can thus be perturbatively included in the leading
contribution to the Green's function that corresponds to the Euclidean zero modes in the massless limit.
Nonetheless, since the Euclidean Dirac operator for a massive fermion with
a general chiral phase is not of definite Hermiticity,
it remains of interest whether such a spectral sum in terms of orthonormal eigenfunctions
is also possible for a complex mass term without resorting to perturbation theory around the
massless configuration, which is what we discuss in the following section.

\subsection{Complex fermion mass in Euclidean space}
\label{sec:cplx:mass:Eucl}

In this section we focus on the Euclidean operator in Eq.~(\ref{Dirac:massive}). The operator $\slashed D^{\E}+m{\rm e}^{{\rm i}\alpha\gamma^5}=\slashed D^\E+m_{\rm R}+\ic\gamma^5m_{\rm I}$ has the following properties in certain simplified cases. For $m=0$, it is anti-Hermitian, while for $m_{\rm I}=0$, it is ``$\gamma^5$-Hermitian'', i.e.
\begin{align}
\label{gamma5:Hermitean}
\left(\slashed D^\E+m_{\rm R}\right)^\dagger=\gamma^5\left(\slashed D^\E+m_{\rm R}\right)\gamma^5\,.
\end{align}
When using the eigenmodes $\hat\psi^\E_\lambda$ from the massless problem~(\ref{eveq:massless}) in the presence of a real mass, these still lead to eigenmodes
with the eigenvalues
\begin{subequations}
\label{gamma5:EV}
\begin{align}
\left(\slashed D^E+m_{\rm R}\right)\hat\psi^\E_\lambda=&(\lambda^\E+m_{\rm R})\hat\psi^E_\lambda\,,
\\
\left(\slashed D^\E+m_{\rm R}\right)\gamma^5\hat\psi^\E_\lambda=&\gamma^5\left(-\slashed D^\E+m_{\rm R}\right)\hat\psi^\E_\lambda=(-\lambda^\E+m_{\rm R})\gamma^5\hat\psi^\E_\lambda\,.
\end{align}
\end{subequations}
Hence, since the real mass term is proportional to the identity matrix in spinor space, a spectral sum can be computed in terms of the same basis vectors as for the massless case.
Moreover, $\hat\psi^\E_\lambda$ and $\gamma^5\hat\psi^\E_\lambda$ are orthogonal for $\lambda^\E\not=0$ because
they correspond to different eigenvalues of the anti-Hermitian operator $\slashed D^\E$.

For a complex mass term, where in addition $m_{\rm I}\not=0$, it it is less obvious that a spectral sum can be constructed in terms of the massless eigenmodes because the mass term is no longer simply proportional to an identity matrix in spinor space. Nonetheless, this can still be accomplished with an additional basis transformation
among the pairs $\hat\psi^\E_\lambda$ and $\gamma^5\hat\psi^\E_\lambda$.
To see this, we note that for a given pair of massless eigenmodes $\hat\psi^\E_\lambda$
and $\gamma^5 \hat\psi^\E_\lambda$ ($\lambda^\E\neq 0$), the Dirac operator takes the matrix form
{\arraycolsep=1pt
\begin{align}
	\label{Dirac:pair}
	\scriptstyle
	\begin{pmatrix}
		\slashed{D}^\E+m_{\rm R} + \ic \gamma^5 m_{\rm I} & 0 \\
		0 & \slashed{D}^\E+m_{\rm R} + \ic \gamma^5 m_{\rm I}
	\end{pmatrix}
	\left(\begin{array}{c}\hat\psi^\E_\lambda\\\gamma^5\hat\psi^\E_\lambda\end{array}\right)	= \left(\begin{array}{cc}\lambda^\E+m_{\rm R} & \ic  m_{\rm I}\\ \ic m_{\rm I} & -\lambda^\E+m_{\rm R} \end{array}\right) \left(\begin{array}{c}\hat\psi^\E_\lambda \\ \gamma^5\hat\psi^\E_\lambda\end{array}\right)\,.
\end{align}}

The eigenvalues of this matrix are
\begin{align}\label{eq:xipm}
\xi^\E_\pm(\lambda^\E)=m_{\rm R}\pm\sqrt{(\lambda^\E)^2-m_{\rm I}^2}\
\end{align}
and the normalized eigenvectors are
\begin{align}
\label{eq:eigenv}
\psi^\E_{\xi\pm}=
\frac{1}{\sqrt{2\lambda^\E}}\left(\frac{m_{\rm I}}{\sqrt{\lambda^\E\mp\sqrt{(\lambda^\E)^2-m_{\rm I}^2}}}\hat\psi^\E_\lambda+{\rm i}\sqrt{\lambda^\E\mp\sqrt{(\lambda^\E)^2-m_{\rm I}^2}}\gamma^5\hat\psi^\E_\lambda\right)\,.
\end{align}
The spinors $\psi^\E_{\xi\pm}$
are pairwise orthogonal, which can be checked explicitly when making use of the fact that $\slashed D^\E$ is anti-Hermitian such that $\lambda^\E$ is purely imaginary. Since the zero mode  is chiral, it is still an eigenfunction $\psi^\E_0\equiv\hat\psi^\E_0$ for the Dirac operator when a complex mass is added.
Altogether, we still have an orthonormal system such that the Green's function in the $\eta=-1$ instanton background is given by
\begin{align}
\label{spectralsum:complexmass}
S^\E(x^\E,x^{\E \prime})=\frac{\psi^\E_0(x){\psi^{\E \dagger}_0}(x^{\E\prime})}{m {\rm e}^{-{\rm i}\alpha}}
+\SumInt\limits_{\lambda^\E/{\rm i}>0}\sum\limits_\pm\frac{\psi^\E_{\xi\pm}(x^\E){\psi^{\E\dagger}_{\xi\pm}}(x^{\E \prime})}{\xi^\E_{\pm}}\,.
\end{align}

In addition, we note that $(\lambda^\E)^2-m_{\rm I}^2<0$ { ($\lambda^{\E}$ is purely imaginary because of the anti-Hermiticity of $\slashed{D}^{\E}$)}, such that the coefficients of
$\hat\psi^\E_\lambda$ and $\gamma^5\hat\psi^\E_\lambda$ in Eq.~\eqref{eq:eigenv} have the same phase. The basis
transformation is thus orthogonal, up to an arbitrary overall phase.
Hence, $\psi^\E_{\xi\pm}$
are also eigenvectors of the Hermitian conjugate operator
{\arraycolsep=1pt
	\begin{align}
	\scriptstyle
	\begin{pmatrix}
		(\slashed{D}^\E+m_{\rm R}+\ic \gamma^5 m_{\rm I})^\dagger & 0 \\
		0 & (\slashed{D}^\E+m_{\rm R}+\ic \gamma^5 m_{\rm I})^\dagger
	\end{pmatrix}
	\left(\begin{array}{c}\hat\psi^\E_\lambda\\\gamma^5\hat\psi^\E_\lambda\end{array}\right) 	=
	\left(\begin{array}{cc}-\lambda^\E+m_{\rm R} & -\ic  m_{\rm I}\\-\ic m_{\rm I} & \lambda^\E+m_{\rm R}\end{array} \right) \left(\begin{array}{c}\hat\psi^\E_\lambda\\\gamma^5\hat\psi^\E_\lambda\end{array}\right),
\end{align}}
with eigenvalues $(\xi^\E_{\pm})^*$ because the above operator acts on the pair $\hat\psi^\E_\lambda$
and $\gamma^5 \hat\psi^\E_\lambda$ as the complex conjugate of the operator in Eq.~(\ref{Dirac:pair}).
(If the coefficients of $\hat\psi^\E_\lambda$ and $\gamma^5\hat\psi^\E_\lambda$ did not have the same phase, the coefficients would have to be complex conjugated in order to obtain the eigenvectors of the complex conjugate matrix.)

The anomalous divergence of the chiral current can now be straightforwardly verified.
We first note that
\begin{align}
\partial^\E_m{\rm tr}\,\gamma^5\gamma^\E_m \psi^\E_{\xi\pm}(x^\E) {\psi^{\E\dagger}_{\xi\pm}}(x^\E)
=&
{\rm tr}\Big\{\gamma^5\left[\left(\slashed D^\E +m {\rm e}^{{\rm i}\alpha\gamma^5}+{\rm i}\gamma^\E_m A^\E_m-m {\rm e}^{{\rm i}\alpha\gamma^5}\right)\psi^\E_{\xi\pm}(x^\E)\right]{\psi^{\E\dagger}_{\xi\pm}} (x^\E)\notag\\
-&\gamma^5 \psi^\E_{\xi\pm}(x^\E)\left[\left(\slashed D^\E -m {\rm e}^{-{\rm i}\alpha\gamma^5}+{\rm i}\gamma^\E_m A^\E_m+m {\rm e}^{-{\rm i}\alpha\gamma^5}\right)\psi^\E_{\xi\pm}(x^\E)\right]^\dagger\Big\}
\notag\\
=&
{\rm tr}\left\{2\gamma^5\xi^\E_\pm \psi^\E_{\xi\pm}(x^\E) {\psi^{\E\dagger}_{\xi\pm}}(x^\E)-2\gamma^5 m{\rm e}^{{\rm i}\alpha\gamma^5} \psi^\E_{\xi\pm}(x^\E) {\psi^{\E\dagger}_{\xi\pm}}(x^\E)\right\}\,,
\end{align}
and that the according relation also holds for the zero mode $\psi^\E_0(x^\E)$.
The trace is understood to run over the spinor indices, and we have substituted the eigenvalues
of the massive Dirac operator and its Hermitian conjugate as discussed above.
Substituting this into Eq.~(\ref{spectralsum:complexmass}), we indeed obtain
\begin{align}\label{eq:anomaly}
\partial^\E_m {\rm tr}\,\gamma^5\gamma^\E_m S^\E(x^\E,x^\E)=
2{\psi^{\E\dagger}_{0}}(x^\E)\gamma^5 \psi^\E_{0}(x^\E)+
\SumInt\limits_{\lambda^\E/{\rm i}>0}\sum_\pm 2{\psi^{\E\dagger}_{\xi\pm}}(x^\E)\gamma^5 \psi^\E_{\xi\pm}(x^\E)
{+2}\langle{\psi^{\E\dagger}}(x^\E)\gamma^5 m {\rm e}^{{\rm i}\alpha\gamma^5} \psi^\E(x^\E)\rangle\,.
\end{align}
We note that the second term on the right-hand side vanishes because the trace of $\gamma^5$ over the nonzero modes is not anomalous. The first term on the right gives the usual anomaly upon integration over spacetime and accounting for the unit norm of the zero modes: For  a $\eta=\pm1$ background with a right (left)-handed zero mode, one gets a change of chirality by $\pm$2 units. The last term in Eq.~\eqref{eq:anomaly} reproduces the classical divergence of the current.

From the spectral decomposition we can also observe that the phase of the determinant of the operator $\slashed D^\E+m_{\rm R}+{\rm i}\gamma^5 m_{\rm I}$ is entirely determined by the zero modes of $\slashed D^\E$.  For a $\eta=\pm1$ instanton background with a right(left)-handed zero mode one has
\begin{align}
\label{eq:EuclideanDet0}
 \det (-\slashed D^\E-m_{\rm R}-{\rm i}\gamma^5 m_{\rm I}) = \det (-\slashed D^\E-m {\rm e}^{{\rm i} \alpha\gamma_5})=-m{\rm e}^{{\rm i}\eta\alpha}\prod_{\lambda^\E/{\rm i}>0}\xi^\E_+(\lambda^\E)\xi^\E_-(\lambda^\E)=-m {\rm e}^{{\rm i}\eta\alpha}\prod_{\lambda^\E/{\rm i}>0}(m^2+|\lambda^\E|^2)\,.
\end{align}
As a consequence, we can write
\begin{align}
\label{eq:EuclideanDet}
 \det (-\slashed D^\E-m{\rm e}^{{\rm i}\alpha\gamma^5}) =-{\rm e}^{{\rm i}\eta\alpha}|\det (-\slashed D^\E-m{\rm e}^{{\rm i}\alpha\gamma^5})|\,, \quad  \eta=\pm1\,.
\end{align}
One can use the fact that the instanton and anti-instanton backgrounds are simply related by parity conjugation to prove that the determinants in both backgrounds are related by the substitution  $\alpha\rightarrow-\alpha$. This is consistent with the phases in Eqs.~\eqref{eq:EuclideanDet0} and~\eqref{eq:EuclideanDet}. Moreover, according to Eq.~\eqref{eq:EuclideanDet0}, $|{\rm det} (-\slashed D^{\rm E}-m{\rm e}^{{\rm i}\alpha\gamma_5})|$ is independent of $\alpha$, and thus it is identical for both backgrounds.
A similar analysis can be done for the operator $-\slashed\partial^{\rm E} -m_{\rm R}-{\rm i}\gamma^5m_{\rm I}$. In this case, since the gauge-field background is trivial with zero winding number, according to the Atiyah–Singer index theorem the number of left-handed zero modes for $\slashed{\partial}^\E$ must equal to the number of right-handed zero modes, ending up with a vanishing chiral phase in the determinant:
\begin{align}
\label{eq:detDiractrivial}
    \det (-\slashed \partial^\E-m{\rm e}^{{\rm i}\alpha\gamma^5}) =|\det (-\slashed \partial^\E-m{\rm e}^{{\rm i}\alpha\gamma^5})|\,.
\end{align}

In preparation for the extension of the spectral decomposition of the propagator~\eqref{spectralsum:complexmass} to arbitrary rotations of the time contour, we consider separately the Euclidean eigenfunctions belonging to the discrete and continuum spectrum and introduce associated notation and properties. The normalizable eigenfunctions belonging to the discrete spectrum are denoted as $\psi^\E_n$ and their eigenvalues as $\xi^\E_n$. These modes have a finite norm and are mutually orthogonal under the usual scalar product,
\begin{align}
 (\psi^\E_m,\psi^\E_n)=\int {\rm d}^4 x^\E \, {\psi^{\E\dagger}_m}(x^\E) \psi^\E_n(x^\E) = \delta_{mn}\,.
\end{align}
In regards to the continuum spectrum, involving improperly normalizable eigenfunctions, it can be constructed from solutions which approach plane waves at $x_4\rightarrow-\infty$, characterized by asymptotic momenta $k_m,m=1,\dots,4$. We will thus denote the eigenfunctions as $\psi^\E_{\{k^\E\}}=\psi^\E_{\{\vec{k},k_4\}}$ and their eigenvalues as $\xi^\E_{\{k^\E\}}=\xi^\E_{\{\vec{k}.k_4\}}$. A difference with the work of Ref.~\cite{Ai:2019fri}, which focuses on differential operators in backgrounds invariant under spatial translations like a planar domain-wall, is that the continuum modes will not be given by a single plane wave for all $x_4$, due to the spatial inhomogeneity of the BPST instanton background. However, one can always choose a basis of modes approaching a single plane wave at $x_4\rightarrow-\infty$ and given by a superposition of plane waves at $x_4\rightarrow \infty$.  Indeed, from the results in this section it follows that generic Euclidean modes $\psi^\E_\xi$ with eigenvalues $\xi^\E$ satisfy
\begin{align}\label{eq:Euclidid}\begin{aligned}
 (\slashed D^\E+m{\rm e}^{{\rm i}\alpha\gamma^5})\psi^\E_\xi(x^\E)=&\,\xi^\E\psi^\E_\xi(x^\E)\,,\\
 (\slashed D^\E+m{\rm e}^{{\rm i}\alpha\gamma^5})^\dagger\psi^\E_\xi(x^\E)=&\,(-\slashed D^\E+m{\rm e}^{-{\rm i}\alpha\gamma^5})\psi_\xi^\E(x^\E)=(\xi^{\E })^*\psi^\E_\xi(x^\E)\,,
\end{aligned}\end{align}
which gives
\begin{align}
 (\slashed D^\E+m{\rm e}^{{\rm i}\alpha\gamma^5}) (-\slashed D^\E+{
m e}^{-{\rm i}\alpha\gamma^5})\psi^\E_\xi=|\xi^E|^2\psi^\E_\xi = (-(\partial_m-{\rm i}A^\E_m)(\partial_m-{\rm i}A^\E_m)+m^2)\psi^\E_\xi\,.
\end{align}
Therefore the Euclidean eigenvalue problem implies
\begin{align}
( (\partial_m-{\rm i}A^\E_m)(\partial_m-{\rm i}A^\E_m)-m^2+|\xi^E|^2)\psi^\E_\xi=0\,.
\end{align}
For a solution going asymptotically as a plane wave in the infinite Euclidean past---thus being improperly normalizable and belonging to the continuum spectrum---one has
\begin{align}
\label{eq:scalareq}
 \psi^\E_{\{k^\E\}}\sim {\rm e}^{{\rm i} k_m x_m}\ { \rm as}\  x_4\rightarrow-\infty\,,
\end{align}
and the Euclidean eigenvalues satisfy (using the fact that the instanton background $A^\E_m$ goes to zero at infinity)
\begin{align}
\label{eq:Euclideancont}
|\xi^\E_{\{k^\E\}}|^2=m^2+k_m k_m\,.
\end{align}
As the background also goes to zero for $x_4\rightarrow+\infty$, the solutions  will tend to a superposition of plane waves with the same value of $k^2=k_m k_m$, fixed in terms of $|\xi^\E_{\{k_m\}}|^2$ as above. In this sense, the eigenvalue equation is analogous to a wave-mechanical scattering problem. We expect that we can form a basis for the continuum spectrum by considering all possible plane waves at
$ x_4\rightarrow-\infty$. As the solutions are eigenfunctions of a Hermitian operator, the   $\psi^\E_{\{k_m\}}$ are orthogonal, and they can be normalized so that the norm is a delta function in $k$-space:
\begin{align}
 (\psi^\E_{\{k^\E\}},\psi^\E_{\{k^{\E\prime}\}})=\delta^{4}(k^\E-k^{\E\prime})\,.
\end{align}

In the massless limit, as discussed above the continuum eigenvalues must become purely imaginary. Denoting these massless eigenvalues as $\lambda^\E_{\{k_m\}}$ and using Eq.~\eqref{eq:Euclideancont} in the massless limit, if follows that
\begin{align}
 \lambda^\E_{\{k^\E\}}=\ic \sqrt{k_m k_m}\,.
\end{align}
Then, the results of Eq.~\eqref{eq:xipm} imply that the continuum Euclidean eigenvalues for a general complex mass have the form
\begin{align}
\label{eq:Euccontinuum}
 \xi^\E_{\pm\{k^\E\}}=m_{\rm R}\pm\ic\sqrt{k_mk_m+m^2_{\rm I}}\,.
\end{align}

\subsection{Complex fermion mass for an arbitrary rotation of the time contour}
\label{sec:cplx:mass:arbitrary}

In this section we generalize the spectral decomposition of the Euclidean propagator to the case of arbitrary rotations of the time contour, using the methods of Ref.~\cite{Ai:2019fri} adapted to complex fermion fields in generic, rather than bosonic planar backgrounds. We use superscripts ``$\vartheta$'' for objects defined for a general time contour. Under the analytic continuation of Eq.~\eqref{Euclideanvector}, the fermionic kinetic term of the Lagrangian involves the operator
\begin{align}
\label{eq:Diraccont}
 -\slashed D^\E-m{\rm e}^{{\rm i}\alpha\gamma^5}\,\rightarrow\ {\rm i}\slashed{D}^\vartheta-m{\rm e}^{{\rm i}\alpha\gamma^5}\,,   \qquad \slashed{D}^\vartheta = \,\gamma^{\vartheta\mu}(\partial_\mu-{\rm i}A^\vartheta_{\mu}(x))\,,
\end{align}
with  the following $\gamma$-matrices and  gauge field components:
\begin{align}\begin{aligned}
 \gamma^{\vartheta0}= & {\rm e}^{{\rm i}\vartheta} \gamma^0\,,&\gamma^{\vartheta i}=&\,\gamma^i\,,\\
A^{\vartheta 0}(x^0,\vec{x})=&\,\ic {\rm e}^{-\ic \vartheta}A^\E_4(\vec{x},x_4={\rm i}{\rm e}^{-{\rm i}\vartheta}x^0)\,,& A^{\vartheta i}(x^0,\vec{x})=&\,A^\E_i(\vec{x},x_4={\rm i}{\rm e}^{-{\rm i}\vartheta}x^0)\, .
\end{aligned}\end{align}
Recall that $x^0$ is meant to be real, parameterizing the rotated time contour; one also has $\partial_\mu=\partial/\partial x^\mu$ with $x^\mu$ the components of the four-vector in Eq.~\eqref{Minkowskivector}.
The matrices $\gamma^{\vartheta\mu}$, which have been defined in terms of their Minkowskian counterparts $\gamma^{\mu}$, satisfy a Clifford algebra $\{\gamma^{\vartheta\mu},\gamma^{\vartheta\nu}\}=g^{\vartheta \mu\nu}$, with the metric $g^{\vartheta \mu\nu}={\rm diag}\{{\rm e}^{2\ic\vartheta},-1,-1,-1\}$. The latter
coincides with the effective metric appearing in the kinetic terms for scalar fields for arbitrary $\vartheta$ in Ref.~\cite{Ai:2019fri}. Note that here we are looking at the analytic continuation between the two operators in Eq.~\eqref{eq:Diraccont}. When taking $\vartheta=\pi/2$, the $\gamma^{\vartheta\mu}$ do not render the Euclidean $\gamma$-matrices but differ from these by a factor of ${\rm i}$. This is due to the signature $(+,-,-,-)$ used in Minkowski spacetime, as opposed to the positive signature in Euclidean spacetime. However ${\rm i}\slashed{D}^\vartheta-m{\rm e}^{{\rm i}\alpha\gamma^5}$ does return to $-\slashed D^\E-m{\rm e}^{{\rm i}\alpha\gamma^5}$ for $\vartheta=\pi/2$.

As in Ref.~\cite{Ai:2019fri}, one can construct (im)properly normalizable eigenfunctions for the  differential operator for arbitrary $\vartheta$ by analytic continuation of the corresponding Euclidean eigenfunctions in the time variable and, for the continuum spectrum, additionally in the asymptotic parameter $k_4$. In order to obtain eigenfunctions $\psi^\vartheta_n$ in the discrete spectrum it suffices to perform the usual analytic continuation, for which one obtains same eigenvalues as in Euclidean space, safe for the minus sign that follows from Eq.~\eqref{eq:Diraccont} and the fact that the Euclidean eigenvalues were defined as corresponding to the operator $\slashed{D}^\E+m{\rm e}^{{\rm i}\alpha\gamma^5}$:
\begin{align}
\label{eq:discrete}
 \psi^\vartheta_{n}(x)=\psi^\vartheta_{n}(x^0,\vec{x})=\sqrt{{\rm i}{\rm e}^{-{\rm i}\vartheta}}\,\psi^\E_{n}(\vec{x},x_4={\rm i}{\rm e}^{-{\rm i}\vartheta}x^0)\, ,\quad \xi^{\vartheta}_{n}=-\xi^\E_n\,, \quad\text{discrete spectrum}.
\end{align}
The factor of $\sqrt{{\rm i}{\rm e}^{-{\rm i}\vartheta}}$ is taken to lie in the principal branch and is necessary to guarantee a unit norm, defined with an inner product that will be described below. For the continuum spectrum, in order to preserve the plane-wave behaviour at $t\rightarrow-\infty$, one needs to rotate the asymptotic parameter $k_4$, and as a result the continuum eigenvalues in Minkowski are $\vartheta$-dependent:
\begin{align}
\label{eq:continuum}
 \psi^{\vartheta}_{\{k\}}(x)= \psi^{\vartheta}_{\{k^0,\vec{k}\}}(x^0,\vec{x})=\,\psi^\E_{\{\vec{k},-{\rm i}{\rm e}^{{\rm i}\vartheta}k_0\}}(\vec{x},x_4={\rm i}{\rm e}^{-{\rm i}\vartheta}x^0)\,, \quad \xi^{\vartheta}_{\{k^0,\vec{k}\}}=-\xi^\E_{\{\vec{k},-{\rm i}{\rm e}^{{\rm i}\vartheta}k_0\}}\quad\text{continuum spectrum}.
\end{align}
In the following we denote a generic eigenfunction with eigenvalue $\xi^\vartheta$---either in the discrete or continuum spectrum---as  $\psi^{\vartheta}_{\xi}$. It turns out that the eigenfunctions constructed as above are  orthogonal and complete with respect to the following inner product,
\begin{align}\label{eq:tildeproduct}
 (\psi^{\vartheta}_{\xi},\psi^{\vartheta}_{\xi'})_\vartheta=\int {\rm d}^4x \,\tilde\psi^{\vartheta}_{\xi}(x)\,\psi^{\vartheta}_{\xi'}(x)\,,
\end{align}
with $\tilde\psi^\vartheta$ defined as
\begin{align}
\label{eq:modes}
\begin{aligned}
\tilde\psi^{\vartheta}_{n}(x^0,\vec{x})=&\,\sqrt{\ic {\rm e}^{-\ic \vartheta}}\left.(\psi^\E_n(\vec{x},x_4))^\dagger\right|_{x_4={\rm i}{\rm e}^{-{\rm i}\vartheta}x^0}=\,\ic {\rm e}^{-\ic \vartheta}\left.(\psi^{\vartheta}_{n}(x^0,\vec{x}))^\dagger\right|_{{x^0}\rightarrow-{\rm e}^{-2{\rm i}\vartheta}x^0}\,, &\text{discrete spectrum}\\
 \tilde\psi^{\vartheta}_{\{k^0,\vec{k}\}}(x^0,\vec{x})=&\,\left.\left(\psi^\E_{\{\vec{k},k_4\}}(\vec{x},x_4)\right)^\dagger\right|_{\scriptsize\begin{array}{ll}
    x_4={\rm i}{\rm e}^{-{\rm i}\vartheta}x^0\\
k_4={-\rm i}{\rm e}^{{\rm i}\vartheta}k^0
 \end{array}
}= \,\left.\psi^{\vartheta}_{\{k^0,\vec{k}\}}(x^0,\vec{x})^\dagger\right|_{\scriptsize\begin{array}{ll}
   { x^0}\rightarrow-{\rm e}^{-2{\rm i}\vartheta}x^0\\
{k^0}\rightarrow{-}{\rm e}^{2{\rm i}\vartheta}k^0
\end{array}}\,,& \text{continuum spectrum.}
\end{aligned}\end{align}
We refer to this operation indicated by a tilde and to the associated inner product in Eq.~\eqref{eq:tildeproduct} as {\it $\vartheta$-adjoint} and {\it $\vartheta$-adjoint inner product}, respectively.
In Eq.~\eqref{eq:modes}, the dagger operation is to be understood assuming that the corresponding coordinates and asymptotic parameters are treated as real, i.e. $\psi^\E_{\{\vec{k},k_4\}}(\vec{x},x_4)^\dagger$ should be calculated assuming $k_m,x_m$ are real, and the same goes for $k^0,\vec{k},x^0,\vec{x}$ when evaluating $\psi^{\vartheta}_{n}(x)^\dagger$. The last equalities in both lines of Eq.~\eqref{eq:modes} follow from the fact that the transformations $x_0\rightarrow-{\rm e}^{-2{\rm i}\vartheta}x^0,\ k_0\rightarrow{-}{\rm e}^{2{\rm i}\vartheta}k^0$ undo the complex conjugation of the combinations $\ic {\rm e}^{-\ic \theta} x^0,- \ic {\rm e}^{\ic \theta} k^0$ corresponding to the Euclidean variables $x_4,k_4$. A consequence of the above definition  is that both $\psi^{\vartheta}$ and $\tilde\psi^{\vartheta}$ are holomorphic functions of $x^0$ and $k^0$. Then one can prove orthogonality and completeness of the $\vartheta$ eigenfunctions constructed as above by relating all integrals over the parameters $x^0,k^0$ to their Euclidean counterparts  $x_4,k_4$ using the Cauchy theorem \cite{Ai:2019fri}. In particular, the discrete modes have the normalization
\begin{align}
 (\psi^\vartheta_m,\psi^\vartheta_n)_\vartheta= \delta_{mn}\, ,
\end{align}
where as advertised earlier the prefactors $\sqrt{{\rm i}{\rm e}^{-{\rm i}\vartheta}}$  in Eqs.~\eqref{eq:discrete} and Eq.~\eqref{eq:modes} cancel the Jacobian from the rotation of the contour to the Euclidean time. On the other hand, for the eigenfunctions in the continuum one has
\begin{align}
 (\psi^\vartheta_{\{k\}},\psi^\vartheta_{\{{k'}\}})_\vartheta=\delta^{4}(k-k')\,,
\end{align}
where in this case the Jacobian from the rotation to Euclidean time is cancelled by the the one arising from the analytic continuation of the Euclidean delta function of the asymptotic momenta.

Proceeding along these lines, and as explained in detail in Ref.~\cite{Ai:2019fri}, the orthogonality and completeness of the basis of eigenfunctions for arbitrary $\vartheta$ follow from the analogous properties of the Euclidean spectrum.  The former implies that one can resolve the operator ${\rm i}\slashed D^\vartheta-m{\rm e}^{{\rm i}\alpha\gamma^5}$ in terms of orthogonal projectors,
\begin{align}
 {\rm i}\slashed D^\vartheta-m{\rm e}^{{\rm i}\alpha\gamma^5}=&\,\SumInt_{\xi^\vartheta} \xi^\vartheta \psi^{\vartheta}_{\xi}(x)\tilde\psi^{\vartheta}_{\xi}(x')= \sum_n\,{\xi^\vartheta_n}\,\psi^{\vartheta}_{n}(x)\tilde\psi^{\vartheta}_ {n}(x')+\int {\rm d}^4 k \,\xi^\vartheta_{\{k\}}\psi^{\vartheta}_{\{k\}}(x)\tilde\psi^{\vartheta}_{\{k\}}(x')\,,
 \end{align}
 and thus its inverse, i.e. the propagator, is given by
 \begin{align} \begin{aligned}
 S^\vartheta(x,x')\equiv\,( {\rm i}\slashed D^\vartheta-m{\rm e}^{{\rm i}\alpha\gamma_5})^{-1}(x,x')=&\,\SumInt_{\xi^\vartheta}\frac{1}{ \xi^\vartheta}\, \psi^{\vartheta}_{\xi}(x)\tilde\psi^{\vartheta}_{\xi}(x')\\
 =&\,\sum_n\,\frac1{\xi^\vartheta_n}\,\psi^{\vartheta}_{n}(x)\tilde\psi^{\vartheta}_{n}(x')+\int {\rm d}^4 k \,\frac1{\xi^\vartheta_{\{k\}}}\psi^{\vartheta}_{\{k\}}(x)\tilde\psi^{\vartheta}_{\{k\}}(x')\,.
\end{aligned}\end{align}
The above propagator is nothing but the analytic continuation of its Euclidean counterpart, up to an overall constant:
\begin{align}\label{eq:Scontinuation}
S^\vartheta(x,x')=-{\rm i} {\rm e}^{-{\rm i}\vartheta}S^\E(x^\E,x^{\E\prime })|_{x_{4}\rightarrow {\rm i}{\rm e}^{-{\rm i}\vartheta}x^0,\, x'_4\rightarrow {\rm i}{\rm e}^{-{\rm i}\vartheta}x^{\prime0}}\,,
\end{align}
The overall minus in Eq.~\eqref{eq:Scontinuation} arises as a result of Eq.~\eqref{eq:Diraccont} (or equivalently from the minus signs in the relations between rotated and Euclidean eigenvalues in Eqs.~\eqref{eq:discrete} and \eqref{eq:continuum}). The constant ${\rm i} {\rm e}^{-{\rm i}\vartheta}$ appears in the contribution from the discrete spectrum due to the different normalization of the modes, see Eqs.~\eqref{eq:discrete}~and ~\eqref{eq:modes}, while for the continuum spectrum the same factor arises when relating the integral over the rotated $k^0$ to its Euclidean counterpart $k_4=-{\rm i}{\rm e}^{{\rm i}\vartheta}k^0$. Note that for $\vartheta=\pi/2$ one recovers the Euclidean result up to a minus sign, arising because the propagator $S^{\vartheta=\pi/2}$ is the inverse of $\slashed D^{\vartheta=\frac{\pi}{2}}-m{\rm e}^{{\rm i}\alpha\gamma^5}=-\slashed D^\E-m{\rm e}^{{\rm i}\alpha\gamma^5}$. {For $\vartheta=0^+$, one recovers the relation~\eqref{eq:analyCon;Greens}.}

As an explicit application of the previous construction for $\vartheta=0$, in \ifarXiv\else{Section}\fi~\ref{app:freeprop} we use a spectral sum involving the $\vartheta$-adjoint inner product to derive the free Minkowskian propagator for a fermion with a complex mass term.

\subsection{Complex fermion mass in Minkowski spacetime}
\label{sec:cplx:mass:Mink}

The results of the previous section can be applied to Minkowski spacetime by taking the limit $\vartheta\rightarrow0^+$. Throughout this section, unless specified otherwise all objects are assumed to be defined in Minkowski spacetime. The relevant differential operator,
\begin{align}\label{eq:DiracM}
{\rm i}\slashed D -m_{\rm R}-{\rm i}\gamma^5m_{\rm I}\,,
\end{align}
is Hermitian when evaluated in a background of real $A_\mu^{a}$  and multiplied by $\gamma^{0}$. This may suggest that for such real backgrounds one could define an  inner product involving Dirac adjoint spinors rather than the inner product of Eq.~\eqref{eq:tildeproduct} defined in terms of the $\vartheta$-adjoint spinors introduced in Eq.~\eqref{eq:modes}.  For the Dirac adjoint inner product the operator ${\rm i}\slashed D -m_{\rm R}-{\rm i}\gamma^5m_{\rm I}$ would remain Hermitian, and one would naively expect orthogonal eigenvectors with real eigenvalues, giving a spectral decomposition of the propagator in terms of projectors of the form $\psi_\xi\bar\psi_\xi$. However, this is not the case because the Dirac adjoint inner product is not positive definite, and thus the  $\psi_\xi\bar\psi_\xi$ operators do not behave as projectors. This is best illustrated by considering the case of the free Minkowskian propagator, which is studied in \ifarXiv\else{Section}\fi~\ref{app:freeprop}; as shown there, when using the Dirac adjoint inner product the eigenfunctions have zero norm and are not orthogonal, while using the $\vartheta$-adjoint inner product one recovers normalizability, orthogonality and completeness, and the usual propagator is recovered from the spectral sum of the tilde projectors. Finally, one could think of defining a propagator from the Hermitian operator $\gamma^0({\rm i}\slashed D -m_{\rm R}-{\rm i}\gamma^5m_{\rm I})$, but this plays no role for $S$-matrix elements, which are constructed from Green's functions involving products of spinors $\psi$, $\bar\psi$ and thus defined in terms of the inverse of the operator in Eq.~\eqref{eq:DiracM}. In any case, in the Minkowskian instanton background the background fields $A^{a}_\mu$ are not real, so that Hermiticity cannot be a guiding principle for the choice of operator or inner product.

From the results of the previous sections we therefore infer a spectral decomposition for the Minkowskian Dirac operator and its associated propagator,
\begin{align}
\label{spectral:sum:Minkowski}
\left({\rm i}\slashed D -m_{\rm R}-{\rm i}\gamma^5 m_{\rm I}\right)_{x,x^\prime}=&\,\SumInt\limits_{\xi}\xi\psi_{\xi}(x)\tilde\psi_{\xi}(x^{\prime})\,,& {\rm i}S(x,x^{\prime})=&\,{\rm i}\SumInt\limits_{\xi}\frac{\psi_{\xi}(x)\tilde\psi_{\xi}(x^{\prime})}{\xi}\,.
\end{align}

An explicit discussion of the analytic continuation of the continuum spectrum
of fermionic and bosonic excitations about instantons would be
of interest in the future. To this end, we only comment on the fermion zero-mode,
that is normalizable in the proper sense and accountable for the effects from the chiral anomaly.  By ``zero mode'' we refer to eigenstates with zero eigenvalue of the {\it massless} Dirac operator. As these modes have well-defined chirality, they are also eigenstates of the general Dirac operator with a complex mass, with eigenvalue $\xi_{0{\rm R}}=-m{
\rm e}^{\ic \alpha}$ for right-handed modes, and $\xi_{0{\rm L}}=-m{\rm e}^{-\ic \alpha}$ for left-handed ones. As follows from the results of the previous section, these discrete zero modes are obtained by analytically continuing the corresponding Euclidean solutions. Then, as in Euclidean spacetime, this gives one right-handed zero-mode for a $\eta=1$ background, and a left-handed zero mode for $\eta=-1$. Applying Eq.~\eqref{eq:discrete} to the Euclidean expression of Eq.~\eqref{eq:zeromodeanti} for the zero mode in the $\eta=-1$ background  gives
\begin{align}
 \psi_{0\rm L}(x^0,\vec{x})\equiv \sqrt{\ic}\,\varphi_{0{\rm L}}(x^0,\vec{x})= \sqrt{\ic}\,\psi^\E_{0\rm L}(\vec{x},\ic x^0)\,,
\end{align}
with
\begin{align}
\label{eq:zeromodeantiM}
\varphi_{0{\rm L}}(x)=\left(\begin{array}{c}\chi_0(x)\\[1mm]\left(\begin{array}{c}0\\0\end{array}\right)\end{array}\right)\,,\quad\text{}\quad
\chi_0(x)=\frac{\varrho  u}{\pi (\varrho^2-{x^{ 2}})^{\frac32}}\,,
\end{align}
where $u$ is defined below Eq.~\eqref{eq:zeromodeanti}. The zero mode satisfies the property
\begin{align}\label{eq:tildeproperty}
 \tilde \psi_{0\rm L}(x)= \sqrt{\ic}\,(\varphi_{0\rm L}(x))^\dagger\,,
\end{align}
as follows from the definition of the $\vartheta$-adjoint operation in Eq.~\eqref{eq:modes} and the invariance of $\varphi^\dagger_{0{\rm L}}(x)$ under time reflections, as can be readily seen from Eq.~\eqref{eq:zeromodeantiM}.

Hence the spectral decomposition of the propagator in Eq.~\eqref{spectral:sum:Minkowski} features a contribution involving $\varphi_{0\rm L}(\varphi_{0\rm L})^\dagger$. Note that this structure indicates anomalous violation of chirality, as it should, which would not be the case if the spectral decomposition were constructed with the Dirac adjoint inner product. Such construction, which was discarded in the previous section, would involve terms of the form $\varphi_{0\rm L}\overline{\varphi}_{0\rm L}$.

Assuming that the zero mode dominates the contributions to the Green's function in the
$\eta=-1$ instanton background close to its centre $x_0$, we thus arrive at the approximation
\begin{align}
{\rm i}S(x,x^{\prime})=&\,{\rm i} S_{\rm cont}(x,x^{\prime}){+}\frac{\varphi_{0{\rm L}}(x-x_0)\,{\varphi^\dagger_{0{\rm L}}}(x^{\prime}-x_0)}{m {\rm e}^{-{\rm i}\alpha}}\notag\\
\approx& \,{\rm i} S_{\rm 0inst}(x,x^{\prime}){+}\frac{\varphi_{0{\rm L}}(x-x_0)\,{\varphi^\dagger_{0{\rm L}}}(x^{\prime}-x_0)}{m {\rm e}^{-{\rm i}\alpha}}\,,
\label{eq:prop}
\end{align}
which captures the dominant contributions from both close to the centre and far away from it. Here, ${\rm i} S_{\rm cont}(x,x^{\prime})$ is the contribution from the continuum spectrum and
\begin{align}
\label{S:0inst}
{\rm i}S_{0\rm inst}(x,x^{\prime})=(-\gamma^{\mu}\partial_\mu+{\rm i}m {
\rm e}^{-{\rm i}\alpha \gamma^5})\int\frac{{\rm d}^4p}{(2\pi)^4} {\rm e}^{-{\rm i}p(x-x^{\prime})}\frac{1}{p^2-m^2+{\rm i}\epsilon}
\end{align}
is the propagator in the trivial background with vanishing gauge fields, whose derivation from a spectral decomposition involving the $\vartheta$-adjoint inner product is presented in \ifarXiv\else{Section}\fi~\ref{app:freeprop}.
Furthermore, we have explicitly inserted the dependence on the translational coordinates $x_0$ of the instanton. Noting that ${\rm i}S_{0\rm inst}$ has a spectral decomposition purely in terms of continuum modes and that  ${\rm i}S_{0\rm inst}(x,x^{\prime})\approx{\rm i}S(x,x^{\prime})$ for $|{x}^2|,|{x^{\prime 2}}|\gg\rho^2$
is an approximation to the Green's function in the instanton background that is valid
at large distances from the centre of the instanton, explains the last equality in Eq.~\eqref{eq:prop}.
In Eq.~(\ref{S:0inst}),
we have chosen the $\epsilon$-prescription corresponding to the Feynman propagator, while of course also other boundary conditions are of interest,
e.g. in view of applications within the Schwinger-Keldysh formalism. The Fourier integral can be straightforwardly evaluated, while the explicit result is not relevant to this end.

The propagator in the $\eta=+1$ instanton background follows from the $\eta=-1$ case
by switching the chiral block of the zero mode in Eq.~\eqref{eq:zeromodeantiM}, using the resulting right-handed zero mode $\varphi_{0{\rm R}}$ in place of $\varphi_{0{\rm L}}$ in Eq.~\eqref{eq:prop}, and replacing $\alpha\to-\alpha$. For a background consisting of a dilute gas of $n$ instantons and $\bar n$ anti-instantons with centres $x_{0,\nu},x_{0,\bar\nu}$, the propagator can be approximated again by the ordinary contribution plus a sum over the zero-mode contributions of the instantons and anti-instantons:
\begin{align}
\label{eq:proptotal}
 {\rm i}S_{n,\bar n}(x,x^{\prime})\approx{\rm i} S_{\rm 0inst}(x,x^{\prime}){+}\sum_{\bar\nu=1}^{\bar n}\frac{\varphi_{0{\rm L}}(x-x_{0,\bar\nu}){\varphi^\dagger_{0{\rm L}}}(x^{\prime}-x_{0,\bar \nu})}{m {\rm e}^{-{\rm i}\alpha}}{+}\sum_{\nu=1}^n\frac{\varphi_{0{\rm R}}(x-x_{0,\nu}){\varphi^\dagger_{0{\rm R}}}(x^{\prime}-x_{0,\nu})}{m {\rm e}^{{\rm i}\alpha}}\,.
\end{align}

To end this section, we may note that, using the results of Ref.~\cite{Ai:2019fri}, the determinant of the Minkowski-space operator ${\rm i}\slashed D-m_{\rm R}-{\rm i}\gamma^5 m_{\rm I}$ can be obtained from the Euclidean result of Eq.~\eqref{eq:EuclideanDet} by analytic continuation of the time interval $T^\E\rightarrow {\rm i} T$ (with $T^\E$ and $T$ referring to the Euclidean and Minkowskian time intervals of the spacetime volume $VT^\E$ and $VT$, respectively),
\begin{align}
\label{eq:detrotation}
     \det ({\rm i}\slashed D-m{\rm e}^{{\rm i}\alpha\gamma^5})=\det (-\slashed D^\E-m{\rm e}^{{\rm i}\alpha\gamma^5})\big|_{T^\E\rightarrow {\rm i}T}\,.
\end{align}
Actually, in physical quantities it is the ratio $\det ({\rm i}\slashed D-m{\rm e}^{{\rm i}\alpha\gamma^5})/\det ({\rm i}\slashed \partial-m{\rm e}^{{\rm i}\alpha\gamma^5})$ (and the corresponding one in Euclidean space) that enters. And it turns out that for such ratios the $T$-dependence cancels out. It is shown in Ref.~\cite{Ai:2019fri} that the $T$-dependence appears only in the integral over the collective time-coordinate of the instanton which originates from the time-translational zero mode of the gauge-field fluctuations in our case (see Eqs.~\eqref{Correlation:PathIntegral},~\eqref{eq:Zn} below). Therefore we simply have
\begin{align}
    \frac{\det ({\rm i}\slashed D-m{\rm e}^{{\rm i}\alpha\gamma^5})}{\det ({\rm i}\slashed \partial-m{\rm e}^{{\rm i}\alpha\gamma^5})}=\frac{\det (-\slashed D^\E-m{\rm e}^{{\rm i}\alpha\gamma^5})}{\det (-\slashed \partial^\E-m{\rm e}^{{\rm i}\alpha\gamma^5})}\,.
\end{align}
This means, in particular, that the only dependence on the chiral phase $\alpha$ is again coming from the zero modes of $\slashed{D}^\E$ alone. We therefore define
\begin{align}
\label{eq:MinkowskiDet}
     \frac{\det ({\rm i}\slashed D-m{\rm e}^{{\rm i}\alpha\gamma^5})}{\det ({\rm i}\slashed \partial-m{\rm e}^{{\rm i}\alpha\gamma^5})}\equiv -{\rm e}^{{\rm i}\eta\alpha}\Theta\,,\qquad\Theta=\left| \frac{\det ({\rm i}\slashed D-m{\rm e}^{{\rm i}\alpha\gamma^5})}{\det ({\rm i}\slashed \partial-m{\rm e}^{{\rm i}\alpha\gamma^5})}\right|=\left|\frac{\det (-\slashed D^\E-m{\rm e}^{{\rm i}\alpha\gamma^5})}{\det (-\slashed \partial^\E-m{\rm e}^{{\rm i}\alpha\gamma^5})}\right|\,,
\end{align}
where $\Theta$ is a positive real number. As follows from the discussion in Section~\ref{sec:cplx:mass:Eucl}, $\Theta$ is the same for both instantons and anti-instantons, hence the omission of a label indicating  $\eta$.

\section{Correlation functions for fermions}
\label{sec:fermi:corr}

\subsection{Path integral in fixed topological sectors}
\label{sec:path:int:fixed:top}

In this section we consider correlation functions for massive fermions with chiral phases, working directly in Minkowski spacetime. We first derive the two-point correlator in a theory with a single fermion and after that, we generalize the result to the cases of multiple fermions and higher-order correlators.

For fluctuations about a given classical background---or about a saddle point on a certain complexified contour of path integration, the Green's function can be identified with the leading order approximation to the two-point correlation function. In the case of the vacuum of a non-Abelian gauge theory, the correlation function is to be computed by summing over contributions coming from fluctuations around backgrounds from different topological sectors, i.e. of different winding number. In a dilute instanton gas approximation, such backgrounds are described by configurations with  all possible numbers of (anti-)instantons, with arbitrary locations in spacetime. The required summation can be carried out along the lines of Ref.~\cite{Diakonov:1985eg}, though here we will track  explicitly the factors of spacetime volume, rather than using instanton densities (which may be phenomenologically more accurate).  In a theory with a single massive Dirac fermion, the two-point correlation function is given by
\begin{align}
\label{correlation:function}
\begin{aligned}
\langle \psi(x) \bar\psi(x^\prime)\rangle
=&\frac{1}{Z}\int{\cal D} A {\cal D}\bar\psi {\cal D}\psi \,\psi(x) \bar\psi(x') {\rm e}^{{\rm i} S}\,,\\
Z=&\int{\cal D} A {\cal D}\bar\psi {\cal D}\psi\, {\rm e}^{{\rm i} S}\,,
\end{aligned}\end{align}
where $S$ is the Minkowskian action and $Z$ the partition function.
In order to relate this to the previously obtained Green's functions in a one-(anti-)instanton background, we denote the numbers of $\eta=-1$ and $\eta=1$ instantons
in the spacetime volume $VT$ under consideration by $\bar n$ and $n$, respectively.

{Requiring that the saddle points of the action take finite values implies vanishing physical fields at the spacetime boundary at infinity~\cite{Coleman:1985rnk}. For the field $A$, this still allows pure gauge configurations while the winding number $\Delta n=n-\bar n$ is topologically restricted to integer values. Consequently, because the topological term is a total divergence, configurations with different values of $\Delta n$ have different boundary conditions for the gauge field configuration.}

These therefore lead to separate contributions to the path integral.
In order to add up these pieces to obtain the partition function or an
observable, we need to take into account the fact that the vacuum state
is a superposition of configurations with
all Chern--Simons numbers, i.e. (up to an irrelevant normalization factor)~\cite{Callan:1976je,Jackiw:1976pf}
\begin{align}
\label{eq:vacuum}
|{\rm vac}\rangle=\sum\limits_{n_{\rm CS}}|n_{\rm CS}\rangle\,.
\end{align}
Here, $|n_{\rm CS}\rangle$ is a state with a fixed Chern--Simons number.
The states are generally also characterized by a vacuum angle, as it is reviewed in Section~\ref{sec:schroedinger}.
The vacuum angle $\theta$ does not explicitly appear here since
we choose to absorb it in the topological Lagrangian term $\theta {\rm tr} F\widetilde F/(16\pi^2)$,
where $F$ denotes the field strength tensor of the gauge field,
$\widetilde F$ its dual and  $\theta$ is the vacuum angle of the gauge theory under consideration. It is easy to see that the following arguments
do not rely on whether the phase is attributed to the state $|{\rm vac}\rangle$
or to the Lagrangian. We choose the latter option such as to simplify notation.

There are then distinct path integrals with different boundary conditions
for each winding number $\Delta n=n-\bar n$
contained in the spacetime volume. This is because in regular gauge, the integral
over the topological term is determined by the configuration
of the gauge field at infinity, where the boundary conditions are imposed.
It also implies that the individual contributions must
be evaluated in the limit $VT\to \infty$, which turns out to be of substantial consequence.
We therefore consider these pieces separately.

First we have to specify the determinant of the Dirac operator in a general background with winding number $\Delta n=n-\bar{n}$. Naively one may write it as
\begin{align}
\label{eq:Diracdetgeneralnaive}
    \det ({\rm i}\slashed D-m{\rm e}^{{\rm i}\alpha\gamma^5})_{n,\bar{n}}=\left(\left.\det ({\rm i}\slashed D-m{\rm e}^{{\rm i}\alpha\gamma^5})\right|_{\eta=1}\right)^n \left(\left.\det ({\rm i}\slashed D-m{\rm e}^{{\rm i}\alpha\gamma^5})\right|_{\eta=-1}\right)^{\bar n}\,.
\end{align}
However, this would lead to an overcounting of the vacuum fluctuations
from the domains of spacetime far away from instantons or anti-instantons, where we recall
that e.g. the propagator reduces to its vacuum form in those regions,
cf. Eq.~\eqref{eq:proptotal}.
In order to count these fluctuations for the trivial background one time
and one time only, instead of Eq.~\eqref{eq:Diracdetgeneralnaive}, the correct contribution is
\begin{align}\label{eq:detDirac}
     \det ({\rm i}\slashed D-m{\rm e}^{{\rm i}\alpha\gamma^5})_{n,\bar{n}}&= \det ({\rm i}\slashed \partial-m{\rm e}^{{\rm i}\alpha\gamma^5})\left(\left. \frac{\det ({\rm i}\slashed D-m{\rm e}^{{\rm i}\alpha\gamma^5})}{\det ({\rm i}\slashed \partial-m{\rm e}^{{\rm i}\alpha\gamma^5})}\right|_{\eta=1}\right)^n\left(\left. \frac{\det ({\rm i}\slashed D-m{\rm e}^{{\rm i}\alpha\gamma^5})}{\det ({\rm i}\slashed \partial-m{\rm e}^{{\rm i}\alpha\gamma^5})}\right|_{\eta=-1}\right)^{\bar n}\notag\\
     &=\left.|\det (-\slashed \partial^{\rm E}-m{\rm e}^{{\rm i}\alpha\gamma^5})|\right|_{T^{\rm E}\rightarrow \ic T}\,{\rm e}^{-{\rm i}(\bar n-n)\alpha}(-\Theta)^{\bar n+n}\,,
\end{align}
which can be seen to follow formally from  Eq.~\eqref{eq:proptotal} and
where we have used Eqs.~\eqref{eq:detDiractrivial},~\eqref{eq:detrotation},~\eqref{eq:MinkowskiDet} and the fact that $\Theta$ is independent of the winding number $\eta$. Similarly for the functional determinant of the gauge and ghost fields, we have
\begin{align}
    {\det}^\prime_{\bar{A}_{n,\bar n}}={\det}_{\bar{A}=0} \left(\frac{{\det}^\prime_{\bar A}}{{\det}_{\bar{A}=0}}\right)^{n+\bar n}\,,
\end{align}
where $\bar{A}$ denotes the background gauge-field configuration and a prime on the determinant indicates that factors from zero eigenvalues have been deleted. Here ${\det}_A$ represents the functional determinant of the gauge and ghost fields in the one-instanton backgrounds. We have used  that the determinants for $\eta=1$ and $\eta=-1$ are identical, as can be seen to follow from the fact that the instanton and anti-instanton backgrounds are related by parity conjugation.

For notational convenience, we define
\begin{align}\label{eq:varpidef}
    \varpi\equiv\frac{1}{\sqrt{{\det}^\prime_{\bar{A}}/{\det}_{\bar{A}=0}}}\,.
\end{align}
Then for a two-point fermionic correlation function, we have to evaluate the contributions
\begin{align}
\langle \psi(x) \bar\psi(x^\prime)\rangle_{\Delta n}
=&\sum_{m}{}_{\rm out}\langle m+\Delta n|\psi(x) \bar\psi(x^\prime)|m\rangle_{\rm in}=\sum\limits_{\bar n,n\geq 0 \atop n-\bar n=\Delta n}
\int{\cal D}A_{\bar n,n} {\cal D}\bar\psi{\cal D}\psi\,
\psi(x)\bar\psi(x^\prime) {\rm e}^{{\rm i}S}
\notag\\
=&\sum\limits_{\bar n,n\geq 0 \atop n-\bar n=\Delta n}\frac{1}{\bar n! n!}
\left(\prod\limits_{\bar\nu=1}^{\bar n}\;\int\limits_{VT} {\rm d}^4x_{0,\bar \nu}{\rm d}\Omega_{\bar \nu} J_{\bar\nu}\right)
\left(\prod\limits_{\nu=1}^{n}\;\int\limits_{VT} {\rm d}^4x_{0,\nu}{\rm d}\Omega_\nu J_\nu\right)
{\rm i}S_{n,\bar{n}}(x,x^\prime)\notag\\
\times&\,
\left.|\det (-\slashed \partial^\E-m{\rm e}^{{\rm i}\alpha\gamma^5})|\,
\right|_{T^{\rm E}\rightarrow\ic T}({\det}_{\bar{A}=0})^{-1/2} \,
{\rm e}^{-S_{\rm E}(\bar n +n)}
{\rm e}^{-{\rm i}(\bar n -n)(\alpha+\theta)}
\varpi^{(\bar n +n)}
(- \Theta)^{\bar n+n}\,.
\label{Correlation:PathIntegral}
\end{align}
Here, $|n\rangle_{\rm in/out}$ are Heisenberg states at times $\mp T/2$, with well-defined Chern--Simons number, ${\cal D}A_{\bar n,n}$ stands for the restriction of the path integrals to fluctuations about the configuration with $\bar n$ instantons with $\eta=-1$ and
$n$ with $\eta=+1$, and the classical Euclidean action is $S_{\rm E}=8\pi^2/g^2$ (before adding the topological term). Note that the classical action for the $\vartheta$-dependent instanton solution is however $\vartheta$-independent, i.e. ${\rm i}S[\bar{A}^\vartheta]=-S^\E[\bar{A}^\E]$, cf. Ref.~\cite{Ai:2019fri}. This is also assumed for the topological contribution to the action. The collective coordinates corresponding to dilatational and gauge-orientation zero modes are integrated through ${\rm d}\Omega_{\bar\nu,\nu}$,
and $J_{\bar\nu,\nu}$ are the Jacobians that arise when trading the zero modes for collective coordinates, which are derived for Euclidean space in Refs.~\cite{tHooft:1976snw,Bernard:1979qt}. For the path integral in Minkowski spacetime, the
Jacobians are purely imaginary because of the analytic continuation of the collective coordinate
corresponding to time-translations~\cite{Ai:2019fri}.
Furthermore, all determinants are understood to be renormalized. In regards to the bosonic fluctuations, one can use here the results of Ref.~\cite{Ai:2019fri}, which show how the integral over the bosonic fluctuations on a thimble (i.e. an appropriately chosen contour for the bosonic path integral) about an analytically continued complex saddle, when the zero modes are separated, is related to the functional determinant evaluated at the corresponding Minkowskian saddle. The combinatorial factor $1/(\bar n! n!)$ is due to the fact that exchanging any two locations $x_{0,\bar\nu}$
or $x_{0,\nu}$ results in the same configuration. {We note that when integrating over fluctuations about all the dilute instanton backgrounds with finite action, we admit contributions from fluctuations that asymptotically take the form of plane waves which, despite having infinite action, do not contribute to the integral of the topological term in the Gau{\ss}ian approximation.  Thus the former integral remains proportional to an integer, and is given by $\theta(n-\bar n)$, as it appears in the result~(\ref{Correlation:PathIntegral}).}

The contribution  $Z_{\Delta n}$ from
the configurations with $\Delta n$ to the partition function, that is necessary for normalization, is computed as in Eq.~(\ref{Correlation:PathIntegral}), just with the factor $\psi(x)\bar\psi(x^\prime)$ deleted from the integrand:
\begin{align}\label{eq:Zn}\begin{aligned}
Z_{\Delta n}=&\sum_{m}{}_{\rm out}\langle m+\Delta n|m\rangle_{\rm in}=\sum\limits_{\bar n,n\geq 0 \atop n-\bar n=\Delta n}
\int{\cal D}A_{\bar n,n} {\cal D}\bar\psi{\cal D}\psi\,
{\rm e}^{{\rm i}S}\\
=&\sum\limits_{\bar n,n\geq 0 \atop n-\bar n=\Delta n}\frac{1}{\bar n! n!}
\left(-{\textstyle\int}\!{\rm d}\Omega\,J\, VT \,
\Theta\,\varpi\, {\rm e}^{-S_{\rm E}}\right)^{(\bar n +n)}\,\left.|\det (-\slashed \partial^\E-m{\rm e}^{{\rm i}\alpha\gamma^5})|\,
\right|_{T^{\rm E}\rightarrow\ic T}({\det}_{\bar{A}=0})^{-1/2}\,
{\rm e}^{-{\rm i}(\bar n -n)(\alpha + \theta)}\,.
\end{aligned}\end{align}
Here, we have carried out the spacetime integrals over the instanton locations, resulting in powers of the spacetime volume. Since we are considering here real time,
$\Delta n$ can be interpreted as the net change in Chern--Simons number
over the time $T$, i.e. each path integral associated with $\Delta n$
corresponds to a transition between states with Chern--Simons number $m$ and $m+\Delta n$, as suggested by the notation in  the first line of Eq.~\eqref{eq:Zn}. The factors $|\det (-\slashed \partial^{\rm E}-m{\rm e}^{{\rm i}\alpha\gamma^5})|_{T^{\rm E}\rightarrow\ic T}$ and $({\det}_{\bar{A}=0})^{-1/2}$ are common for all $Z_{\Delta n}$ and the correlation functions in backgrounds with any fixed $\Delta n$. They are thus total factors that cancel out in any physical quantities. To clean up notation, we will simply drop these factors below.

In order to evaluate the fermion correlation~(\ref{Correlation:PathIntegral}),
we first notice that for dilute instantons in a fixed configuration, as discussed around Eq.~\eqref{eq:proptotal}, the
correlation agrees with its form in the zero-instanton background
almost everywhere, except near the locations of the anti-instantons and
instantons.

Now for fixed $x$ and $x^\prime$, each spacetime integral
${\rm d}x_{0,\bar\nu}$ and ${\rm d}x_{0,\nu}$ sweeps over the point $(x+x^\prime)/2$ once, thus leading to $\bar n$ contribution with $\eta=-1$ and $n$ with $\eta=+1$. For a single of these integrals, e.g. for the location of a $\eta=-1$ instanton, this yields anomalous terms of the type
\begin{align}
\begin{aligned}
\int\limits_{VT}{\rm d}^4 x_{0,\bar\nu}\ {\rm i}S_{n,\bar{n}}(x,x^\prime)\approx&\,
\int\limits_{VT}{\rm d}^4 x_{0,\bar\nu} \left[{\rm i}S_{0{\rm inst}}(x,x^\prime)
{+}\frac{\varphi_{0{\rm L}}(x-x_{0,\bar\nu})\varphi_{0{\rm L}}^\dagger(x^\prime-x_{0,\bar\nu})}{m {\rm e}^{-{\rm i}\alpha}}+\cdots\right]
\\
=&\,VT\,({\rm i}S_{0{\rm inst}}(x,x^\prime)+\cdots){ +}m^{-1} {\rm e}^{{\rm i}\alpha} h(x,x^\prime)P_{\rm L}\,,
\end{aligned}\end{align}
where the dots represent the contributions to the propagator from the zero modes of the (anti)-instantons  whose centres were not integrated over (see Eq.~\eqref{eq:proptotal}), and $h(x,x')$ is defined as a block-diagonal matrix (with two identical blocks) satisfying
\begin{align}
\label{eq:hdef}
h(x,x^\prime)P_{\rm L}=&\, \int\limits_{VT}{\rm d}^4 x_{0,\bar\nu}\,\varphi_{0{\rm L}}(x-x_{0,\bar\nu}) \varphi_{0 {\rm L}}^\dagger(x^\prime-x_{0,\bar\nu})\,,& h(x,x^\prime)P_{\rm R}=& \, \int\limits_{VT}{\rm d}^4 x_{0,\nu}\,\varphi_{0{\rm R}}(x-x_{0,\nu}) \varphi_{0{\rm R}}^\dagger(x^\prime-x_{0,\nu})\,.
\end{align}
Unfortunately, we do not find an analytic expression for this matrix-valued function that depends on the invariant distance $(x-x^\prime)^2$ only.
Note though that this function
is independent of $VT$ as we take this spacetime volume to infinity. The overlap integral $h(x,x')$ as defined above depends on other collective coordinates of the instanton, e.g. the scale $\rho$. As such, insertions of $h(x,x')$ do not factor out of the integration over the collective coordinates. We choose then to approximate $h(x,x')$ by its average over the collective coordinates, defined as
\begin{align}
\label{eq:hav}
 \bar h(x,x')\equiv\frac{\int d\Omega \,h(x,x')}{\int d\Omega}.
\end{align}
This approximation allows to carry out all spacetime integrals over the instanton locations and collective coordinates. Neglecting contributions for which two or more of these locations coincide, the result is
\begin{align}
&\langle \psi(x) \bar\psi(x^\prime)\rangle_{\!\Delta n}\notag\\
=&
\sum\limits_{\bar n,n\geq 0 \atop n-\bar n=\Delta n}\frac{1}{\bar n! n!}
\Big[
{\,\bar h(x,x^\prime)}\left(\frac{\bar n}{m {\rm e}^{-{\rm i}\alpha}} P_{\rm L}+\frac{n}{ m {\rm e}^{{\rm i}\alpha} }P_{\rm R}\right) \left(VT\right)^{\bar n+n -1}
+{\rm i} S_{0{\rm inst}}(x,x^\prime) \left(VT\right)^{\bar n+n}
\Big]
({\rm i}\kappa)^{\bar n +n}
(-1)^{n+\bar n}
{\rm e}^{{\rm i}\Delta n(\alpha + \theta)}
\notag\\
=&\left[\left({\rm e}^{{\rm i}\alpha} I_{\Delta n+1}(2{\rm i}\kappa VT)P_{\rm L}+{\rm e}^{-{\rm i}\alpha} I_{\Delta n-1}(2{\rm i}\kappa VT) P_{\rm R}\right)\frac{{\rm i}\kappa}{m}\,\bar h(x,x^\prime)
+I_{\Delta n}(2{\rm i}\kappa VT){\rm i} S_{0{\rm inst}}(x,x^\prime)\right]
(-1)^{\Delta n}
{\rm e}^{{\rm i}\Delta n(\alpha + \theta)}
\,,
\label{correlation:sum}
\end{align}
where
\begin{align}\label{eq:kappadef}
{\rm i}\kappa={\textstyle\int}\!{\rm d}\Omega\,J\,
\,\Theta\, \varpi\,{\rm e}^{-S_{\rm E}}\,,
\end{align}
and $I_\alpha(x)$ is the modified Bessel function. Recall that the Jacobian $J$ contains an imaginary factor ${\rm i}$ and that $\Theta$ is a positive real number so that $\kappa$ is defined to be a positive number as well.
Correspondingly, the contributions to the partition function are found to be
\begin{align}
\label{Z:Delta:n}
Z_{\Delta n}=I_{\Delta n}(2 {\rm i}\kappa VT)\,(-1)^{\Delta n}{\rm e}^{{\rm i}\Delta n(\alpha+\theta)}\,.
\end{align}

Notice that all terms appearing in the fermion correlation~\eqref{correlation:sum} as well
as the partition function~\eqref{Z:Delta:n} are multiplied by the same global phase
$\exp({\rm i}\Delta n(\alpha+\theta))$. This is illustrated in Figure~\ref{fig:phase:attribution} and can be attributed to the fact that the fermion determinants and topological phases multiply all operators computed in the path integral, no matter whether these are fermionic or not or whether they are induced by instantons.

\subsection{Boundary conditions in the saddle point expansion}
\label{sec:bc}

For an infinite spacetime volume, in order to have saddle points with finite action,
one should impose on the path integral boundary conditions with vanishing physical fields at infinity~\cite{Coleman:1985rnk}. This is a standard procedure. As we will see in Section~\ref{subsec:summation}, its correct implementation turns out however to be decisive for the calculation of correlation functions, in particular when $\alpha+\theta\not=0$. We therefore consider here some aspects of boundary conditions in the present context of the saddle point expansion around instantons.

In order to appreciate the necessity of using boundary conditions at infinity, we consider in contrast how we would need to proceed when restricting the path integral to a finite region of spacetime. Fixing the boundary conditions up to gauge transformations on its finite surface, which is homeomorphic  to a three sphere, leads to the quantization of topological sectors.  However, while this is the procedure that would lead to the conclusion that $CP$ violation is present in the strong interactions (see Section~\ref{subsec:summation}), there is no physical principle that would allow for only considering a single boundary configuration, for example the vanishing configuration, for the physical fields around a finite spacetime volume. This matter is reviewed in Section~\ref{sec:schroedinger}, where we argue instead that one should sample over a weighted range of boundary conditions that can theoretically be obtained by projecting the Schrödinger wave functional on field eigenstates. Practically, in the presence of interactions, the wave functional is not known in most of its details but yet some essential properties may be inferred. In the present case, this feature is the vacuum angle $\theta$.

Now, when sampling over a range of boundary conditions, the winding number is no longer restricted to integer values, such that it may not be clear why $\theta$ should behave as an angular variable. However, since the wave functional is not known in detail, the only practicable way of evaluating the path integral again goes via taking the boundaries to infinity. Then, even for a continuum of boundary conditions weighted by the wave functional, the only saddle points of the Euclidean action occur for vanishing physical field configurations, for which the winding number again takes integer values. We note that taking the volume of Euclidean spacetime to infinity projects on the lowest accessible energy eigenstate. (After all, given that only vanishing physical fields on the boundary at infinity lead to saddle points, it should be clear that in this approximation we cannot calculate correlation functions for different states, so that this projection does not amount to an additional restriction.)  Nonetheless, the configurations of fixed $\Delta n$ do not change the vacuum angle $\theta$ such that this quantum number of the vacuum state remains conserved also in the limit of an infinite spacetime volume.

Another point of view on the boundary conditions on finite spacetime volumes is to leave these open and to integrate out the fluctuations over the remainder of spacetime. We carry this out in Section~\ref{sec:CDC1} and~\ref{sec:CDC}, where it is seen that $CP$ phases from the two partitions of spacetime cancel. This means that after all the boundary condition imposed on the remainder at infinity remains relevant.

From these considerations, we conclude that when using the saddle point expansion to compute path integrals in the presence of gauge theory instantons---there appear to be no viable alternatives as far as analytical approximations are concerned---the boundary conditions should be given by vanishing physical fields, and for consistency these should be imposed at spacetime infinity. In the following subsection, we show that this has material consequences for the question of $CP$ violation in the strong interactions.

\subsection{Summation over topological sectors}
\label{subsec:summation}

The total partition function, given by the transition amplitude from the vacuum $|\rm vac\rangle$ onto itself, is given by
\begin{align}
 Z(N,VT)=\sum_{{m,n}\atop{|m-n|\leq N}}{}_{\rm out}\langle m| n\rangle_{\rm in}=\sum\limits_{\Delta n=-N}^{N}\sum_{m}{}_{\rm out}\langle m+\Delta n| m\rangle_{\rm in}=\sum_{\Delta n=-N}^{N}Z_{\Delta n}(VT).
\end{align}
Above, we have regulated the sum over topological sectors by introducing a cutoff $N$.
While we occasionally suppress the arguments of the partition functions,
we have made here explicit that $Z_{\Delta n}$ is a function of $VT$ as per Eq.~(\ref{eq:Zn}). Eventually, $N$ and $VT$ are to be taken to infinity. This can lead to singular behaviour when considering $Z$ in isolation, as we discuss at the end of Section~\ref{sec:several:flavours}, but can be carried out for normalized correlation functions. In particular, the fermion correlator in the vacuum~(\ref{eq:vacuum}) is given by
\begin{align}\begin{aligned}
 \langle\psi(x)\bar\psi(x')\rangle \equiv&\,\,\lim_{N\to\infty \atop N\in \mathbbm N} \lim_{VT\to\infty} \frac{1}{Z(N,VT)} \sum_{{m,n}\atop{|m-n|\leq N}}\,{}_{\rm out}\langle m|\psi(x)\bar\psi(x')|n\rangle_{\rm in}\\=&\lim_{N\to\infty \atop N\in \mathbbm N} \lim_{VT\to\infty}\frac{\sum_{\Delta n=-N}^{N}\sum_{n}{}_{\rm out}\langle n+\Delta n|\psi(x)\bar\psi(x')| n\rangle_{\rm in}}{\sum_{\Delta n=-N}^{N} Z_{\Delta n}}\\
 =&\,\lim_{N\to\infty \atop N\in \mathbbm N} \lim_{VT\to\infty}\frac{\sum_{\Delta n=-N}^{N}\langle \psi(x) \bar\psi(x^\prime)\rangle_{\!\Delta n}}{\sum_{\Delta n=-N}^{N} Z_{\Delta n}}
={\rm i}S_{0\text{inst}}(x,x^\prime)
+{\rm i}\kappa
\bar h(x,x^\prime) m^{-1} {\rm e}^{-{\rm i}\alpha\gamma^5}\,.
\label{correlation:effective}
\end{aligned}\end{align}

We emphasize that the order of the limits follows from the fact that integer winding numbers $\Delta n$ are a consequence of the requirement of finite saddle-point actions in infinite spacetime volumes $VT\to\infty$~\cite{Coleman:1985rnk}, as it is explained in Section~\ref{sec:bc}. To meet this, the physical gauge fields must vanish at infinity, leaving the possibility of pure gauge configurations on the boundary. Topologically, the boundary of the spacetime 
is homeomorphic to $S^3$. The maps between the latter and the gauge group 
can be characterized by equivalence classes corresponding to the winding numbers $\Delta n$.
Discrete sectors $\Delta n$ may also arise when imposing instead boundary conditions on some finite subvolume, but then there is no reason (such as finite action at the saddle points) to fix these constraints which are hence unphysical. (Instead, one should sum over all possible boundary conditions for the subvolume as done in Section~\ref{sec:clustering:inf:vol}.) Therefore, only the restriction to field configurations with vanishing physical boundary conditions in infinite spacetime volumes leads to integer $\Delta n$.

For more details on the derivation of Eq.~\eqref{correlation:effective}, we note that the connected correlator is constructed as the limit of infinite $N$ for a series of normalized expectation values. The latter are given by the
  ratios of corresponding partial sums for given $N$ and $VT\to\infty$.
 (This is stated for definiteness---one may choose to use alternative decompositions of the sums.)  In effect, we take the limit of $N$ going to infinity simultaneously in the numerator and the denominator  in order to make sense of the quantities whose limit in isolation is ill-defined. Equation~\eqref{correlation:effective} with
its prescription of a common limit also follows more formally when introducing fermionic currents $\j,\bar \j$ in the partition function for fixed $N$ and define the connected fermion correlators in terms of the derivatives of its logarithm with respect to the sources:
\begin{align}
\langle\psi(x)\bar\psi(x')\rangle=\lim_{N\to\infty \atop N\in \mathbbm N} \lim_{VT\to\infty} \left.\frac{\delta^2}{\delta \bar\j(x)\delta  \j(x')}\log Z(N,VT,\j,\bar \j)\right|_{\j=\j'=0}.
\end{align}

In regards to the limit of infinite spacetime volume,in Eq.~\eqref{correlation:effective} we have used the result $\lim_{x\to\infty}I_{\Delta n}({\rm i} x \,{\rm e}^{-{\rm i} 0^+})/I_{\Delta n^\prime}({\rm i}x \,{\rm e}^{-{\rm i} 0^+})=1$. The factor ${\rm e}^{-{\rm i}0^+}$ is due to the rotation $T^\E\rightarrow {\rm i}{\rm e}^{-{\rm i} 0^+} T$ so that the Jacobian $J$ actually contains a factor of ${\rm i}{\rm e}^{-{\rm i}0^+}$. The limit however also holds
for real positive arguments in the modified Bessel functions such that the steps presented here can also be applied in Euclidean space.
With the leading asymptotic behaviour of the modified Bessel functions then being independent of $\Delta n$, the remaining terms with exponents of $\Delta n$ in Eqs.~(\ref{correlation:sum}) and~(\ref{Z:Delta:n}) lead to a geometric series that cancel between numerator and denominator. For some values where $(\alpha+\theta) = 2\pi q$, with $q$ a rational number, there occur partial sums in the numerator and denominator where the geometric series evaluate to zero. In this case it is not appropriate to proceed by going beyond the leading asymptotic behaviour of the modified Bessel functions, which is suppressed by an extra power of $\kappa VT$ (cf. Eq.~(\ref{eq:Inas2})). For each sector $\Delta n$, this corresponds to an arbitrarily small (as $VT\to\infty$) correction
to the leading contribution and therefore cannot be reliably calculated in any approximation. Rather, the problematic partial sums should be dealt with by taking $VT\to \infty$ first for general $(\alpha+\theta)$ and then taking the limit of the problematic rational value for the ratio of the partial sums. Around that rational value, there is always a neighbourhood where the partial sum is well defined using
the leading asymptotic behaviour of the modified Bessel functions, such that the limit for the ratio of the partial sums indeed exists. Another point to note is that the fermion determinant included in $\kappa$ contains a leading factor $m$ that
cancels with the explicit occurrence of $m^{-1}$ in the final expression in Eq.~\eqref{correlation:effective}, leading to a finite nonperturbative correction even in the massless limit.

We next consider what would happen if we summed over the topological sectors in a finite spacetime volume first and took the latter to infinity in the last step.
As discussed in Section~\ref{sec:bc}, on finite surfaces one should integrate over a set of physical boundary conditions given by the wave functional. When  instead using  vanishing physical fields as boundary conditions, one is forced to impose these at infinity.
If nonetheless the order of the limits were immaterial, one could yet consider taking the subvolume to infinity after the summation over $\Delta n$.
However, the order is of crucial relevance for the form of the final result because
if we were not taking $VT\to\infty$ first, we would instead obtain
\begin{align}
&\sum\limits_{\bar n,n\geq 0}\frac{1}{\bar n! n!}
\Big[
{\,\bar h(x,x^\prime)}(\bar n\, m^{-1} {\rm e}^{{\rm i}\alpha} P_{\rm L}+n\, m^{-1} {\rm e}^{-{\rm i}\alpha} P_{\rm R}) \left(VT\right)^{\bar n+n -1}
+{\rm i} S_{0{\rm inst}}(x,x^\prime) \left(VT\right)^{\bar n+n}
\Big]
(-\ic\kappa)^{\bar n +n}
{\rm e}^{{\rm i}\Delta n(\alpha + \theta)}
\notag\\
&\hskip2.5cm=\left[-\left({\rm e}^{-{\rm i}\theta} P_{\rm L}+{\rm e}^{{\rm i}\theta}  P_{\rm R}\right)\frac{\ic \kappa}{m}\bar h(x,x^\prime)
+{\rm i} S_{0{\rm inst}}(x,x^\prime)\right]{\rm e}^{-2 \ic\kappa VT\cos(\alpha+\theta)}
\,.
\label{correlation:limtsexchanged}
\end{align}
Analogously, taking the $VT\rightarrow\infty$ limit in the end, the total partition function would be
\begin{align}
\label{Z:wrongorderoflimits}
 Z\rightarrow\sum_{n,\bar n}\frac{1}{n!\bar n!}(-\ic\kappa V T)^{\bar n+n}{
\rm e}^{-\ic (\bar n-n)(\alpha+\theta)}={
\rm e}^{-2\ic\kappa VT \cos(\alpha+\theta)}.
\end{align}
(From this expression, one may read the $\theta$-dependence of the vacuum energy density as $E(\theta)/V=2\kappa\cos\bar\theta$, where $\kappa>0$. For pure gauge theories without fermions the energy of the $\theta$-vacuum is $E(\theta)/V=-2\kappa^\prime\cos\theta$ with $\kappa^\prime>0$. The respective minus sign is due to the minus sign attached to the Dirac mass term, see Eqs.~\eqref{eq:EuclideanDet} and~\eqref{eq:MinkowskiDet}. The sign can be removed by shifting either $\theta$ or $\alpha$ by a value of $\pi$ in their definitions.)
For the two-point function,
we see that different phases are multiplying the left and right anomalous terms when compared to Eq.~(\ref{correlation:effective}).
One may notice here that in the limit $|\Delta n|\ll\bar n+ n$, which
gives the dominant contributions to the binomial distribution for $VT\to\infty$~\cite{Diakonov:1985eg}, there are no relative chiral phases
between the anomalous terms involving $\bar{h}$ and the term containing ${\rm i} S_{0{\rm inst}}(x,x^\prime)$.
This would indicate that any $CP$-violating contribution from
a background with $|\Delta n|\ll\bar n+ n$, that can e.g.
be measured by an observer in the same background, is suppressed by the volume.
The fact that in Eq.~(\ref{correlation:limtsexchanged}) the $CP$-violation
is enhanced follows from a cancellation of phases that
is a consequence of the exchange of limits in Eq.~(\ref{correlation:effective}).
We comment on the relevance of the different phases appearing in
Eqs.~(\ref{correlation:effective}) and~(\ref{correlation:limtsexchanged})
in the following.

We observe that in Eq.~(\ref{correlation:effective}) the chiral phase multiplying
the anomalous term proportional to $\bar h$ is the same as the
one that appears together with ${\rm i}S_{0{\rm inst}}$ (see Eq.~\eqref{S:0inst}).
Furthermore, the anomalous
term has the expected exponential suppression compared to the contributions
corresponding to regions that are not influenced by the instantons.
As a consequence, this correlation function does not exhibit $CP$ violation.
The instanton effects are often approximated in terms
of an effective operator~\cite{tHooft:1976rip,tHooft:1976snw}, which in our case,
based on Eq.~(\ref{correlation:effective}) reads
\begin{align}
\label{L:eff}
{\cal L}\to{\cal L}-\bar \psi(x) \Gamma {\rm e}^{{\rm i}\alpha\gamma^5} \psi(x)\,,
\end{align}
where at leading order in a gradient expansion $\Gamma$ is a real number that can in principle be inferred from Eq.~(\ref{correlation:effective}), in particular after an appropriate
treatment of the dilatations, where the symmetry is broken radiatively.
This corresponds to an effective mass with a chiral phase that is aligned
with the one in the Dirac operator~(\ref{Dirac:massive}).
This alignment then leads to the absence of relative $CP$-odd phases in different contributions to an amplitude, as illustrated in Figure~\ref{fig:phase:attribution}.
Further when using the
operator~(\ref{L:eff}) together with the Dirac mass in order to build an
effective theory valid below the scale of chiral symmetry breaking, as we outline in Section~\ref{sec:chiral_Lagrangian}, there is
only one $CP$-odd phase that can be removed by a field redefinition. Consequently,
the theory explains the absence of $CP$-violating observables, such as the vanishing
permanent electric dipole moment of the neutron~\cite{Baluni:1978rf,Crewther:1979pi} or the nonobservation of the
decay of an $\eta^\prime$-meson into two pions.
This is to be compared with what one would infer from
Eq.~(\ref{correlation:limtsexchanged}),
\begin{align}
\label{L:eff:old}
{\cal L}\to{\cal L}+\bar \psi(x) \Gamma {\rm e}^{-{\rm i}\theta\gamma^5} \psi(x)\,.
\end{align}
Here, the difference between the phase $-\theta$ and the phase $\alpha$ from
a perturbative insertion of the mass $m$ in a fermion line would indicate
a $CP$-odd phase that cannot be removed by a field redefinition.
We emphasize that for Eqs.~(\ref{L:eff})
and~(\ref{L:eff:old}), no assumption about the values of $\theta$ and $\alpha$
are made, which of course transform under chiral rotations of the
fermion fields while leaving the sum $\alpha+\theta$ invariant. It should be noted that the phase
in the operator in Eq.~\eqref{L:eff} is compatible with the following selection rule implied by the anomalous Ward identity: The theory should be invariant under a chiral transformation supplemented with changes in $\alpha,\theta$ going as follows:
\begin{align}
\label{eq:selection}
\psi\rightarrow &\,{\rm e}^{{\rm i}\beta\gamma^5}\psi, & \bar\psi\rightarrow &\,\bar\psi\, {\rm e}^{{\rm i}\beta\gamma^5}, &
 \alpha\rightarrow &\,\alpha-2\beta, &\theta\rightarrow \theta+2\beta,
\end{align}
where $\beta$ is the parameter of the transformation.
The previous selection rule is usually invoked as a justification of an effective operator involving the $\theta$ parameter as in Eq.~\eqref{L:eff:old}; however, this is not the only possibility, and the result of \eqref{L:eff} is equally compliant with the selection rule. We stress again that, given our results for the fermionic fluctuation determinants, our expressions capture the full dependence on the chiral angle $\alpha$.
It can also be observed that while Eq.~(\ref{correlation:effective}) shows that the breaking of the axial ${\rm U}(1)$ symmetry due to the fermion mass is enhanced by the effect of the instantons in a way that is independent of the absolute value of the mass, this still leaves open the question of how the correlations and the low-energy effective theory behave in the massless limit.

{We emphasize that the results obtained by taking the limit of infinite spacetime volume before the sum over topological sectors, as {it is in order for spacetimes that have boundaries at infinity}, applies both for finite and infinite {\it spatial} volumes. Thus, our results hold not only for Minkowski spacetime, but also for spacetimes like $\mathbb{R}\times S_3$ in which space is the compact hypersphere. Crucially, the results are valid as long as there is an infinite extent of time, as required when considering  the $\theta$-vacua as in and out states.}

Nonetheless,
topological sectors with fixed winding number $\Delta n$ are well-defined
within finite spacetimes with periodic boundary conditions~\cite{Leutwyler:1992yt}, i.e. without boundaries. The periodicity and finiteness remove the necessity
of specifying vacuum boundary conditions of a certain Chern--Simons number. This precludes the interpretation of the path integral within a sector of fixed $\Delta n$ as a transition amplitude between vacua with Chern--Simons numbers differing by $\Delta n$. Because of this, there is no principle (save for some correspondence
with the infinite-volume limit) that requires certain weights
for the contributions to the path integral from different $\Delta n$.
We may note that the quantum equations of motion (that may also
apply to an observer) are separately independent for each sector $\Delta n$ in
the periodic spacetime, i.e. $\Delta n$ will appear fixed within each
such sector. {For an observer that can be understood as a local excitation of fields, one expects a unique value of $\Delta n$, which could be measured } e.g. through the correlation
function~(\ref{correlation:sum}). This leads for example to a permanent electric dipole moment of the neutron that depends on $\Delta n$ but is independent of $\bar\theta$ {and whether there are additional topological sectors and whatever happens in these}. Interferences between the different sectors will therefore
not be seen by observers defined through local quantum fields; rather, they would require ``super-observers'' with access to all topological sectors.
From the absence of material effects from interference it follows that the predictions for
the $\theta$-vacuum in a spacetime that has boundaries at infinity, where the limit $VT\to\infty$ is to be taken before
the summation over $\Delta n$, coincide with what is seen
in a large but finite periodic spacetime with $\Delta n=0$. In contrast to common expectations, the path integration over a finite spacetime within a single topological sector still complies with cluster decomposition up to volume suppressed effects, as we discuss in detail in Section~\ref{sec:cluster_decomposition_finite_VT}.
Under the assumption of a finite volume of a spacetime without boundaries, the condition $\Delta n=0$ should be imposed in agreement with the observation that there is no spontaneous $CP$-violation, i.e.
$\langle \Delta n\rangle=0$
in the vacuum for any subvolume of physical spacetime.

\subsection{Several fermion flavours}
\label{sec:several:flavours}

The previous conclusions can be extended to correlation functions in theories with more fermion flavours. In a theory with $N_f$ Dirac fermions $\psi_j\,,\,j=1,\dots,N_f,$ in the fundamental representation of the gauge group and with complex masses $m_j {\rm e}^{{\rm i}\alpha_j\gamma_5}$, one can consider correlation functions of the form
\begin{align}
\langle \prod_{j=1}^N(\psi_{\sigma(j)} \bar\psi_{\sigma(j)})\rangle
=&\frac{1}{Z}\int{\cal D} A \prod_{k=1}^{N_f}\left({\cal D}\bar\psi_k {\cal D}\psi_k\right) \,\prod_{j=1}^N(\psi_{\sigma(j)} \bar\psi_{\sigma(j)}) {\rm e}^{{\rm i} S}\,,
\end{align}
where $\sigma=\{\sigma(1),\dots,\sigma(N)\}$ is a set containing $N$ flavour indices (e.g. the list of all indices, a subset thereof {or other variants,} {in case of which some indices may be repeated}), and  we have not specified spacetime indices or the different possible Lorentz contractions in order to simplify the notation. As before, we construct the correlation function by summing over contributions from topological sectors with fixed winding number $\Delta n$:
\begin{align}
\langle\prod_{j=1}^N( \psi_{\sigma(j)}\bar\psi_{\sigma(j)})\rangle_{\Delta n}
=&\sum\limits_{\bar n,n\geq 0 \atop n-\bar n=\Delta n}\frac{1}{\bar n! n!}
\left(\prod\limits_{\bar\nu=1}^{\bar n}\;\int\limits_{VT} {\rm d}^4x_{0,\bar \nu}{\rm d}\Omega_{\bar \nu} J_{\bar\nu}\right)
\left(\prod\limits_{\nu=1}^{\bar n}\;\int\limits_{VT} {\rm d}^4x_{0,\nu}{\rm d}\Omega_\nu J_\nu\right)
\prod_{j=1}^N\left({\rm i}S_{\sigma(j)}\right)\notag\\
\times&
{\rm e}^{-S_{\rm E}(\bar n +n)}
{\rm e}^{{\rm i}\Delta n(\bar\alpha+\theta)}
\varpi^{(\bar n +n)}
\,\bar\Theta^{(\bar n +n)}\,(-1)^{N_f(\bar n+n)}\,,
\label{Correlation:PathIntegral2}
\end{align}
where  $\bar\alpha$ denotes the argument of the determinant of the fermionic mass matrix,
\begin{align}\label{eq:alphabar}
 \bar\alpha=\sum_j^{N_f}\alpha_j
\end{align}
and
\begin{align}
\bar\Theta=\prod_{j=1}^{N_f}\Theta_j\,,
\end{align}
where $\Theta_j$ is defined for each flavour in analogy with Eq.~(\ref{eq:MinkowskiDet}). { As before, we have dropped factors involving the determinants of the free fermionic and bosonic fluctuation operators { since they appear in both the (unnormalized) correlators and partition functions}.} Note that $\bar{\Theta}$ is also a positive real number.

The partition functions $Z_{\Delta n}$, on the other hand, are now given by
\begin{align}\label{eq:ZNf}\begin{aligned}
Z_{\Delta n}
=&\sum\limits_{\bar n,n\geq 0 \atop n-\bar n=\Delta n}\frac{1}{\bar n! n!}
\left({\textstyle\int}\!{\rm d}\Omega\,J\, VT \,
\bar\Theta\,\varpi\, {\rm e}^{-S_{\rm E}}\right)^{(\bar n +n)}\, (-1)^{N_f(\bar n+n)}
{\rm e}^{{\rm i}\Delta n(\bar\alpha+ \theta)}\\
\,\equiv &\,\sum\limits_{\bar n,n\geq 0 \atop n-\bar n=\Delta n}\frac{1}{\bar n! n!}({\rm i}\kappa_{N_f} VT)^{\bar n+n}\,(-1)^{N_f\Delta n}\,{\rm e}^{{\rm i}\Delta n(\bar\alpha+ \theta)}=I_{\Delta n}(2 \ic\kappa_{N_f}VT)\,(-1)^{N_f\Delta n}\,{\rm e}^{{\rm i}\Delta n(\bar\alpha+ \theta)}\,,
\end{aligned}\end{align}
where we partly abbreviate the factors in the round bracket by $\ic\kappa_{N_f}$.

Using propagators of the form of Eq.~\eqref{eq:proptotal} and approximating nontrivial integrals over the translational coordinates $x_{0,\nu},x_{0,\bar\nu}$ by their averages over the remaining collective coordinates, as in Eqs.~\eqref{eq:hdef}, \eqref{eq:hav}, we have the following types of contributions:
\begin{itemize}
 \item terms with only propagators as in the zero-instanton background,

 \item ``diagonal'' terms, which are obtained by summing over terms in which all zero modes correspond to a common (anti-)instanton,

 \item ``off-diagonal" contributions which mix zero modes from different (anti-)instantons.
\end{itemize}
For contributions with only propagators as in the  zero-instanton background, the integrals over the centres are trivial and simply lead to $Z_{\Delta n}\prod_j {\rm i}S_{\sigma(j),0\rm inst}$, so that the ensuing contributions to the full correlator are simply given by products of these propagators. The ``diagonal'' contributions involve overlap integrals over varying numbers of zero-modes of a single (anti-)instanton. When summing over (anti-)instantons, one always gets a factor of $n$  ($\bar n$), exactly as in the two-point function case analyzed before, resulting in contributions that go  schematically as (for the case of instantons)
\begin{align}\label{eq:correlatorsNf}\begin{aligned}
 &\left(\prod_{m=1}^p {\rm i}S_{\sigma_p(m),0\rm inst}\right)\left(\prod\limits_{j=1}^q m^{-1}_{\sigma_q(j)}{\rm e}^{{-\rm i}\alpha_{\sigma_q(j)}} P_{{\rm R}\sigma_q(j)}\right)\bar h_{q}\sum\limits_{\bar n,n\geq 0 \atop n-\bar n=\Delta n}\frac{n}{\bar n! n!}\, (VT)^{\bar n+n-1}(\ic\kappa_{N_f})^{\bar n+n}\,(-1)^{N_f\Delta n}\,{\rm e}^{{\rm i}\Delta n(\bar\alpha+ \theta)}\\
 =&\left(\prod_{m=1}^p {\rm i}S_{\sigma_p(m),0\rm inst}\right)\left(\prod\limits_{j=1}^q m^{-1}_{\sigma_q(j)}{\rm e}^{-{\rm i}\alpha_{\sigma_q(j)}}P_{{\rm R}\sigma_q(j)}\right) \bar h_{q} \, (\ic\kappa_{N_f})\,I_{\Delta n-1}(2\ic\kappa_{N_f}VT)\,(-1)^{N_f\Delta n}\,{\rm e}^{{\rm i}\Delta n(\bar\alpha+ \theta)}\,.
 \end{aligned} \end{align}
 In this equation $\sigma_{p/q}=\{\sigma_{p/q}(1),\dots,\sigma_{p/q}(p/q)\}$  are  subsets of the set $\sigma$ defined above, with $p+q=N$, $\sigma_{p}\cup\sigma_{q}=\sigma$. $P_{{\rm R}\sigma_q(j)}$ are right-handed projectors for the flavour $\sigma_q(j)$, while $\bar h_{q}$ denotes a generalized tensor-valued overlap integral constructed from a product of $q$ instanton zero-mode projectors, averaged over the collective coordinates of the instanton. As before, when computing contributions to the fermion correlation by taking the infinite volume limit, summing over $\Delta n$ and dividing by the partition function, the phases proportional to $\bar\alpha+\theta$ drop out, and one ends up with contributions to the correlator of the form
 \begin{align}
 \label{eq:diagcorr}
 \langle\prod_{j=1}^N( \psi_{\sigma(j)}\bar\psi_{\sigma(j)})\rangle \supset  \left(\prod_{m=1}^p {\rm i}S_{\sigma_p(m),0\rm inst}\right)\left(\prod\limits_{j=1}^q m^{-1}_{\sigma_q(j)}{\rm e}^{-{\rm i}\alpha_{\sigma_q(j)}}P_{{\rm R}\sigma_q(j)}\right) \bar h_{q} \, (\ic\kappa_{N_f})\,.
 \end{align}
As in the single-flavour case, all the phases of the correlators are determined by the chiral phases in the mass matrices, and similar results hold for the diagonal anti-instanton contributions. The contributions to the correlators can be captured by effective operators whose $\alpha_j$-dependent phases are in accordance with the generalization of the selection rule of Eq.~\eqref{eq:selection} for $N_f$ flavours, which reads
\begin{align}
\label{eq:selectionNf}
\psi_j\rightarrow &\,{\rm e}^{{\rm i}\beta\gamma^5}\psi_j, & \bar\psi_j\rightarrow &\,\bar\psi_j\, {\rm e}^{{\rm i}\beta\gamma^5}, &
 \alpha_j\rightarrow &\,\alpha_j-2\beta, &\theta\rightarrow \theta+2N_f\beta\,.
\end{align}

In particular, the 't Hooft interactions with $N_f$ flavours induced by (anti-)instantons correspond to diagonal contributions to correlators with $N=N_f$ pairs of fermions, $p=0$ and $q=N_f$, with the resulting effective vertices having the form
\begin{align}
\label{eq:tHooft}
 {\cal L}\rightarrow {\cal L}-\Gamma_{N_f}{\rm e}^{-{\rm i}\bar\alpha}\prod_{j=1}^{N_f}(\bar\psi_j P_{\rm L}\psi_j)-\Gamma_{N_f}{\rm e}^{{\rm i}\bar\alpha}\prod_{j=1}^{N_f}(\bar\psi_j P_{\rm R}\psi_j),
\end{align}
where at leading order in a gradient expansion the $\Gamma_{N_f}$ are constant.
Note how the dependence on the chiral phases is such that all of these can be removed by the same redefinitions that get rid of the phases in the tree-level mass terms.
Once again, had we done the summation over $\Delta n$ before taking the infinite volume limit, we would have obtained different phases, with $\bar\alpha$  replaced by $-\theta$. For these 't Hooft interactions, the $q=N_f$ factors of $m^{-1}_{\sigma_q(j)}$ in Eq.~\eqref{eq:diagcorr} are canceled with the factor of $\prod_{j=1}^{N_f} m_j$  associated with the fermionic zero modes implicit in $\kappa_{N_f}\propto\bar \Theta$. Diagonal correlators with $p=0$ but $N<N_f$ yield additional interaction vertices with fewer fermions, higher powers of $m_i$ and phases compatible again with the selection rule, confirming the symmetry arguments put forth for example in the context of $\rm SU(2)$ instantons in Ref.~\cite{Cerdeno:2018dqk}.
Finally, the off-diagonal terms involve contributions to the fermionic propagators coming from
different instantons. These can be classified according to the number of different (anti-)instantons involved and the number of propagators corresponding to each (anti-)instanton. Each class has an associated combinatorial factor for the number of terms in the class contained in the product of fermion propagators of the form of Eq.~\eqref{eq:proptotal}. For example, as we have seen, the diagonal class of single-(anti-)instanton contributions has an associated combinatorial factor of $n(\bar n)$.
Now for the off-diagonal term, suppose we consider a
class where $m$ different instantons are involved. This amounts to $m$ combinations from a set of size $n$ and gives a
combinatorial factor $n!/(m!(n-m)!)$. In this case the integrals over the translational collective coordinates give now contributions proportional to
\begin{align}
 \sum\limits_{\bar n,n\geq 0 \atop n-\bar n=\Delta n}\frac{1}{\bar n! m!(n-m)!} (VT)^{\bar n+n-m}{(\ic \kappa_{N_f})^{\bar n+n}}(-1)^{N_f\Delta n}{\rm e}^{{\rm i}\Delta n(\bar\alpha+ \theta)}=\frac{(\ic\kappa_{N_f})^m}{m!}I_{\Delta n-m}(2\ic\kappa_{N_f}VT)(-1)^{N_f\Delta n}{\rm e}^{{\rm i}\Delta n(\bar\alpha+ \theta)}\,.
\end{align}
Since $\kappa_{N_f}\propto {\rm e}^{-S_\E}$, we see that these contributions have a higher suppression factor and are expected to be subdominant. Nevertheless, taking the limit of $VT\rightarrow\infty$ before summing over $\Delta n$ and dividing by the partition function, the dependence on $\theta$ drops from the corresponding contribution to the correlator. Analogous results hold for other contributions involving anti-instantons, or mixed instantons and anti-instantons: In general one obtains modified Bessel functions multiplied by extra factors of $\kappa_{N_f}$ and inverse powers of $VT$. This makes the terms subleading but also in such a way that the $\theta$-dependence disappears from the final contributions to the correlators.

Our closing remark for this section is that the partition function
\begin{align}
 Z=\lim_{N\to\infty \atop N\in \mathbbm N} \lim_{VT\to\infty} \sum\limits_{\Delta n=-N}^N Z_{\Delta n},
\end{align}
with $Z_{\Delta n}$ given in Eq.~\eqref{eq:ZNf}, exhibits a non-analytic behaviour in $\theta$. Indeed, as pointed out above for $VT\rightarrow\infty$, one has $I_{\Delta n}(2\ic \kappa_{N_f} VT)\sim I_0(2\ic \kappa_{N_f} VT)$, so that
\begin{align}\begin{aligned}
Z=I_0(2\ic \kappa_{N_f} VT)\lim_{N\to\infty \atop N\in \mathbbm N} \sum_{|\Delta n|\leq N}\,{\rm e}^{{\rm i}\Delta n(\bar\alpha+ \theta+N_f\pi)}\,.
\end{aligned}\end{align}
With $\theta'\equiv \bar\alpha+\theta+N_f\pi$, the dependence on $\theta'$  becomes proportional to a periodic delta function maximized at $0$ and $2\pi$. The corresponding Euclidean partition function is thus maximized at $\theta'=0$, which complies with the general expectations  argued for in Ref.~\cite{Vafa:1984xg}, yet with a nonanalytic dependence on $\theta'$. This is to be contrasted with the standard result
\begin{align}\label{eq:Zusual}
 \lim_{VT\to\infty} \lim_{N\to\infty \atop N\in \mathbbm N} \sum\limits_{\Delta n=-N}^N Z_{\Delta n}={\rm e}^{2\ic\kappa_{N_f} VT\cos(\bar\alpha+\theta+N_f\pi)}
\end{align}
which, while having an Euclidean counterpart maximized at $\theta'=0$, retains analyticity in $\theta$. As discussed in Section~\ref{subsec:suceptibility}, several standard results in the literature linking $\theta$ with $CP$ violation through the topological susceptibility $\chi_\Omega\equiv\langle\Delta n^2\rangle/\Omega$, where $\Omega$ denotes a spacetime volume, have been derived assuming an analytic dependence on $\theta$ in the partition function and thus do not necessarily apply given the ordering of limits proposed in this paper. Let us remark that the partition function itself is not observable, and the singular limit is not pathological because the resulting {correlation} functions have a well defined limit. The possibility of a non-analytic dependence on $\theta$ has been succinctly considered in Ref.~\cite{Witten:1979vv}, where nonanalyticity was linked to periodicity in $\theta$. We emphasize again that the latter is related to the quantization of the topological charge, which as argued in Section~\ref{sec:bc} is required for an infinite spacetime.

\subsection{More general correlation functions}
\label{sec:general:correlations}

The $2N_f$-point functions
discussed in Section~\ref{sec:fermi:corr} correspond to expectation
values of observables,
up to their gauge-covariant nature and the fact that they transform
under redefinitions of the fermion fields, in particular under chiral
rotations of these. A gauge-invariant observable can  be obtained e.g.
by taking the trace of the gauge indices. We have obtained these
correlations to tree-level accuracy in an expansion around multi-instanton backgrounds (based on additional approximations
spelled out in Section~\ref{sec:fermi:corr}). Here, we comment
on how to obtain any correlation function in addition to the $2N_f$-point fermion correlation, where also loop corrections may be included. {As an example for such
a general correlation function, one may take the stress-energy tensor. It is the source of gravitational fields and should therefore be used to infer the vacuum energy, in contrast to the logarithm of the partition function $Z$ that is sometimes used for this purpose in the context of $\theta$-vacua.}

For simplicity, we consider again the case of a single fermion flavour.
To compute the  expectation value of an observable ${\cal O}$ to
some approximation, Eq.~(\ref{Correlation:PathIntegral}) generalizes
to { (dropping as before the contributions from the free determinants)}
\begin{align}
\langle {\cal O}(z_1&,\ldots,z_t)\rangle_{\Delta n}
=\sum\limits_{\bar n,n\geq 0 \atop n-\bar n=\Delta n}\frac{1}{\bar n! n!}
\left(\prod\limits_{\bar\nu=1}^{\bar n}\;\int\limits_{VT} {\rm d}^4x_{0,\bar \nu}{\rm d}\Omega_{\bar \nu} J_{\bar\nu}\right)
\left(\prod\limits_{\nu=1}^{n}\;\int\limits_{VT} {\rm d}^4x_{0,\nu}{\rm d}\Omega_\nu J_\nu\right)\notag\\
\times&{\,
{\rm e}^{-S_{\rm E}(\bar n +n)}
\varpi^{(\bar n +n)}
(- \Theta)^{\bar n+n}
\int
{\rm d}^4z^\prime_1\cdots{\rm d}^4 z^\prime_u
{\cal F}\left(z^\prime_1,\ldots,z^\prime_u;z_1,\ldots,z_t \right)
{\rm e}^{{\rm i}\Delta n(\alpha+\theta)}}
\notag\\
=&
\sum\limits_{\bar n,n\geq 0 \atop n-\bar n=\Delta n}\frac{1}{\bar n! n!}
\int{\rm d}^4z^\prime_1\cdots{\rm d}^4 z^\prime_u
\Big[\left(\bar n \,\bar{\cal G}_{\bar 1}\left(z^\prime_1,\ldots,z^\prime_u;z_1,\ldots,z_t \right)+n \,{\bar{\cal G}}_{1}\left(z^\prime_1,\ldots,z^\prime_u;z_1,\ldots,z_t \right)\right)(VT)^{\bar n+n-1}\notag\\
&+{\cal G}_{0\rm inst}\left(z^\prime_1,\ldots,z^\prime_u;z_1,\ldots,z_t \right)(VT)^{\bar n+n}\Big]
({\rm i}\kappa)^{\bar n +n}
(-1)^{n+\bar n}
{\rm e}^{{\rm i}\Delta n(\alpha + \theta)}\notag\\
=& \int {\rm d}^4 z_1^\prime\cdots {\rm d}^4 z_u^\prime \left[\left(I_{\Delta n+1}(2{\rm i}\kappa VT)\bar{\cal G}_{\bar 1}+I_{\Delta n-1}(2{\rm i}\kappa VT) {\bar{\cal G}}_{1}\right){\rm i}\kappa
+I_{\Delta n}(2{\rm i}\kappa VT)\, {\cal G}_{0{\rm inst}}\right]
(-1)^{\Delta n}
{\rm e}^{{\rm i}\Delta n(\alpha + \theta)}
\label{expectation:value:general}
\,.
\end{align}
The function ${\cal F}$ can be represented by a sum of Feynman diagrams,
i.e. as a sum of products of two-point Green's functions and their derivatives in the
multi-instanton background. For the fermions, these Green's functions
may be approximated by ${\rm i}S_{n,\bar n}$ given in Eq.~(\ref{eq:proptotal}), but other species, e.g. gauge bosons,
can contribute as well. For these additional fields we assume that, in analogy to the fermionic propagators, their two-point functions can be approximated by the free contribution plus a sum over contributions peaking at the centres of each (anti-)instanton.
Each of the two-point functions is evaluated
at a given pair of the spacetime arguments of ${\cal F}$. (For gauge bosons
and self-interacting scalars, these arguments may coincide according to
the Feynman rules.) The integrations over the coordinates $z_i^\prime$
correspond to loop integrals.
In the second step, we have carried out the
integrations over the collective coordinates in analogy with Eq.~(\ref{correlation:sum}).
Organized in powers of $VT$, this defines the contributions
${\cal G}_{0\rm inst}$ from the bulk of the spacetime volume where there are
no instantons,  as well as $\bar{\cal G}_{1}$ and $\bar{\cal G}_{\bar 1}$ which
are obtained for one instanton sweeping over ${\cal F}$. This gives $\bar{\cal G}_{\bar 1/1}$ as generalized overlap integrals averaged over the collective coordinates $\Omega$, involving products of free propagators times one or more contributions to two-point functions---either fermionic or bosonic---arising from  a single (anti-)instanton.
Contributions of lower order in $VT$, corresponding to more than one instanton sweeping over ${\cal F}$ at a time, are suppressed exponentially and have thus been omitted in Eq.~\eqref{expectation:value:general}. In the third
step, we have carried out the summation and suppressed the spacetime arguments
of ${\cal G}_{0\rm inst}$, $\bar{\cal G}_{1}$ and $\bar{\cal G}_{\bar 1}$.
Note also that $\cal O$ in general has a spinor structure. In contrast to
Eq.~(\ref{correlation:sum}), for which this structure is presented,
we do not explicitly show the chiral phases, which are ${\rm e}^{\pm{\rm i}\alpha}$ for left and right-chiral contributions, respectively, because
the only phases in ${\cal G}_{0\rm inst}$, $\bar{\cal G}_{1}$ and $\bar{\cal G}_{\bar 1}$ can originate from the mass term.

In order to evaluate the expectation value by first taking
$VT\to\infty$ and then summing over the topological sectors $\Delta n$,
we note that the volume-dependence of the loop integrand can be
isolated as
\begin{align}\label{eq:In_asymp}
I_n(x)\sim\frac{{\rm e}^{x}}{\sqrt{2\pi x}}\quad\text{for}\quad|x|\to\infty\quad\text{and}\quad |{\rm arg}(x)|<\frac{\pi}{2}\,,
\end{align}
and we can apply the same arguments as in Section~\ref{sec:fermi:corr}. (Recall that the time interval is to be understood as $Te^{-\ic 0^+}$, so that we can apply the asymptotic expansion of Eq.~\eqref{eq:In_asymp}.)
Taking limits in this order thus leads to
\begin{align}
 \langle{\cal O}(z_1,\ldots,z_t)\rangle
 =&\,\lim_{N\to\infty \atop N\in \mathbbm N} \lim_{VT\to\infty}\frac{\sum_{\Delta n=-N}^{N}\langle{\cal O}(z_1,\ldots,z_t)\rangle_{\!\Delta n}}{\sum_{\Delta n=-N}^{N} Z_{\Delta n}}
=\int
{\rm d}^4z^\prime_1\cdots{\rm d}^4 z^\prime_u
\left[{\ic\kappa}\,\bar{\cal G}_{\bar 1}+{\ic\kappa}\,\bar{\cal G}_{1} +{\cal G}_{0\rm inst}\right]
\,.
\end{align}
Again, we observe that interferences from contributions from
different topological sectors $\Delta n$ cancel when normalizing
with the partition function. The only phases for the terms in square
bracket are ${\rm e}^{\pm{\rm i}\alpha}$, and they appear in accordance with
the breaking of chiral symmetry by the mass term. However, unless
additional $CP$-odd phases are introduced in the theory, the phase
$\alpha$ can be removed by field redefinitions and is unobservable.

This cancellation does not hold when taking the limit $VT\to \infty$
after the summation over the topological sectors $\Delta n$, which leads
to
\begin{align}
\langle{\cal O}(z_1,\ldots,z_t)\rangle
 =&\,
\int{\rm d}^4z^\prime_1\cdots{\rm d}^4 z^\prime_u
\left[{\ic\kappa}\,\bar {\cal G}_{\bar 1}{\rm e}^{-{\rm i}(\theta+\alpha)}+{\ic\kappa}\,\bar{\cal G}_1{\rm e}^{{\rm i}(\theta+\alpha)} +{\cal G}_{0\rm inst}\right]\,.
\end{align}
Here, in addition the phase $\theta+\alpha$ appears which is
independent of field redefinitions, according to Eq.~(\ref{eq:selection}),
and generally leads to $CP$-violating observables.

Under certain conditions, these general correlation functions
can be obtained using the effective operators from Eqs.~(\ref{L:eff}) and~(\ref{L:eff:old}).
To see this
explicitly, we assume that the loop integrals are not ultraviolet
sensitive in the sense that only contributions with $|(z^\prime_i-z_j^\prime)^2|\gg \varrho^2$ are relevant.
We can then assume the arguments of the Green's functions to be
sufficiently separated such that we do not have to account for
contributions where two of the Green's functions ${\rm i}S_{n,\bar n}$
are to be evaluated close to the same instanton. Recalling that $\cal F$ depends on the two-point fermionic Green's functions, we denote ${\cal F}={\cal F} (\{{\rm i}S^{(i)}\},...)$ where the dots represents all other arguments. {Here the superscripts $(i)$ (and $(j)$ below) are used to denote the different two-point functions appearing in expansions of $\mathcal{F}$.}
Furthermore, close to an (anti-)instanton we only collect the contributions from the corresponding fermionic zero-mode.
In this case,
within Eq.~(\ref{expectation:value:general}), we can identify
\begin{align}
\,{\cal G}_0=&{\cal F}(\{{\rm i}S^{(i)}\},...)\;\;\text{where}\;\;{{\rm i} S^{(i)}={\rm i} S_{0 \rm inst}\;\;\forall i}\,,\notag\\
\,\bar{\cal G}_{\bar 1}=&\sum\limits_j {\cal F} (\{{\rm i}S^{(i)}\},...)\;\text{where}\; \;{{\rm i} S^{(i)}=\frac{\bar h\,P_{\rm L}}{m{\rm e}^{-{\rm i}\alpha}}\;\;\text{for}\;i=j\;\;\text{and}\;\;{\rm i} S^{(i)}={\rm i} S_{0 \rm inst}\;\;\text{for}\;i\not=j}\,,\notag\\
\,{\bar{\cal G}}_{1}=&\sum\limits_j {\cal F}(\{{\rm i}S^{(i)}\},...)\;\text{where}\;\;{{\rm i} S^{(i)}=\frac{\bar h\,P_{\rm R}}{m{\rm e}^{{\rm i}\alpha}}\;\;\text{for}\;i=j\;\;\text{and}\;\;{\rm i} S^{(i)}={\rm i} S_{0 \rm inst}\;\;\text{for}\;i\not=j}\,,
\end{align}
where $\bar h$ is given in Eq.~(\ref{eq:hav}).
Then, following Section~\ref{sec:fermi:corr},
all anomalous contributions can be approximated to linear order
in $\kappa$ as
\begin{align}
\notag
\langle {\cal O}(z_1,\ldots,z_t)\rangle\approx
\int
{\rm d}^4z^\prime_1\cdots{\rm d}^4 z^\prime_u
\left[{\cal G}_0+{\cal G}_1\right]
\,.\\[-5.5mm]
\label{O:linear}
\end{align}

The term with ${\cal G}_0$ is just the contribution that
would arise in a background without instantons, while the term
with ${\cal G}_1$ represents the leading instanton-effects. When taking $VT\to\infty$
first, we are led to substitute
\begin{align}
{\cal G}_1=&\sum\limits_j {\cal F}(\{{\rm i}S^{(i)}\},...)\;\text{where}\;{{\rm i} S^{(i)}=\frac{{\rm i}\kappa { \bar h}}{m}{\rm e}^{-{\rm i}\alpha\gamma^5}\;\;\text{for}\;i=j\;\;\text{and}\;\;{\rm i} S^{(i)}={\rm i} S_{0 \rm inst}\;\;\text{for}\;i\not=j}\,,
\end{align}
while, when summing over $\Delta n$ first, we take
\begin{align}
{\cal G}_1=&\sum\limits_j {\cal F}(\{{\rm i}S^{(i)}\},...)\;\text{where}\;{{\rm i} S^{(i)}=\frac{{\rm i}\kappa {\bar h}}{m}{\rm e}^{{\rm i}\theta\gamma^5}\;\;\text{for}\;i=j\;\;\text{and}\;\;{\rm i} S^{(i)}={\rm i} S_{0 \rm inst}\;\;\text{for}\;i\not=j}\,.
\end{align}
Now, we can indeed observe that the result~(\ref{O:linear}) can be obtained
by using effective operators of the form~(\ref{L:eff}) or, respectively,
of~(\ref{L:eff:old}) to linear order. The effective operators cannot be used for higher-order calculations in $\kappa$. For example, the effective operator~\eqref{L:eff:old} would imply that the chiral phases $\e^{\pm{\rm i}\theta\gamma^5}$ are additive in quantities in higher power of $\kappa$. However in case of summing over $\Delta n$ before taking $VT\to\infty$ and when replacing more than one fermion line with the interaction induced by the same (anti-)instanton, one obtains only one phase factor $\exp{(\pm{\rm i}\theta\gamma^5)}$. Only in case a contribution to the correlation function involves the effect of more than one (anti-)instanton, the phases are additive.  When aiming to go beyond linear order
in $\kappa$, one should note however that any explicit dependence
on $\theta$ can only enter via the global phases $\Delta n(\alpha+\theta)$
that appear for the path integrals over the individual topological sectors. (We reemphasize that these global phases are immaterial when taking $VT\rightarrow\infty$ before summing over $\Delta n$. Then, no misalignment of chiral phases occurs, which also holds beyond the linear order in $\kappa$.)

Beyond linear order, one should therefore go back to
 Eq.~(\ref{expectation:value:general}) as a starting point. The same applies when considering values of $z_i$ that are not well separated, such that one cannot neglect contributions in which more than one Green's function are evaluated close to the same instanton. Similarly, using the 't~Hooft vertices of Eqs.~\eqref{eq:tHooft} in diagrams with ordinary propagators would only capture a restricted set of contributions of order $\kappa$ in which $N_f$ fermion propagators are evaluated close to the instanton. For higher-order in $\kappa$ or for capturing contributions with more propagators close to the instanton, the use of the effective vertex cannot be justified.

\subsection{Chiral Lagrangian, the $\theta$-angle and the $\eta'$-mass\label{sec:chiral_Lagrangian}}

At low energies, QCD becomes fully nonperturbative and confines. The physical degrees of freedom are mesons and baryons. Their dynamics can be captured by an effective theory defined by the chiral Lagrangian, see e.g. Refs.~\cite{Pich:1995bw,Scherer:2002tk} for reviews. The lightest mesons can be embedded into a  matrix-valued field $U$---with the matrix indices associated with the light quark flavours $u,d,s$---which can be written as
\begin{align}
 U= U_0 {\rm e}^{\frac{\ic}{f_\pi}\Phi},\quad \Phi=\left[\begin{array}{ccc}
     \pi^0+\frac{1}{\sqrt{3}}\eta+\sqrt{\frac23}\eta^\prime & \sqrt{2}\,\pi^+ & \sqrt{2}\,K^+\\
     \sqrt{2}\,\pi^- & -\pi^0+\frac{1}{\sqrt{3}}\eta+\sqrt{\frac23}\eta^\prime & \sqrt{2}\,K^0\\
     \sqrt{2}\,K^- & \sqrt{2}\,\bar K^0 & -\frac{2}{\sqrt{3}}\,\eta+\sqrt{\frac23}\eta^\prime
    \end{array}\right],
\end{align}
where $f_\pi$ is the pion decay constant. Here we are neglecting mixing effects among the mesons $\eta_{8,1}$ and readily approximate these with $\eta$ and $\eta^\prime$. The lowest-order terms in the chiral Lagrangian are given by
\begin{align}\label{eq:LC}
 {\cal L} = \frac{f_\pi^2}{4}{\rm Tr} \,\partial_\mu U\partial^\mu U^\dagger+\frac{f_\pi^2 B_0}{2}\,{\rm Tr} (MU+U^\dagger M^\dagger)+|\lambda| {\rm e}^{-\ic\xi} f_\pi^4\,{\rm det}\, U+|\lambda| {\rm e}^{\ic\xi} f_\pi^4\,{\rm det}\, U^\dagger\,.
\end{align}
It is understood here that $U_0$ is a unitary field expecation value and $\Phi$ describes the meson expectaions about $U_0$ so that $\langle U\rangle =U_0$.
Above, $M$ is a diagonal matrix with ${\rm diag}{M}=\{m_u {\rm e}^{\ic \alpha_u},m_d {\rm e}^{\ic \alpha_d},m_s {\rm e}^{\ic \alpha_s}\}$ being fixed by the light quark masses. The parameter $B_0$ in Eq.~\eqref{eq:LC} is real, while the matrices $U$ and $M$ inherit the following transformations under the selection rule of Eq.~\eqref{eq:selectionNf}:
\begin{align}
\label{chiral:trafo:EFT}
 U\rightarrow {\rm e}^{2 \ic\beta}U\,\Rightarrow\,\det U\rightarrow {\rm e}^{2 {\rm i}N_f\beta}\det U,  \qquad M\rightarrow {\rm e}^{-2\ic\beta}M, \qquad {\theta\rightarrow\theta+2N_f\beta}\,,
\end{align}
where the relation to the quark fields of the ultraviolet theory is given by $\arg U=\arg\langle \bar\psi(x) P_{\rm R} \psi(x)\rangle$.
The effective Lagrangian~(\ref{eq:LC}) is thus constructed in such a way that its properties under chiral transformations are determined by the quark mass terms and the `t~Hooft operator from Eq.~(\ref{digest:eff:operator}).

It does not appear to be widely acknowledged that invariance of the chiral Lagrangian under the former transformations can be achieved for two choices of the phase $\xi$: $\xi=\theta$---the standard choice, leading to $CP$-violating effects \cite{Baluni:1978rf,Crewther:1979pi}---and $\xi=-\bar\alpha=-\alpha_u-\alpha_d-\alpha_s$, as it is required by the topological quantization of the winding number in infinite spacetime volumes. (Recall that $\theta$ and $\bar\alpha$ transform as given in Eq.~\eqref{eq:selectionNf}.) The latter result for $\xi$ leads to no $CP$-violating effects, as in this case it can be easily seen that all the phases  can be removed by a field redefinition. Thus, our results for the `t~Hooft vertices in QCD implying $\xi=-\bar\alpha$, leads to no $CP$ violation  and no electric dipole moment for the neutron. {It should be pointed out that in some of the literature assuming the standard choice $\xi=\theta$, the anomalous terms is not written in terms of ${\rm det} \,U,\;{\rm det} \,U^\dagger$ but keeping the first terms in an expansion of the determinants of Eq.~\eqref{eq:LC} in terms of ${\rm tr}\log U,\; {\rm tr}\log U^\dagger$. Such expressions are obtained after integrating out an auxiliary field related to the topological charge density \cite{DiVecchia:1980yfw}.

Let us further comment that the `t Hooft operator~(\ref{eq:tHooft}) does not vanish in the limit of massless quarks---due to the cancellation of powers of the fermion masses in the fluctuation determinants and the zero mode contributions to the propagators, as discussed with the details on Eq.~\eqref{correlation:effective}---a property that is shared with the chiral Lagrangian above.
The absence of $CP$ violation inherent in the result $\xi=-\bar\alpha$ is therefore not in conflict with the observed enhancement in the mass of the $\eta'$-boson. From the Lagrangian~(\ref{eq:LC}), one still gets a nonzero mass for $\eta^\prime$ even for massless quarks, going as  $m^2_{\eta'}=8|\lambda|\,f_\pi^2$.

Furthermore, as discussed in Section~\ref{subsec:suceptibility}, our results are not in conflict with the relation between the mass of the $\eta'$  boson and the infrared regulated topological susceptibility $\chi_\Omega=\langle\Delta n^2\rangle/\Omega$  in the pure gauge theory, which is found in the limit of a large number of colours in Refs.~\cite{Witten:1979vv,DiVecchia:1980yfw}. Matching the coefficient $|\lambda|$ in Eq.~\eqref{eq:LC} to the instanton calculations, we do find proportionality between the mass of the $\eta'$ meson and the topological susceptibility defined in finite subvolumes of the pure gauge theory for arbitrary number of colours. Crucially, a nonzero value of $\chi_\Omega$ for finite subvolumes does not imply $\theta$-dependence of the partition function in the full volume. This is due to the nonanalytic dependence of $Z$ on $\theta$, as discussed at the end of Section~\ref{sec:several:flavours}.

Given the potential from Eq.~(\ref{eq:LC}), the field $U$ and accordingly $\Phi$ acquire vacuum expectation values. Depending on whether $\xi=-\bar\alpha$ or $\xi=\theta$, the determinant of $U$ is aligned with the determinant of the mass matrix $M$ or not. Just as for the ultraviolet theory, this implies the absence or respectively the presence of $CP$-violating effects.

To see this in more detail, we note that $M$ being diagonal implies that $\langle U\rangle$ is diagonal as well. We therefore parametrize the expectation values as
\begin{align}
\langle U \rangle=U_0=\text{diag}\left({\rm e}^{{\rm i}\varphi_u},{\rm e}^{{\rm i}\varphi_d},{\rm e}^{{\rm i}\varphi_s}\right)\,.
\end{align}
The minimization of the potential in Eq.~(\ref{eq:LC}) leads to the system of equations
\begin{align}
B_0 f_\pi^2 m_i \sin \varphi_i - 2 |\lambda| \sin(\xi+\alpha_u+\alpha_d+\alpha_s-\varphi_u-\varphi_d-\varphi_s)=0
\end{align}
for $i=u,d,s$, which implies that $m_u \sin \varphi_u=m_d \sin \varphi_d=m_s \sin \varphi_s$. Assuming that $|\lambda|\gg B_0 f_\pi^2 m_i$, a leading order solution is obtained when setting $\xi+\alpha_u+\alpha_d+\alpha_s-\varphi_u-\varphi_d-\varphi_s=0$. Neglecting the contributions from the strange quark, one finds the solution~\cite{Cheng:1987gp,Srednicki:2007qs}
\begin{align}
m_{u,d} \sin \varphi_{u,d}=\frac{\sin(\xi+\alpha_u+\alpha_d)}{\sqrt{1/m_u^2+1/m_d^2+2\cos(\xi+\alpha_u+\alpha_d)/(m_u m_d)}}\,.
\end{align}
For three flavours, provided $\xi+\alpha_u+\alpha_d+\alpha_s\ll 1$, one can approximate $m_u \varphi_u=m_d  \varphi_d=m_s \varphi_s$, what leads to~\cite{Baluni:1978rf}
\begin{align}
\label{mphi}
m_i \varphi_i=\frac{m_u m_d m_s (\xi+\alpha_u+\alpha_d+\alpha_s)}{m_u m_d + m_d m_s + m_s m_u}=\tilde m (\xi+\alpha_u+\alpha_d+\alpha_s)\,.
\end{align}
This result serves in Refs.~\cite{Baluni:1978rf,Crewther:1979pi} as the input to calculate $CP$-violating observables using current algebra techniques. Here, we discuss the consequences in the framework of effective chiral perturbation theory, which is presented in view of $CP$-violation in the strong interactions in Ref.~\cite{Srednicki:2007qs}.

Substituting $U$ with these angles $\varphi_i$ into the term involving $B_0$ in Eq.~(\ref{eq:LC})
yields the $CP$-odd effective interactions
\begin{align}
\label{EFT:CPV:mesons}
-\frac{{\rm i} \sqrt{2} B_0}{\sqrt{3} f_\pi} \tilde m (\xi+\alpha_u+\alpha_d+\alpha_s) \eta^\prime (\pi^0\pi^0+ 2\pi^+ \pi^-)\,.
\end{align}
These lead to $CP$-violating decays $\eta^\prime\to 2\pi$.

The main observable of interest is the permanent electric dipole moment of the neutron~\cite{Baluni:1978rf,Crewther:1979pi}. Considering the transformation of the nucleon multiplet $N$, the quark mass matrix $M$ and the meson fields $U$ under chiral field redefinitions leads to an effective theory of the interactions of mesons and nucleons that is consistent with these symmetries and the way in which these are broken. Given the quark mass matrix $M$ and the expectation value $U_0$, it can be shown that this includes a $CP$-odd operator~\cite{Srednicki:2007qs}
\begin{align}
\label{EFT:CPV:nucleons}
\frac{c_+\tilde m (\xi+\alpha_u+\alpha_d+\alpha_s)}{2 f_\pi}\bar N \Phi N\,,
\end{align}
where $c_+$ is a parameter of the effective theory. This includes an interaction between the neutrons, protons and charged pions that leads to a permanent electric dipole moment of the neutron.

Setting $\xi=\theta$ in Eq.~(\ref{mphi}) and consequently in the interactions~(\ref{EFT:CPV:nucleons}) and~(\ref{EFT:CPV:mesons}), these results are proportional to $\bar \theta$ when identifying $\bar\alpha=\alpha_u+\alpha_d+\alpha_s$. In general, $CP$-violating effects would then follow. Taking instead $\xi=-\bar\alpha$, as it is indicated by taking the spacetime volume to infinity before summing over topological sectors, these interactions vanish in the effective theory so that there are no $CP$-violating effects.

\section{General correlation functions from cluster decomposition}
\label{sec:CDC1}

In this section it is shown that the above conclusions regarding the  phases of general correlation functions can be derived without resorting to the dilute instanton approximation. Rather, working in Euclidean space for simplicity, in a theory with $N_f$ flavours in the fundamental representation, we constrain the form of the partition functions $Z_{\Delta n}$ from arguments based on  cluster decomposition, the index theorem and parity. From this one can derive integrated fermionic correlation functions, which are sensitive to the { constant phases  of the correlators} discussed in the previous sections. Again, when the infinite volume limit is taken before the sum over topological sectors, no $CP$-violating relative phases remain.

\subsection{\label{subsec:thetacluster}The $\theta$-angle and cluster decomposition}
We start by recalling that, alternative to a coupling constant in the Lagrangian, the $\theta$-angle can be understood in terms of relative weights $f(\Delta n)$ among the contributions from the different topological sectors to the path integral, constrained by the requirement of cluster decomposition~\cite{Weinberg:book:vol2:1996}.

Let us consider the expectation value of an operator $\cal O$ in an infinite spacetime volume $\Omega=VT$, and interfere different topological sectors $\Delta n$ as
\begin{align}\label{eq:Ocluster}
 \langle {\cal O}\rangle_\Omega= \lim_{N\to\infty \atop N\in \mathbbm N} \lim_{\Omega\to\infty} \frac{\sum\limits_{\Delta n=-N}^N f(\Delta n)\int\limits_{\Delta n} {\cal D}\phi\,{\cal O}\,{\rm e}^{-S_\Omega[\phi]}}{\sum\limits_{\Delta n=-N}^N f(\Delta n)\int\limits_{\Delta n} {\cal D}\phi\,{\rm e}^{-S_\Omega[\phi]}}
\,.
\end{align}
Here, the action $S_\Omega$ is defined to arise from integrating the Lagrangian over $\Omega$. The Lagrangian does not include the topological term $\theta {\rm tr} F \tilde F/(16\pi)$ which, as it turns out, can be attributed to the weights
$f(\Delta n)$ of the individual topological sectors. Further, writing
$\Delta n$ under the integration symbol, we specify the topologically conserved winding number which can
be imposed on vanishing physical gauge fields at the boundary of $\Omega$ at infinity.

Next, we divide the volume of the spacetime as $\Omega=\Omega_1\cup\Omega_2$ according to Figure~\ref{fig:locality}.
Accordingly, the winding numbers behave additively $\Delta n(\Omega)=\Delta n_1(\Omega_1)+\Delta n_2(\Omega_2)$. Now suppose we consider a local operator ${\cal O}_1$ whose spacetime arguments are restricted to lie within $\Omega_1$.
Then, we can separate off the contribution from $\Omega_2$ as
\begin{align}\label{eq:O1}
 \langle {\cal O}_1\rangle_\Omega= \lim_{N_2\to\infty \atop N_2\in \mathbbm N}
 \lim_{N_1\to\infty \atop N_1\in \mathbbm N} \lim_{\Omega\to\infty}\frac{\sum\limits_{\Delta n_1=-N_1}^{N_1}\sum\limits_{\Delta n_2=-N_2}^{N_2} \!\!\!\! f(\Delta n_1+\Delta n_2)\int\limits_{\Delta n_1} {\cal D}\phi\,{\cal O}_1\,{\rm e}^{-S_{\Omega_1}[\phi]}\int\limits_{\Delta n_2} {\cal D}\phi\,{\rm e}^{-S_{\Omega_2}[\phi]}}{\sum\limits_{\Delta n_1=-N_1}^{N_1}\sum\limits_{\Delta n_2=-N_2}^{N_2} \!\!\!\! f(\Delta n_1+\Delta n_2)\int\limits_{\Delta n_1} {\cal D}\phi\,{\rm e}^{-S_{\Omega_1}[\phi]}\int\limits_{\Delta n_2} {\cal D}\phi\,{\rm e}^{-S_{\Omega_2}[\phi]}}
\,.
\end{align}
We note that for this partition, $\Delta n_{1,2}$ cannot strictly be assumed to
be integers because the field configurations at the boundary between the two subvolumes do not give rise to topologically conserved winding numbers
within each of $\Omega_{1,2}$. We proceed nonetheless, taking above expression
as a suitable approximation for sparse populations of instantons.

Independence of $\langle {\cal O}_1\rangle_\Omega$ from the fluctuations in $\Omega_2$ is achieved if both the numerator and the denominator factorize into contributions that separately depend on $\Delta n_1$, $\Omega_1$ and  $\Delta n_2$, $\Omega_2$, respectively. Then the fluctuations within the volume $\Omega_2$ cancel. For this to occur, without using particular properties of the path integral factors to this end, $f(\Delta n)$ needs to satisfy the following functional relation:
\begin{align}
\label{eq:ftheta}
 f(\Delta n_1+\Delta n_2)=f(\Delta n_1)f(\Delta n_2)\Rightarrow f(\Delta n)={\rm e}^{{\rm i} \Delta n \theta}\,.
\end{align}
We have attributed here the phase $\theta$ such that the topological term
in the action is indeed recovered. {In Section~\ref{sec:CDC}, we present more aspects of cluster decomposition based on the correlation and partition functions that are derived in Section~\ref{sec:fermi:corr} using the dilute instanton gas approximation or, alternatively, from the constraints derived in the present section.}


\subsection{Constraining the partition functions from cluster decomposition, the index theorem and parity}
\label{subsec:constrain:Z}

From the denominator of Eq.~\eqref{eq:Ocluster} one recovers the partition function in the volume $\Omega=VT$ as
\begin{align}
 Z(\Omega,N)=\sum_{{\Delta n}=-N}^N Z_{\Delta n}(\Omega),\quad Z_{\Delta n}(\Omega)= f({\Delta n}) \int_{\Delta n} {\Dcal}\phi\,{\rm e}^{-S_\Omega[\phi]}.
\end{align}
The above factorization assumption gives
\begin{align}
 Z_{\Delta n}(\Omega)=f({\Delta n}) \int_{\Delta n} {\Dcal}\phi\,{\rm e}^{-S_\Omega[\phi]}= \sum_{{\Delta n}_1=-\infty}^\infty f({\Delta n}) \int_{{\Delta n}_1} {\Dcal}\phi\,{\rm e}^{-S_{\Omega_1}[\phi]} \int_{{\Delta n}-{\Delta n}_1} {\Dcal}\phi\,{\rm e}^{-S_{\Omega_2}[\phi]}\,,
\end{align}
and the property \eqref{eq:ftheta} of the weight factors simply leads to
\begin{align}\label{eq:clusterZ}
 Z_{{\Delta n}}(\Omega=\Omega_1+\Omega_2)=\sum_{{\Delta n}_1=-\infty}^\infty Z_{{\Delta n}_1}(\Omega_1) Z_{{\Delta n}-{\Delta n}_1}(\Omega_2)
\end{align}
with
\begin{align}\label{eq:Ztheta}
 Z_{\Delta n}(\Omega)={\rm e}^{{\rm i}{\Delta n}\theta}g_{\Delta n}(\Omega), \quad g_{\Delta n}(\Omega)=\int_{{\Delta n}} {\Dcal}\phi\,{\rm e}^{-S_{X}[\phi]}.
\end{align}
In the following we use Eq.~\eqref{eq:clusterZ}, which can be thought of as a formulation of the cluster decomposition principle at the level of the partition function, to constrain $g_{\Delta n}(\Omega)$.

First, we isolate the possible complex phases in $g_{\Delta n}(\Omega)$. The latter can be understood in terms of fluctuation determinants of gauge fields and fermions about a gauge field background with topological charge ${\Delta n}$, where we do not assume here a construction from a dilute gas of instantons and anti-instantons. The Euclidean gauge determinants  are real, while the fermion fluctuation determinants pick up a phase from unpaired right-handed and left-handed zero modes of the massless Dirac operator. Indeed, as discussed in  Section~\ref{sec:cplx:mass:Eucl}, eigenvalues of the massive Dirac operator come in pairs with mutually conjugate eigenvalues (see Eq.~\eqref{sec:cplx:mass:Eucl}) whose product is always real; this applies to arbitrary backgrounds, not just for $\Delta n=\pm1$). The Atiyah-Singer index theorem relates the topological charge $\Delta n$ to the difference in the number of left-handed and right-handed zero modes. This implies that the product of the fermionic fluctuation determinants for the flavours $j=1,\dots,N_f$ with complex masses
\begin{align}\label{eq:mathfrakm}
m_j {\rm e}^{\ic \alpha_j\gamma_5}=m_j {\rm e}^{\ic \alpha_j}P_{\rm R}+m_j {\rm e}^{-\ic \alpha_j}P_{\rm L}\equiv \mathfrak{m}P_{\rm R}+\mathfrak{m}^* P_{\rm L}
\end{align}
in a sector with fixed $\Delta n$ carries a phase given by ${\Delta n}\,\bar\alpha$, with $\bar\alpha$ defined as in Eq.~\eqref{eq:alphabar}.
Therefore we can write
\begin{align}\label{eq:Zgtilde}
 g_{\Delta n}(\Omega)= {\rm e}^{{\rm i}{\Delta n}\bar \alpha}\tilde g_{\Delta n}(\Omega), \quad \tilde g_{\Delta n}(\Omega)\in\mathbb{R}.
\end{align}
The requirement of cluster decomposition as in Eq.~\eqref{eq:clusterZ}, together with Eq.~\eqref{eq:Ztheta} and Eq.~\eqref{eq:Zgtilde} implies now
\begin{align}\label{eq:clustertilde}
 \tilde g_{{\Delta n}}(\Omega_1+\Omega_2)=\sum_{{\Delta n}_1=-\infty}^\infty\tilde g_{{\Delta n}_1}(\Omega_1) \tilde g_{{\Delta n}-{\Delta n}_1}(\Omega_2).
\end{align}

{Next}, we  assume that, as in standard instanton calculations, parity relates the sectors of opposite charges $\pm {\Delta n}$. The functions $\tilde g_{\Delta n}(\Omega)$ capture the fluctuation determinants for real fermion masses, since when $\alpha_i=0$ one has $g_{\Delta n}(\Omega)|_{\alpha_i\rightarrow0}=\tilde g_{\Delta n}(\Omega)$ (see \eqref{eq:Zgtilde}). But for real fermion masses parity is conserved, so that
\begin{align}\label{eq:parity}
 \tilde g_{-{\Delta n}}(\Omega)=\tilde g_{{\Delta n}}(\Omega).
\end{align}
In order to find a solution for {$\tilde g_{\Delta n}(\Omega)$ that complies with cluster decomposition through satisfying Eq.~(\ref{eq:clustertilde})}, we consider first the limiting case $\Omega_1=\Omega_2=0$ that implies
\begin{align}\label{eq:cond0}
 \tilde g_{\Delta n}(0)=\sum_{{\Delta n}_1=-\infty}^\infty \tilde{g}_{{\Delta n}_1}(0) \tilde{g}_{{\Delta n}-{\Delta n}_1}(0)\Rightarrow \tilde{g}_{{\Delta n}}(0) = \delta_{{\Delta n} 0}.
\end{align}
This brings us to the following ansatz:
\begin{align}
\label{eq:Ansatz}
 \tilde g_{\Delta n}(\Omega)=\tilde g_{|{\Delta n}|}(\Omega)=\Omega^{|{\Delta n}|} f_{|{\Delta n}|}(\Omega^2), \quad f_{|{\Delta n}|}(0)\neq0.
\end{align}
The dependence on $|{\Delta n}|$ follows from the parity constraint~(\ref{eq:parity}), while factoring out $\Omega^{|{\Delta n}|}$ guarantees that condition~\eqref{eq:cond0} is met. The latter condition implies now
\begin{align}
 f_0(0)=1.
\end{align}
The Ansatz further yields
\begin{align}\label{eq:der0}
 \tilde g'_{\Delta n}(\Omega)=|{\Delta n}|\, \Omega^{|{\Delta n}|-1}f_{|{\Delta n}|}(\Omega^2)+2\,\Omega^{|{\Delta n}|+1}f'_{|{\Delta n}|}(\Omega^2)\quad\Rightarrow\quad \tilde g'_{\Delta n}(0)=\delta_{|{\Delta n}| 1}f_1(0).
\end{align}
Taking the derivative of the cluster decomposition relation \eqref{eq:clustertilde} with respect to $\Omega_1$ gives
\begin{align}
 \tilde g'_{\Delta n}(\Omega_1+\Omega_2)=\sum_{{\Delta n}_1=-\infty}^\infty \tilde g'_{{\Delta n}_1}(\Omega_1) \tilde g_{{\Delta n}-{\Delta n}_1}(\Omega_2).
\end{align}
Setting now $\Omega_1=0$ gives
\begin{align}
 \tilde g'_{\Delta n}(\Omega_2)=\sum_{{\Delta n}_1=-\infty}^\infty \tilde g'_{{\Delta n}_1}(0)\, \tilde g_{{\Delta n}-{\Delta n}_1}(\Omega_2).
\end{align}
Applying Eq.~\eqref{eq:der0} one gets
\begin{align}
  \tilde g'_{\Delta n}(\Omega_2)=f_1(0)(\,\tilde g_{{\Delta n}+1}(\Omega_2)+\tilde g_{{\Delta n}-1}(\Omega_2)).
\end{align}
This allows to solve recursively for all the higher order derivatives of $\tilde g_{\Delta n}(\Omega_2)$ in terms of functions without derivatives. Renaming $\Omega_2\rightarrow\Omega$, one has
\begin{align}
 \frac{{\rm d}^n}{{\rm d}\Omega^n}\tilde g_{\Delta n}(\Omega)=(f_1(0))^n\sum_{m=0}^{n}\left(\begin{array}{c}n\\ m\end{array}\right)\tilde g_{{\Delta n}-n+2m}(\Omega).
\end{align}
In particular, setting $\Omega=0$ and using Eq.~\eqref{eq:cond0} gives
\begin{align}\label{eq:derszero}
 \frac{{\rm d}^n}{{\rm d}\Omega^n}\,\tilde g_{\Delta n}(0)=\left\{\begin{array}{cc}
                                                (f_1(0))^n \left(\begin{array}{c} n\\ \frac{n-{\Delta n}}{2}\end{array}\right) & \text{if} \quad n-{\Delta n}=0,2,4,\dots\\
                                                0& \text{otherwise.}
                                               \end{array}\right.
\end{align}
Using the analyticity of $\tilde g_{\Delta n}(\Omega)$, knowing all the derivatives at the origin allows to  recover the function  from its Taylor expansion around $\Omega=0$:
\begin{align}
 \tilde g_{\Delta n}(\Omega)=\sum_{n=0}^\infty \frac{1}{n!}\frac{{\rm d}^n}{{\rm d}\Omega^n}\,\tilde g_{\Delta n}(0)\,\Omega^n.
\end{align}
With Eq.~\eqref{eq:derszero} implying nonzero derivatives only for $n=|{\Delta n}|+2k,\,k =0,1,2,\dots$, one has
\begin{align}
 \tilde g_{\Delta n}(\Omega)=\sum_{k=0}^\infty \frac{1}{(|{\Delta n}|+2k)!}  (f_1(0))^{|{\Delta n}|+2k}\left(\begin{array}{c} |{\Delta n}|+2k\\ k\end{array}\right) \Omega^{|{\Delta n}|+2k}=(f_1(0) \Omega)^{|{\Delta n}|}\,\sum_{k=0}^\infty\frac{(f_1(0) \Omega)^{2k}}{k!(|{\Delta n}|+k)!}.
\end{align}
We can thus verify that, while we did not impose the parity property beyond using it to derive Eq.~\eqref{eq:der0}, the solution we find satisfies the parity property~(\ref{eq:parity}) indeed. Renaming $f_1(0)\equiv \beta$, we can rewrite the solution as:
\begin{align}\label{eq:gtilde}
 \tilde g_{\Delta n}(\Omega)=\sum_{k=0}^\infty\frac{1}{k!(|{\Delta n}|+k)!}\left(\frac{2\beta\,\Omega}{2}\right)^{|{\Delta n}|+2k}= I_{{\Delta n}}(2\beta\Omega).
\end{align}
Thus, we recover the modified Bessel functions of the first kind that have been found in the computations with the dilute instanton gas. The partition function for a fixed topological sector is then
\begin{align}\label{eq:Zbeta}
 Z_{\Delta n} = {\rm e}^{\ic \theta\Delta n} g_{\Delta n}(\Omega) = {\rm e}^{\ic (\theta+\bar\alpha)\Delta n} \tilde g_{\Delta n}(\Omega) = I_{\Delta n}(2\beta\Omega)   \,{\rm e}^{\ic (\theta+\bar\alpha)\Delta n} .
\end{align}
This matches the result of Eq.~\eqref{eq:ZNf} from the dilute instanton gas approximation, up to a redefinition $\theta\rightarrow\theta+N_f\pi$.
Note also that the solutions for $\tilde g_{\Delta n}$ satisfy indeed the desired property under parity transformations as $I_{\Delta n}(x)=I_{|{\Delta n}|}(x)$ for integer ${\Delta n}$.

Since in the derivation we have used the trick of setting $\Omega_1=0$, it may remain to check that the modified Bessel functions satisfy the full requirement of the cluster decomposition principle. But the Bessel functions are readily known to satisfy the required identity
\begin{align}\label{eq:clusterbeta0}
 I_{\Delta n}(2\beta(\Omega_1+\Omega_2))=\sum_{{\Delta n}_1=-\infty}^\infty I_{{\Delta n}_1}(2\beta \Omega_1) I_{{\Delta n}-{\Delta n}_1}(2\beta \Omega_2).
\end{align}
This relation can be proven e.g. using analyticity and the following two properties of the Bessel functions: $I_{{\Delta n}}(0)=\delta_{\Delta n 0}$, $ \frac{d}{dx}I_{\Delta n}(x)=\frac{1}{2}\,(I_{{\Delta n}+1}(x)+I_{{\Delta n}-1}(x))$. Using the former, it can be seen that Eq.~\eqref{eq:clusterbeta0} and the identities obtained by taking derivatives with respect to $\Omega_1$ to arbitrary order are always satisfied at $\Omega_1=0$. Analyticity then implies that eq.~\eqref{eq:clusterbeta0} holds for arbitrary $\Omega_1$.

To derive the fermion correlation functions, we note that  $\beta$ can still depend on the masses of the quarks.
As discussed earlier $\tilde g_{\Delta n}$ is real, and corresponds to the powers of $m_k=\sqrt{\mathfrak{m}_k\mathfrak{m}^*_k}$ coming from the zero modes plus contributions from nonzero modes of the massless Dirac operator to the product of fermion determinants. As discussed earlier, these nonzero modes come in pairs with mutually conjugate eigenvalues, whose product only depends on  $m_k^2=\mathfrak{ m}_k \mathfrak{ m}_k^*$. Then, it follows that $\beta$ in Eqs.~\eqref{eq:gtilde}, \eqref{eq:Zbeta} can only be a function of $\mc_k \mc_k^*$, $\beta=\beta(\mc_k \mc_k^*)$. That is,
\begin{align}\label{eq:Znumm0}
  Z_{\Delta n}(\Omega)=&\,{\rm e}^{\ic{\Delta n}(\theta+\bar\alpha)} I_{\Delta n}(2\beta(\mc_k \mc^*_k)\, \Omega).
\end{align}
Noting that $\bar\alpha$ depends on the masses as $\bar\alpha=-\ic/2 \sum_k \log({\mc_k/\mc^*_k})$, one can write
\begin{align}\label{eq:Znumm}
  Z_{\Delta n}(\Omega)=&\,{\rm e}^{\ic{\Delta n}(\theta-\ic/2\sum_k\log(\mc_k/\mc_k^*))}I_{\Delta n}(2\beta(\mc_k \mc^*_k)\, \Omega).
\end{align}
Given the mass terms in the Euclidean Lagrangian,
\begin{align}\label{eq:Lm}
 {\cal L}\supset\sum_j\bar\psi_j(m_j {\rm e}^{\ic\alpha_j\gamma_5})\psi_j=\sum_j\bar\psi_j(\mathfrak{ m}_j P_{\rm R}+\mathfrak{ m}^*_j P_{\rm L})\psi_j,
\end{align}
one can interpret $\mc_i$ as a ``current'' pertaining to the correlator $\bar\psi_i P_{\rm R} \psi_i$, and $\mc^*_i$ to $\bar\psi_i P_{\rm L} \psi_i$, where the correlations are evaluated at coincident points. From the Euclidean path integral formulation it follows that, within a topological sector ${\Delta n}$,
\begin{align}
 \int {\rm d}^4 x\,\langle \bar\psi_i P_{\rm R} \psi_i \rangle_{\Delta n}=&\,-\frac{\partial}{\partial \mc_i}Z_{\Delta n}, & \int {\rm d}^4 x\,\langle \bar\psi_i P_{\rm L} \psi_i \rangle_{\Delta n}=&\,-\frac{\partial}{\partial \mc^*_i}Z_{\Delta n}.
\end{align}
Applying this to Eq.~\eqref{eq:Znumm}, noting that $\bar\alpha$ depends on the masses as $\bar\alpha=-\ic \sum_i \log(\sqrt{\mc_i/\mc^*_i})$, gives
\begin{align}\begin{aligned}
 \int {\rm d}^4 x\,\langle \bar\psi_i P_{\rm R} \psi_i \rangle_{\Delta n}=&\,- {\rm e}^{\ic{\Delta n}(\theta+\bar\alpha)}\left(\frac{{\Delta n}}{2\mc_i}I_{\Delta n}(2\beta\Omega)+ 2\Omega\,\mc^*_iI_{\Delta n}'(2\beta\Omega)\frac{\partial}{\partial (\mc_i\mc^*_i)}\, \beta(\mc_k \mc^*_k)\right)\,,\\
\int {\rm d}^4 x\,\langle \bar\psi_i P_{\rm L} \psi_i \rangle_{\Delta n}=&\,- {\rm e}^{\ic{\Delta n}(\theta+\bar\alpha)}\left(-\frac{{\Delta n}}{2\mc^*_i}I_{\Delta n}(2\beta\Omega)+ 2\Omega\,\mc_iI_{\Delta n}'(2\beta\Omega)\frac{\partial}{\partial (\mc_i\mc^*_i)}\, \beta(\mc_k \mc^*_k)\right)\,.
\end{aligned}\end{align}
Using the identities
\begin{align}
\label{eq:nuI} \frac{{\rm d}}{{\rm d}z}I_{\Delta n}(z)=&\,\frac{1}{2}\,(I_{{\Delta n}+1}(z)+I_{{\Delta n}-1}(z)), &{\Delta n} I_{\Delta n}(z)=&\,-\frac{z}{2}\,(I_{{\Delta n}+1}(z)-I_{{\Delta n}-1}(z)),
\end{align}
and dividing by $\Omega=VT$, we get the following spacetime averages of the fermionic correlators:
\begin{align}\begin{aligned}
  \frac{1}{VT}\int {\rm d}^4 x\,\langle \bar\psi_i P_{\rm R} \psi_i \rangle_{\Delta n}=&\,- {\rm e}^{\ic{\Delta n}(\theta+\bar\alpha)}\left(-\frac{\beta}{2\mc_i}(I_{{\Delta n}+1}(2\beta\Omega)-I_{{\Delta n}-1}(2\beta\Omega))\right.\\
  &\left.+\,\mc^*_i(I_{{\Delta n}+1}(2\beta\Omega)+I_{{\Delta n}-1}(2\beta\Omega))\frac{\partial}{\partial (\mc_i\mc^*_i)}\, \beta(\mc_k \mc^*_k)\right)\,,\\
 \frac{1}{VT}\int {\rm d}^4 x\,\langle \bar\psi_i P_{\rm L} \psi_i \rangle_{\Delta n}=&\,- {\rm e}^{\ic{\Delta n}(\theta+\bar\alpha)}\left(\frac{\beta}{2\mc^*_i}(I_{{\Delta n}+1}(2\beta\Omega)-I_{{\Delta n}-1}(2\beta\Omega))\right.\\
  &\left.+ \,\mc_i(I_{{\Delta n}+1}(2\beta\Omega)+I_{{\Delta n}-1}(2\beta\Omega))\frac{\partial}{\partial (\mc_i\mc^*_i)}\, \beta(\mc_k \mc^*_k)\right)\,.
\end{aligned}\end{align}
Note that the correlators carry the correct amount of $\pm2$ units of spurious chiral charge, where the latter is defined from the transformation rules of Eq.~\eqref{eq:selectionNf}, which imply $\mathfrak{m}_j\rightarrow {\rm e}^{-2\ic \beta}\mathfrak{m}_j, \,\mathfrak{m}^*_j\rightarrow {\rm e}^{2\ic \beta}\mathfrak{m}^*_j$.

From the previous expressions it is not clear how to exactly recover our previous results found in the dilute instanton gas approximation---e.g. identify the free piece vs. instanton-like corrections---but it should be kept in mind that $\beta$ is meant to capture perturbative as well as non-perturbative results. Furthermore, we are computing integrated and coincident correlators, rather than the correlators evaluated at some arbitrary $x,x'$. The final values of the spacetime averaged correlators are:
\begin{align}
 \frac{1}{VT}\int {\rm d}^4 x\,\langle \bar\psi_i P_{{\rm }R/{\rm L}} \psi_i \rangle=\frac{1}{VT\sum_{\Delta n} Z_{\Delta n}}\sum_{\Delta m}\int {\rm d}^4 x\,\langle \bar\psi_i P_{R/L} \psi_i \rangle_{\Delta m}.
\end{align}
Taking the limit $\Omega\rightarrow \infty$ before the sum over ${\Delta n}$, as corresponds to an infinite flat spacetime with topological sectors defined by the boundary conditions at infinity that are required to have a finite action, we have that
$I_{\Delta n}(2\beta\Omega)=I_0(2\beta\Omega)(1+O(1/\Omega))$. Then the contributions to the correlators proportional to $\beta$ vanish, while the ones proportional to the derivatives of $\beta$ survive, giving
\begin{align}\begin{aligned}
  \frac{1}{VT}\int d^4 x\,\langle \bar\psi_i P_{\rm R} \psi_i \rangle =\frac{{-}\sum_{\Delta m} {\rm e}^{\ic{\Delta m}(\theta+\bar\alpha)}2\mc_i^* I_0(2\beta\Omega)\,\partial_{\mc_i \mc^*_i} \beta(1+O(1/\Omega))}{\sum_{\Delta n} {\rm e}^{\ic{\Delta n}(\theta+\bar\alpha)}I_0(2\beta\Omega)(1+O(1/\Omega))}\rightarrow {-}2\mc^*_i\,\partial_{\mc_i \mc^*_i} \beta(\mc_k \mc^*_k),\\
  \frac{1}{VT}\int d^4 x\,\langle \bar\psi_i P_{L} \psi_i \rangle =\frac{{-}\sum_{\Delta m} {\rm e}^{\ic{\Delta m}(\theta+\bar\alpha)}2\mc_i I_0(2\beta\Omega)\,\partial_{\mc_i \mc^*_i} \beta(1+O(1/\Omega))}{\sum_{\Delta n} {\rm e}^{\ic{\Delta n}(\theta+\bar\alpha)}I_0(2\beta\Omega)(1+O(1/\Omega))}\rightarrow {-}2\mc_i\,\partial_{\mc_i \mc^*_i} \beta(\mc_k \mc^*_k).
\end{aligned}\end{align}
Thus, the spacetime average of the full coincident correlator has no $\theta$-dependent phases in the infinite volume limit. The { constant} phases of the correlators  (i.e. insensitive to the integration over spacetime) are all set by the tree-level masses, and therefore there is no $CP$ violation.

By taking higher derivatives of $Z_{\Delta n}$, the previous results can be generalized to products of spacetime-averages of coincident two-point correlation functions for different flavours. These correspond to particular spacetime averages of arbitrary correlation functions, and should display the same constant phases as the full space-time dependent correlators. The relations~\eqref{eq:nuI} allow to link derivatives of the $I_{\Delta n}$ to the $I_{\Delta n}$ themselves, which in the infinite volume limit all match $I_0$ asymptotically. Analogously, contributions going as ${\Delta n}^m I_{\Delta n}$ arising from derivatives with respect to $\bar\alpha$ can be traded for linear combinations of Bessel functions without ${\Delta n}$ factors. In the infinite volume limit, again one has that the  ${\Delta n}^m I_{\Delta n}$ are zero or proportional to $I_0$. With all contributions proportional to $I_0$,  the interferences of the $\theta$-dependent phases disappear again and one ends up with $\theta$-independent correlators.

The fact that the phase of the correlation functions is aligned with the phase of the quark masses when the infinite volume limit is taken before the sum over topological sectors has also been noted in Ref.~\cite{Verbaarschot:2014upa}. The argument given in that work does not rely on the dilute instanton gas approximation either. However, it is restricted to the case of real masses such that the phases can only be multiples of $\pi$. That result is discarded however since it is further assumed in Ref.~\cite{Verbaarschot:2014upa} that the correlation function should be aligned with $\theta$ instead. The latter however is derived there from a partition function of the form~(\ref{Z:wrongorderoflimits}) that, according to the calculations in this work, is valid only if the spacetime volume is taken to infinity last. The argument of Ref.~\cite{Verbaarschot:2014upa} advocating for the opposite order of limits as proposed here is therefore circular.

\section{Cluster decomposition with or without summation over topological sectors}
\label{sec:CDC}

In this section we show that the factorization properties of the path integration imply that one can recover the results for the correlators in an infinite volume $\Omega$ by carrying out path integrals in a finite volume $\Omega_1\subset\Omega$. While the topological term can be cast into a boundary term, we need to carefully track its effects in order to check whether these are physical or not. In order to do so, one cannot only calculate the path integral in $\Omega_1$ and ignore the volume complement $\Omega\setminus \Omega_1$---if boundary terms are important for $\Omega_1$ this should be the case for the larger volume $\Omega$ as well. That is, one must integrate the fluctuations in the complement in order to arrive at an effective path integral for the subvolume (cf. Figure~\ref{fig:locality}).
In this effective path integral over $\Omega_1$ that enters the correlation functions, the usual chiral phases from the fermion determinants and the $\theta$-angle turn out to cancel and therefore there are again no $CP$-violating effects, regardless of whether or not one sums  over the topological sectors in the infinite volume $\Omega$.

Moreover, it will be seen that the previous results can be generalized to a large (i.e. $\kappa\Omega\gg 1$) but finite volume $\Omega\supset\Omega_1$, as long as the path integration in $\Omega$ is restricted to a single topological sector. Such a restriction is motivated by the  considerations on local observers made at the end of Section~\ref{subsec:summation}, and the results are in keeping with the expectation that the observables should not depend on whether
the calculation is carried out in an infinite or a very large but finite
spacetime volume. In particular, the cluster-decomposition property is maintained in such a setup
with large $\Omega$.

For simplicity, { the calculations in this  section are also carried out}
in Euclidean spacetime.

\subsection{Cluster decomposition within an infinite volume}
\label{sec:clustering:inf:vol}

Equation \eqref{eq:O1} can be rewritten as
\begin{align}\label{eq:O1Omega}
 \langle {\cal O}_1\rangle_\Omega=&\lim_{N_1\to\infty \atop N_1\in \mathbbm N}
 \lim_{N\to\infty \atop N\in \mathbbm N} \lim_{\Omega\to\infty}\ \frac{\sum\limits_{\Delta n=-N}^N \sum\limits_{\Delta n_1=-N_1}^{N_1} \!\!\! f(\Delta n)\int\limits_{\Delta n_1} {\cal D}\phi\,{\cal O}_1\,{\rm e}^{-S_{\Omega_1}[\phi]}\int\limits_{\Delta n_2=\Delta n-\Delta n_1} {\cal D}\phi\,{\rm e}^{-S_{\Omega_2}[\phi]}}{\sum\limits_{\Delta n=-N}^N \sum\limits_{\Delta n_1=-N_1}^{N_1} \!\!\! f(\Delta n)\int\limits_{\Delta n_1} {\cal D}\phi\,{\rm e}^{-S_{\Omega_1}[\phi]}\int\limits_{\Delta n_2=\Delta n -\Delta n_1} {\cal D}\phi\,{\rm e}^{-S_{\Omega_2}[\phi]}}\,.
\end{align}
The contributions from the volume $\Omega_2$ are given by the Euclidean version of the corresponding partition functions~(\ref{eq:ZNf}),  which gives the following expression in terms of modified Bessel functions:
\begin{align}
\langle {\cal O}_1\rangle_\Omega=&\lim_{N_1\to\infty \atop N_1\in \mathbbm N}
 \lim_{N\to\infty \atop N\in \mathbbm N} \lim_{\Omega_2\to\infty} \frac{\sum\limits_{\Delta n=-N}^N\sum\limits_{\Delta n_1=-N_1}^{N_1} \!\!\! f(\Delta n)\,I_{\Delta n-\Delta n_1}(2\kappa\Omega_2){(-1)^{N_f(\Delta n-\Delta n_1)}{\rm e}^{\ic\,\bar \alpha(\Delta n-\Delta n_1)}}\int\limits_{\Delta n_1} {\cal D}\phi\,{\cal O}_1\,{\rm e}^{-S_{\Omega_1}[\phi]}}{\sum\limits_{\Delta n=-N}^N\sum\limits_{\Delta n_1=-N_1}^{N_1} \!\!\! f(\Delta n)\,I_{\Delta n-\Delta n_1}(2\kappa\Omega_2){(-1)^{N_f(\Delta n-\Delta n_1)}{\rm e}^{\ic\,\bar \alpha(\Delta n-\Delta n_1)}}\int\limits_{\Delta n_1} {\cal D}\phi\,{\rm e}^{-S_{\Omega_1}[\phi]}}\,.
\label{eq:O1Omega:Bessel}
\end{align}
Note that we have made explicit here the phase factors from the fermion determinant that have not been absorbed in $\kappa$. Since we assume $\Omega_1$ to be finite and $\Omega\to\infty$, we need to take here $\Omega_2\to\infty$. As indicated, we take
again the limit of infinite $\Omega_2$ before summing over $\Delta n$. Then, the Bessel functions with an argument proportional to $\Omega_2$ tend to a common limit, independent of $\Delta n_1$. As a result the sum over $\Delta n$ factorizes  out and one is left with an expression for $\langle {\cal O}_1\rangle$ in terms of a path integration within the volume $\Omega_1$ that is given by
\begin{align}
\label{exp:val:local}
\langle {\cal O}_1\rangle_\Omega=&\frac{\sum\limits_{\Delta n_1=-\infty}^\infty \int\limits_{\Delta n_1} {\cal D}\phi\,{{(-1)^{-N_f\Delta n_1}{\rm e}^{-\ic \,\bar \alpha \Delta n_1}}}{\cal O}_1\,{\rm e}^{-S_{\Omega_1}[\phi]}}{\sum\limits_{\Delta n_1=-\infty}^\infty \int\limits_{\Delta n_1} {\cal D}\phi\,{{(-1)^{-N_f\Delta n_1}{\rm e}^{-\ic\,\bar \alpha    \Delta n_1}}}{\rm e}^{-S_{\Omega_1}[\phi]}}
\,.
\end{align}
We note that in this expression the $\theta$-angle as well as the phases from the fermion determinant proportional to $\Delta n$ have disappeared. Additionally, when carrying out the remaining piece of path integration, the phase factors ${(-1)^{-N_f\Delta n_1}{\rm e}^{-\ic\,\bar \alpha    \Delta n_1}}$ will exactly cancel the phases from the fermion determinants in $\Omega_1$ within each topological sector. Thus, no $\Delta n_1$-dependent phases remain, and there is no interference between topological sectors.
{ The cancellation of global phases in Eq.~\eqref{exp:val:local} can be understood as follows. As has been shown before, the global phases are fixed by the topological sectors of the total spacetime; however, for an infinite spacetime the different topological sectors effectively do not interfere, and the global phases thus drop from observables.}

It follows that when calculating the observable in the subvolume $\Omega_1$ there is no $CP$ violation, which is consistent with the calculation in the full, infinite volume. In particular, applying Eq.~(\ref{exp:val:local}), one can in fact redo the calculation of the fermionic correlation functions in the finite volume $\Omega_1$ and in the dilute instanton approximation, accounting for the lack of $\theta$-dependence and the insertions of the extra phases, and recover exactly the same result as in the underlying case of infinite volume: The phases of the instanton contributions are aligned with the tree-level phases from the fermion mass terms.

As a clarification, we recall that the justification of taking the limit of infinite $\Omega_2$ before the sum over $\Delta n$ is related to the fact that the classification into topological sectors labelled by integers is only necessarily enforced for infinite volume, in order for the action of the saddle points to remain finite. As it has been mentioned before, for the volume $\Omega_1$ the quantity $\Delta n_1$ is not enforced to be an integer, and the restriction to integer values  should be understood as an approximation in which one neglects contributions coming from particular fluctuations near the boundary of $\Omega_1$ (e.g. instantons with centres close to the boundary of $\Omega_1$). We expect the approximation to be accurate whenever $\Omega_1$ is large but finite and embedded into a much larger volume, such that the region in which the additional fluctuations are ignored has a small relative weight.

In spite of the periodic boundary conditions, lattice simulations sample over topological sectors within their volume unless the continuum limit is approached. In the present context, we can interpret such sampling of sectors in a finite volume as an evaluation of the path integral where the contributions from $\Omega_2$ to the action are discarded, see Refs.~\cite{Guo:2015tla,Shindler:2015aqa} for such lattice calculations of $CP$ violation in finite volumes. Crucially, this includes here the phases $f(\Delta n_2)\exp({\rm i}\bar\alpha\Delta n_2)$. If it were possible to circumvent the sign problem on the lattice associated with $CP$ phases, one would therefore find $CP$-odd expectation values. However, one should then keep in mind that the result would not be based on the action integral over the full spacetime but only on an arbitrary subset of it. Appropriately integrating out the contributions from $\Omega_2$ one arrives instead at Eq.~(\ref{exp:val:local}). To evaluate that expression on the lattice, one should set $\bar\theta=0$. This automatically accounts for the phases from $\Omega_2$, irrespective of the value of $\bar\theta$ that follows from the Lagrangian of the theory.

Finally, we note that the result of Eq.~\eqref{exp:val:local} can be equally recovered when keeping the infinite volume limit but removing the sum over topological sectors, i.e. for an infinite spacetime in a fixed topological sector, as would correspond to an observer arising from localized field excitations in a particular sector. In this sense, the property of cluster decomposition---i.e. that the expectation values of local operators only depend on local fluctuations---does not require a summation over the topological sectors of the total spacetime volume $\Omega$. In the next section it will be seen that the previous property also holds for finite spacetimes, up to parametrically small corrections.

\subsection{Cluster decomposition within a finite volume
\label{sec:cluster_decomposition_finite_VT}}

It is apparent from Eq.~\eqref{exp:val:local} that for local observables in the volume $\Omega_1$ the information about the boundary of {$\Omega$}  at infinity---including the possible effect of the $\theta$-angle and the associated $CP$-violating phenomena---is lost, regardless whether or not one sums over topological sectors in the full volume $\Omega$, cf. Figure~\ref{fig:locality}. Hence, in contrast to the usual argument given in Section~\ref{subsec:thetacluster}, the cluster decomposition principle does not strictly require summing over topological sectors when spacetime is infinite.
The same can in fact be argued for a finite $\Omega$, as long as $\Omega_2\gg\Omega_1$ and the path integration on $\Omega_1\cup\Omega_2$ is restricted to a single topological sector $\Delta n$, which can be realized for periodic boundary conditions as in spacetimes with the topology of a torus. In principle, one can also apply the arguments for bounded spacetimes with discrete topological sectors
but one should keep in mind that there is no first principle that would require vanishing fields at a finite boundary of $\Omega$.

To see that cluster decomposition also holds indeed in large (i.e. $\kappa \Omega\gg 1$) but finite volumes, one can  follow the steps in the previous section with the summation over $\Delta n$ omitted, which leads to
\begin{align}\label{eq:clusterfinite}\begin{aligned}
 \langle {\cal O}\rangle_{\Delta n\,\Omega}
=&\,\frac{\sum\limits_{\Delta n_1=-\infty}^\infty \!\!\! I_{\Delta n-\Delta n_1}(2\kappa\Omega_2){{}}\int\limits_{\Delta n_1} {\cal D}\phi\,{{(-1)^{-N_f \Delta n_1}\e^{-\ic \,\bar \alpha \Delta n_1}}}{\cal O}_1\,{\rm e}^{-S_{\Omega_1}[\phi]}}{\sum\limits_{\Delta n_1=-\infty}^\infty \!\!\! I_{\Delta n-\Delta n_1}(2\kappa\Omega_2){{}}\int\limits_{\Delta n_1} {\cal D}\phi\,{{(-1)^{-N_f \Delta n_1}\e^{-\ic \,\bar \alpha \Delta n_1}}}{\rm e}^{-S_{\Omega_1}[\phi]}}.
\end{aligned}\end{align}
In the previous expression it is not readily apparent that the terms that depend on  $\Omega_2$ can be factored out, which would lead again to an expression of $\langle {\cal O}\rangle_{\Delta n\,\Omega}$ in terms of a path integration in $\Omega_1$, in accordance with the expectations of the cluster decomposition principle. It turns out, however, that this factorization works up to corrections suppressed by inverse powers of $\Omega_2$. The idea is that the $\Delta n_1$ dependence of the Bessel functions $I_{\Delta n-\Delta n_1}(2\kappa_2\Omega_2)$ is only relevant for very large $\Delta n_1$, for which the path integration in $\Omega_1$ becomes exponentially suppressed. Hence, in the dominant contributions to $\langle {\cal O}\rangle_{\Delta n\,\Omega}$ one can indeed factorize out the $\Omega_2$ dependence and cluster decomposition is recovered.
To show this in a bit more formal detail, we first note that the factor in the numerator from the integration over $\Omega_1$ can be written as
\begin{align}
\label{factor:Omega1}
\int\limits_{\Delta n_1}{\cal D}\phi\, {\cal O}_1 {\rm e}^{-S_{\Omega_1}[\phi]}
=\sum\limits_r B_r
(-1)^{N_f\Delta n_1}\e^{\ic\,\bar \alpha (\Delta n_1+m_r)}
 I_{\Delta{n_1}+m_r}(2\kappa\Omega_1)\,,\quad m_r\in\mathbbm Z\,,
\end{align}
where $B_r$ are coefficients (that may have a tensor structure) and $m_r$ depend on the chiral fermionic contributions that appear within ${\cal O}_1$, cf. Eqs.~(\ref{correlation:sum}) and \eqref{eq:correlatorsNf}. Analogously, the $\Omega_1$ integration in the denominator of \eqref{eq:clusterfinite} gives the Euclidean generalization of Eq.~\eqref{eq:ZNf}, with the $\theta$-dependence omitted,
\begin{align}
\label{factor:Omega12}
\int\limits_{\Delta n_1}{\cal D}\phi\,  {\rm e}^{-S_{\Omega_1}[\phi]}
= (-1)^{N_f\Delta n_1}\e^{\ic\,\bar \alpha \Delta n_1}
 I_{\Delta{n_1}}(2\kappa\Omega_1)\,.
\end{align}
 Note how the $\Delta n_1$-dependent  phases  in Eqs.~\eqref{factor:Omega1} and \eqref{factor:Omega12} are exactly cancelled by those present in Eq.~\eqref{eq:clusterfinite}.

 Next, we note the asymptotic expansion
\begin{align}\label{eq:Inas2}
 I_n(z)\sim\frac{\e^z}{\sqrt{2\pi z}}\sum_{k=0}^\infty(-1)^k\frac{a_k(n)}{z^k}, \quad z\gg1,
\quad a_0(n)=1, \quad a_k(n)=\frac{(4n^2-1^2)(4n^2-3^2)\dots(4n^2-(2k-1)^2)}{k!8^k}\,,
\end{align}
from which it follows that
\begin{align}
 I_{\Delta n-\Delta n_1}(2\kappa\Omega_2) \,I_{\Delta n_1+m_r}(2\kappa\Omega_1)= &\,I_0(2\kappa\Omega_2) \left(1+\frac{I_{\Delta n-\Delta n_1}(2\kappa\Omega_2)-I_0(2\kappa\Omega_2)}{I_0(2\kappa\Omega_2)}\right) \,I_{\Delta n_1+m_r}(2\kappa\Omega_1)\notag\\
 =&\,I_{0}(2\kappa\Omega_2)\left(1-\frac{4(\Delta n-\Delta n_1)^2}{16\kappa\Omega_2}+{O}\left(\frac{1}{\Omega_2^2}\right)\right) \,I_{\Delta n_1+m_r}(2\kappa\Omega_1)\,.
\label{eq:approx0}
\end{align}
This means that for finite $\Delta n-\Delta n_1$ it is always possible to take $\Omega_2$ large enough such that
\begin{align}
 I_{\Delta n-\Delta n_1}(2\kappa\Omega_2) \,I_{\Delta n_1+m_r}(2\kappa\Omega_1)\sim I_{0}(2\kappa\Omega_2) \,I_{\Delta n_1+m_r}(2\kappa\Omega_1)\,.
\end{align}
However, there is no bound on $\Delta n_1$. We therefore need to show that, as anticipated earlier, the contributions from large $\Delta n_1$ can be neglected. This can be accomplished when considering the asymptotic expansion
\begin{align}\label{eq:Inas}
I_n(z)\sim\frac{1}{\sqrt{2\pi |n|}}\left(\frac{\e z}{2|n|}\right)^{|n|}, \quad |n|\gg 1,\,\,n\in\mathbb{Z}\,,
\end{align}
that implies an exponential suppression of the factor~(\ref{factor:Omega1}) from the integration over $\Omega_1$ for large $\Delta n_1$.

To obtain an upper bound on the magnitude of the contributions that arise for large $\Delta n_1$, note the inequalities
\begin{align}
\sum\limits_{\Delta n_1=\overline{\Delta n_1}}^{\infty} I_{\Delta n_1}(2\kappa\Omega_1)\sim\sum\limits_{\Delta n_1=\overline{\Delta n_1}}^\infty
\frac{1}{\sqrt{2\pi \Delta n_1}}\left(\frac{{\rm e}2\kappa\Omega_1}{2\Delta n_1}\right)^{\Delta n_1}<&\frac{1}{\sqrt{2\pi \overline{\Delta n_1}}}\sum\limits_{\Delta n_1=\overline{\Delta n_1}}^\infty
\left(\frac{{\rm e}\kappa\Omega_1}{\overline{\Delta n_1}}\right)^{\Delta n_1}\notag\\=&\frac{1}{\sqrt{2\pi \overline{\Delta n_1}}}\frac{\left(\frac{{\rm e}\kappa\Omega_1}{\overline{\Delta n_1}}\right)^{\overline{\Delta n_1}}}{1-\frac{{\rm e}\kappa\Omega_1}{\overline{\Delta n_1}}}\,,
\label{est:sum:Omega1}
\end{align}
where we neglect corrections due to $m_r$, assuming $|m_r|\ll \overline{\Delta n_1}$.

With the help of these relations, we see that
the sum over $\Delta n_1$ in the numerator of Eq.~(\ref{eq:clusterfinite}) can be split into a piece that is independent of $\Delta n$ and a remainder that depends on $\Delta n$ but goes to zero as $\Omega_2\to\infty$, which is a prerequisite for factorization:
\begin{align}
&\sum\limits_{\Delta n_1=-\infty}^\infty
\sum\limits_r B_r \,{\rm e}^{\ic\bar\alpha m_r} I_{\Delta n-\Delta n_1}(2\kappa \Omega_2) I_{\Delta n_1 +m_r}(2\kappa\Omega_1)\notag\\
=&\sum\limits_{\Delta n_1=-\infty}^{\infty}
\sum\limits_r B_r {\rm e}^{\ic\bar\alpha m_r} I_{0}(2\kappa \Omega_2) I_{\Delta n_1 +m_r}(2\kappa\Omega_1)\notag\\
+&\sum\limits_{\Delta n_1=-\overline{\Delta n_1}}^{\overline{\Delta n_1}}
\sum\limits_r B_r {\rm e}^{\ic\bar\alpha m_r}\left(I_{\Delta n-\Delta n_1}(2\kappa \Omega_2)-I_{0}(2\kappa \Omega_2)\right) I_{\Delta n_1 +m_r}(2\kappa\Omega_1)\notag\\
+&\sum\limits_{|\Delta n_1|>\overline{\Delta n_1}}
\sum\limits_r B_r {\rm e}^{\ic\bar\alpha m_r}\left(I_{\Delta n-\Delta n_1}(2\kappa \Omega_2)-I_{0}(2\kappa \Omega_2)\right)I_{\Delta n_1 +m_r}(2\kappa\Omega_1)
\notag\\
=&\!\!\!\sum\limits_{\Delta n_1=-\infty}^{\infty}\!\!\!
\sum\limits_r B_r {\rm e}^{\ic\bar\alpha m_r} I_{0}(2\kappa \Omega_2) I_{\Delta n_1 +m_r}(2\kappa\Omega_1)\left(1\!+\!O\left(\frac{(|\Delta n|+\overline{\Delta n_1})^2}{\kappa\Omega_2}\right)\right)
+I_{0}(2\kappa \Omega_2)\, O\left(\left[\frac{{\rm e}\kappa\Omega_1}{\overline{\Delta n_1}}\right]^{\overline{\Delta n_1}}\right)\,.
\end{align}
To estimate the size of the first of the remainder terms, we have used Eq.~(\ref{eq:approx0}) and for the second one Eq.~(\ref{est:sum:Omega1}) together with the relation $0\leq I_{0}(2\kappa \Omega_2) -I_{\Delta n-\Delta n_1}(2\kappa \Omega_2)\leq I_{0}(2\kappa \Omega_2)$. The terms in the denominator of Eq.~(\ref{eq:clusterfinite}) can be rearranged in an analogous way.

With this information, we can put upper bounds on the difference between
$\langle {\cal O} \rangle_{\Delta n\, \Omega}$ for a truncation $N$ of the sum over topological sectors and finite $\Omega_2$ and the limit that arises for $\Omega_2\to\infty$ while keeping $N$ finite:
\begin{align}
\langle {\cal O}_1\rangle_{\Delta n\,\Omega}=&
\Bigg[
\left(1+{O}\left(\frac{(|\Delta n|+\overline{\Delta n_1})^2}{\kappa\Omega_2}\right)\right)\notag\\
&\hskip-.2cm\times\left(
\sum\limits_{\Delta n_1=-\infty}^{\infty}
\sum\limits_r B_r \e^{\ic\,\bar \alpha m_r} I_0(2\kappa\Omega_2) I_{\Delta n_1+m_r}(2\kappa\Omega_1)
+I_{0}(2\kappa \Omega_2)\,{O}\left(\left[\frac{{\rm e}\kappa\Omega_1}{\overline{\Delta n_1}}\right]^{\overline{\Delta n_1}}\right)
\right)\Bigg]\Bigg/\notag\\
&
\Bigg[
\left(1+{O}\left(\frac{(|\Delta n| -\overline{\Delta n_1})^2}{\kappa\Omega_2}\right)\right)
\notag\\&\hskip2.8cm\times
\left(\sum\limits_{\Delta m_1=-\infty}^{\infty}
I_0(2\kappa\Omega_2) I_{\Delta m_1}(2\kappa\Omega_1)
+I_{0}(2\kappa \Omega_2)\,{O}\left(\left[\frac{{\rm e}\kappa\Omega_1}{\overline{\Delta m_1}}\right]^{\overline{\Delta m_1}}\right)\right)
\Bigg]\,.
\label{eq:O1Omega:cutoff}
\end{align}
As the value of $\overline{\Delta n_1}$ can be chosen freely, we can now show that the remainders go to zero as $\Omega_2\to\infty$. The exponentially suppressed corrections in Eq.~(\ref{eq:O1Omega:cutoff}) are under control when choosing $\overline{\Delta n_1}=A\kappa \Omega_1$ where $A\gg1$. In case $\overline{\Delta n_1}> |\Delta n|$, the correction due to ignoring the index of the Bessel function arising from $\Omega_2$ then is of order $A^2\kappa \Omega_1^2/\Omega_2$  which should also be much smaller than one. Both corrections are then suppressed simultaneously when $\Omega_2/(\kappa\Omega_1^2)\gg 1$.  When $\overline{\Delta n_1}\leq |\Delta n|$, one simply takes $\Omega_2\gg|\Delta n|/\kappa$ in order to keep the correction from ignoring the index of the Bessel function arbitrarily small.

Therefore, as advertised before, $\Omega_2$ can always be chosen large enough such that the path integral restricted to $\Omega_2$ is independent of $\Delta n_1$. Neglecting the remainder of higher order in Eq.~\eqref{eq:O1Omega:cutoff} the result is:
\begin{align}\label{eq:Ocluster_fixed_sector}
 \langle {\cal O}_1\rangle_{\Delta n\,\Omega}\approx&\,
\frac{\sum\limits_{\Delta n_1=-\infty}^{\infty}
\sum\limits_r B_r \e^{\ic\,\bar \alpha m_r} I_{\Delta n_1+m_r}(2\kappa\Omega_1)
}{
\sum\limits_{\Delta m_1=-\infty}^{\infty}
I_{\Delta m_1}(2\kappa\Omega_1)}=\frac{\sum\limits_{\Delta n_1=-\infty}^\infty \int\limits_{\Delta n_1} {\cal D}\phi\,{{(-1)^{-N_f\Delta n_1}{\rm e}^{-\ic \,\bar \alpha \Delta n_1}}}{\cal O}_1\,{\rm e}^{-S_{\Omega_1}[\phi]}}{\sum\limits_{\Delta m_1=-\infty}^\infty \int\limits_{\Delta m_1} {\cal D}\phi\,{{(-1)^{-N_f\Delta m_1}{\rm e}^{-\ic\,\bar \alpha    \Delta m_1}}}{\rm e}^{-S_{\Omega_1}[\phi]}}.
\end{align}
In the last step, we used Eqs.~\eqref{factor:Omega1} and \eqref{factor:Omega12} and recovered the factorization result of the previous section, Eq.~\eqref{exp:val:local}. Hence, cluster decomposition and the absence of $CP$ violation also hold in finite volumes, as long as the path integration is restricted to a single topological sector.

\subsection{Comparing with quantum mechanical systems with degenerate vacua}

The results~(\ref{exp:val:local}) and~(\ref{eq:Ocluster_fixed_sector}) may be used to relate the present discussion with one-dimensional periodic potentials in quantum mechanics (see e.g. Ref.~\cite{Coleman:1985rnk}). For these, one considers the evolution on a time interval ${\cal I}_1\subset{\cal I}$, where ${\cal I}$ is the time axis. It is useful to introduce basis states $|i\rangle$, where $i=-\infty,\ldots\infty$ is a label for the $i$th minimum of the periodic potential. Locally, such a state $|i\rangle$ approximately takes the form of a ground state about the $i$th minimum.

Note that due to the exponentially suppressed  instanton transitions between the minima, the states $|i\rangle$ are not time independent. Energy eigenstates are those whose transition amplitudes have an exponential dependence on $T_1$, the length of ${\cal I}_1$. These states can be parametrized by a phase $\theta_{\cal I}$ and therefore correspond to the $\theta$-vacua, i.e. they have the form $|\theta_{\cal I}\rangle=\sum_i \exp({\rm i}\, \theta_{\cal I}\, i)|i\rangle$, and their energy depends on $\theta_{\cal I}$ through a term proportional to $-\cos{\theta_{\cal I}}$. For gauge theory, this is often stated as a reason for constructing the $\theta$-vacua $|\theta \rangle=\sum_{n_{\rm CS}} \exp({\rm i}\, \theta\, n_{\rm CS})|n_{\rm CS}\rangle$, besides their gauge invariance and their cluster decomposition properties. (For the sake of the discussion of the relation with periodic potentials in quantum mechanics, we attribute here $\theta$ to the vacuum state rather than the topological term in the Lagrangian or the weight factors $f(\Delta n)$.)
Now in the dilute gas approximation, one can compute the amplitude for the transition from $|i\rangle$ to $|i+\Delta i\rangle$ for some integer $\Delta i$ within a time $T_1$. Due to the above energy dependence, for large $T_1$, this amplitude is dominated by the contribution from the energy eigenstate with $\theta_{\cal I}=0$,
\begin{align}\label{eq:thetaprojection}
 _{\rm out}\langle i+\Delta i | i\rangle_{\rm in} \propto\,_{\rm out}\langle \theta_{\cal I} | \theta_{\cal I}\rangle_{\rm in} |_{\theta_{\cal I}=0}.
\end{align}

Back to gauge-theory instantons, we note that in the dilute gas approximation the dependence on the spacetime volume of the amplitude for evolving from a state $|n_{\rm CS}\rangle$ to a state $|n_{\rm CS}+\Delta n\rangle$ coincides (up to an overall factor) with that of the amplitude computed for any linear combination of states $|n_{\rm CS}\rangle$ (i.e. in particular also for $\theta$-vacua) in a fixed topological sector $\Delta n$.
For large but finite spacetime volumes, it can therefore be argued that the partition function $Z_{\Delta n}$ for fixed $\Delta n$ yields correlation functions that agree with those obtained when interfering the different topological sectors in finite spacetime volumes for $\theta=0$~\cite{Brower:2003yx}. This corresponds to an alternative explanation of the results~(\ref{exp:val:local}) and~(\ref{eq:Ocluster_fixed_sector}) when applied to fixed topological sectors and finite spacetime volumes $\Omega$.
Nonetheless, we emphasize that the evolution in a fixed topological sector over large or infinite spacetime volumes $\Omega$ does not project a given state $|\theta\not=0\rangle$ onto $|\theta=0\rangle$ (as one may suspect because of the spacetime dependence of the amplitude), in contrast with what we have just stated about periodic potentials, simply because the time evolution changes the state only by an overall factor as $\Delta n$ is fixed. Thus, considering a fixed topological sector in an infinite spacetime volume is not at odds with $\theta$ being a good quantum number.

As for the energy, in periodic potentials it can be inferred from the logarithm of the partition function on ${\cal I}_1$. For gauge theory, the corresponding quantity in the present case is therefore the partition function in the finite subvolume $\Omega_1$, i.e. the denominator in Eq.~(\ref{exp:val:local}), from which we can infer the approximate energy density for fixed topological sectors in large volumes. In that expression, as discussed above, the phases incurred by fermion determinants cancel the explicit phases. In addition, the $\theta$-dependent phases from $f(\Delta n)$ have been cancelled as well. The reason is that the total phases are fixed by $\Delta n$, i.e. the boundary conditions imposed on the spacetime volume $\Omega$, such that the phases in $\Omega_1$ and $\Omega \setminus \Omega_1$ are not independent. It is therefore crucial that for periodic potentials one can obtain observables by just considering a finite interval ${\cal I}_1$, while in gauge theory for finite $\Omega_1$ one must not ignore phases incurred in the complement $\Omega \setminus \Omega_1$.

To explain this, the relevant difference between the quantum mechanical case and gauge theory is that the parameter $\theta_{\mathcal{I}}$ in the quantum-mechanical example, which is related to the crystal momentum of a particle, is not exactly conserved. This is because the finite size of the crystal necessarily breaks periodicity. States with a given $\theta_{\mathcal{I}}\not=0$ therefore have a finite lifetime. For a circular crystal, i.e. a system with periodic minima and periodic boundary conditions, there are states with conserved angular momentum. Then however, there is only a finite number of sectors $\Delta i$ given by the number of degenerate minima, such that there are no material consequences when changing the order of their sum with the integral over ${\cal I}$.

Another case of interest is a quantum mechanical potential with a finite number of degenerate minima. While for the above reasons, it is not possible to find exact eigenstates of crystal momentum or $\theta_{\mathcal{I}}$, closely analogous are energy eigenstates corresponding to standing wave configurations. The archetypical example of such a potential is the double well. In contrast to gauge theory, instead of an infinite number of equivalence classes of boundary conditions, here there are effectively only two, corresponding to trajectories that either start and end in the same well or do so in different wells. We label these two classes by \scalebox{1}[0.7]{$=$} and \scalebox{1}[0.7]{${\times}$}, respectively. The partition functions for these sectors are then given by~\cite{Coleman:1985rnk}
\begin{align}
Z_{\overset{\scalebox{0.71}[0.5]{=}}{\scalebox{0.71}[0.5]{\text{$\times$}}}}=\frac{1}{2}\sqrt{\frac{\omega}{\pi}}{\rm e}^{-\frac{\omega T}{2}}\left({\rm e}^{\kappa \exp(-S_{\rm E}) T}\pm {\rm e}^{-\kappa \exp(-S_{\rm E})T}\right)\,,
\end{align}
where $\omega$ denotes the oscillator frequency to quadratic order around the classical minima and $\kappa$ arises from the functional determinant, with an analogous parameter appearing in the calculation for gauge theory instantons. Note that, in contrast to the partition function~(\ref{Z:Delta:n}) for instantons in a fixed sector $\Delta n$, the configurations sum here to exponentials rather than modified Bessel functions. In analogy to the $CP$ violation we look for in the fermion correlation function, we may consider here the expectation value for parity $P$. The possible states now correspond to even and odd wave functions, for which we obtain (when summing over the path integrals in the sectors  \scalebox{1}[0.7]{$=$} and \scalebox{1}[0.7]{${\times}$} with coefficients set by projection on the wave functions of the even and odd states at the minima of the well)
\begin{align}
\langle P \rangle_{\rm even/odd}
=\frac{ \pm Z_{\scalebox{0.8}[0.56]{$=$}} + Z_{\scalebox{0.8}[0.56]{${\times}$}}}{Z_{\scalebox{0.8}[0.56]{$=$}}\pm Z_{\scalebox{0.8}[0.56]{${\times}$}}}=\pm 1\,.
\end{align}
Corresponding to the procedure in Section~\ref{subsec:summation}, we may evaluate the ratios of partial sums in the limit $T\to\infty$. The main difference is that in the present case the sum is finite so that there arises no question about at which point $T$ is to be taken to infinity.

In contrast to the possible nonconservation of $\theta_{\cal I}$ in the quantum mechanical case, for gauge theory $\theta$ is a good quantum number protected by a superselection rule due to gauge invariance. The latter requires boundary conditions on spatial hypersurfaces that lead to the conservation of $\theta$, whereas such strict conservation does not apply to the quantum-mechanical system.
In a crystal, states with  $\theta_{\mathcal{I}}\not=0$ are therefore observable after finite $T_1$ through
their spontaneous decay or through some measurement that does not conserve $\theta_{\mathcal{I}}$. It may also be possible to determine  $\theta_{\mathcal{I}}$ by a measurement in a finite spacetime volume, e.g. when switching on and off a measurement device. Finally, for systems with a finite number of degenerate minima, only a finite number of boundary conditions needs to be considered such that the question of the correct order of limits does not arise in first place.
Neither of these possibilities is viable in gauge theory: The parameter $\theta$, once chosen, is a constant of the theory, it is not possible to switch on the gauge couplings of quarks in a finite spacetime volume only and off outside of it and gauge invariance requires to sum over an infinite number of boundary conditions.

\vskip\baselineskip
\subsection{Topological susceptibility, instanton density and average topological charge}
\label{subsec:suceptibility}
Moments and cumulants of the topological charge density, such as the chiral susceptibility, are important quantities that allow to relate results for the regime of nonperturbative couplings to physical observables. It is therefore of importance to cross check the correct qualitative behaviour of these quantities in the dilute instanton gas. In terms of the partition function $Z$, the topological susceptibility can be expressed as
\begin{align}
\label{topological:susceptibility}
\chi_\Omega=\frac{1}{\Omega}\,\left.\langle \Delta n^2\rangle\right|_{\bar\theta=1/2(1-(-1)^{N_f})\pi}=\frac{1}{\Omega}\,\left.\left\langle\left(\int_\Omega d^4x \,q(x)\right)^2\right\rangle\right|_{\bar\theta=1/2(1-(-1)^{N_f})\pi}.
\end{align}
The requirement $\bar\theta = 1/2(1-(-1)^{N_f})\pi\equiv\theta_0$ above is imposed to ensure that when performing the sum over topological sectors for a finite spacetime volume, the vacuum energy is minimal and $\chi$ remains positive, and  $q$ stands for the topological charge density,
\begin{align}
 q=\frac{1}{32\pi^2} F^a_{mn}\tilde F^a_{mn}.
\end{align}

Now obviously, for a fixed topological sector $\chi_\Omega$ becomes arbitrarily small as $\Omega\to\infty$. For the case where we sum over the topological sectors after taking the spacetime volume to infinity, this also implies a globally vanishing topological susceptibility. This argument applies regardless of the inclusion of fermion fluctuations. One may see this as a problem given the relation between the topological susceptibility in the pure gauge theory and the mass of the $\eta^\prime$-meson that is derived in the limit of a large number of colours in Refs.~\cite{Witten:1979vv,Veneziano:1979ec}. However, as we have seen in Section~\ref{sec:chiral_Lagrangian}, the chiral Lagrangian matched to the `t Hooft vertices of Eq.~\eqref{eq:tHooft} leads to a nonzero mass of $\eta'$ that is equivalent to the standard result from using $\theta$-dependent phases in the chiral Lagrangian. Hence there is in principle no conflict between the vanishing of $\chi_\Omega$ for infinite $\Omega$ and a nonzero value of the $\eta'$ mass. The apparent contradiction with the results of Refs.~\cite{Witten:1979vv,Veneziano:1979ec} can be resolved by noting that the topological susceptibility considered in e.g.~\cite{Witten:1979vv} is constructed with an infrared regulator in momentum space, which acts as a large length cutoff. Hence, the observable should correspond to an operator of the form of Eq.~\eqref{topological:susceptibility} but defined in a finite subvolume $\Omega_1\subset\Omega$. While $\Delta n$
is conserved for the full volume $\Omega$ within each topological sector, this is not the case for the subvolume $\Omega_1$, across whose boundaries
topological currents may float freely. As a consequence, a nonzero value of $\chi_{\Omega_1}$ can arise. Indeed, when considering the expectation value of the square of the integral of the topological charge density $q$ over a finite subvolume $\Omega_1$, we can
apply Eq.~(\ref{exp:val:local}) to obtain
\begin{align}\label{eq:chi_omega_1}
\chi_{\Omega_1}\equiv \frac{1}{\Omega_1}\left\langle\left(\int_{\Omega_1} d^4x \,q(x)\right)^2\right\rangle&=\frac{1}{\Omega_1}
\frac{\sum\limits_{{\Delta n_1=-\infty}}^\infty\int{\cal D}A_{\Delta n_1}{ {\cal D}\bar\psi {\cal D}\psi}\int\limits_{\Omega_1}{\rm d}^4 z\, q(z) \int\limits_{\Omega_1}{\rm d}^4 z^\prime\, q(z^\prime){\rm e}^{-S_{\Omega_1}[A_\mu]}{{(-1)^{-N_f \Delta n_1}\e^{-\ic\,\bar \alpha    \Delta n_1}}}}{\sum\limits_{{\Delta n_1=-\infty}}^\infty\int{\cal D}A_{\Delta n_1}{ {\cal D}\bar\psi {\cal D}\psi}{\rm e}^{-S_{\Omega_1}[A_\mu]}{{(-1)^{-N_f \Delta n_1}\e^{-\ic\,\bar \alpha   \Delta n_1}}}}\notag\\
=&\frac{1}{\Omega_1}\frac{\sum\limits_{n_1=0}^\infty \sum\limits_{\bar n_1=0}^\infty { \frac{1}{n_1!\bar n_1!}}(n_1-\bar n_1)^2 (\kappa \Omega_1)^{n_1+\bar n_1}}{\sum\limits_{n_1=0}^\infty \sum\limits_{\bar n_1=0}^\infty { \frac{1}{n_1!\bar n_1!}}(\kappa \Omega_1)^{n_1+\bar n_1}}=2\kappa\,.
\end{align}
Here, to simplify notation, we write $\kappa_{N_f}$ as $\kappa$.
In the second line we have used the fact that the insertions of $(-1)^{-N_f \Delta n_1}{\rm e}^{-\ic\,\bar \alpha  \Delta n_1}$ cancel the phases of the fermion determinants.
The result precisely coincides with the one that is obtained when applying the
differential relation from Eq.~(\ref{topological:susceptibility}) to the standard result
$Z=\exp((-1)^{N_f}2\kappa\Omega \cos{\bar\theta})$ that is obtained when interfering the different topological sectors before taking the spacetime volume to infinity. Therefore
both variants of the calculation give rise to the same locally observed topological susceptibility. Note that the nonzero result for $\chi_{\Omega_1}$ does not contradict the vanishing of $\chi_{\Omega}$ when the infinite volume limit is taken before the sum over topological sectors. The reason is that the two quantities correspond to averages of different operators, an operator defined in a local volume for $\chi_{\Omega_1}$, and  a global operator corresponding to a topologically conserved quantity requiring integration over an infinite volume for $\chi_{\Omega}$. Only the first operator corresponds to the infrared regulated topological susceptibility considered in Ref.~\cite{Witten:1979vv}.

While the relation between the topological susceptibility in the pure gauge theory and the $\eta'$ mass obtained in Refs.~\cite{Witten:1979vv,Veneziano:1979ec} is derived in the limit of a large number of colours $N_c$,  our results suggest that the proportionality holds for arbitrary $N_c$. Indeed, the result of Eq.~\eqref{eq:chi_omega_1} relates the topological mass to the factor $\kappa$, which for a theory with $N_f$ flavours in Euclidean space is given by
\begin{align}
 \kappa = \int d\Omega \,J^{\rm E} \varpi\, {\rm e}^{-S_{\rm E}}\,\prod_{j=1}^{N_f}\Theta_j.
\end{align}
The previous equation follows from the definitions in Eqs.~\eqref{eq:kappadef}, \eqref{eq:ZNf} plus the fact that the ratios of determinants  $\Theta_j$ and $\varpi$ (defined in Eqs.~\eqref{eq:MinkowskiDet}~\eqref{eq:varpidef}) are equal in Minkowski and Euclidean space and the property that the Jacobian $J$ of zero modes in Minkowski space is related to the real Euclidean Jacobian $J^{\rm E}$ as $J=\ic J^{\rm E}$. It is reasonable to expect that the Dirac operator in the instanton background has a single discrete zero mode. On the other hand, the continuum spectrum should match that of the free Dirac operator. In this case, given the definition of Eq.~\eqref{eq:MinkowskiDet} one can approximate
\begin{align}
 \Theta_i\approx m_i,
\end{align}
where the $m_i$ are moduli of the fermion masses. Hence we can write
\begin{align}
  \kappa \approx \kappa^{\rm gauge} \,\prod_{j=1}^{N_f}m_j =\frac{1}{2} \,\chi_{\Omega_1}^{\rm gauge}\,\prod_{j=1}^{N_f}m_j.
\end{align}
In this equation we have isolated the contributions $\kappa^{\rm gauge}$ for the pure gauge theory (i.e. omitting the flucutation determinants of the fermions) and related these to the corresponding finite-volume topological susceptibility $\chi^{\rm gauge}_{\Omega_1}$ (i.e. the topological susceptibility for pure gauge theory) as in Eq.~\eqref{eq:chi_omega_1}. Finally, we note that matching the ${\rm det} U, \, {\rm det} U^\dagger$ terms in the chiral Lagrangian in Eq.~\eqref{eq:LC} with the `t Hooft vertices of Eq.~\eqref{eq:tHooft} (which reproduce the correlators with $p=0$, $q=N_f$ in Eq.~\eqref{eq:diagcorr}) gives
\begin{align}
 |\lambda|\propto \Gamma\propto \frac{\kappa}{\prod_j m_j}=\frac{1}{2} \chi^{\rm gauge}_{\Omega_1},
\end{align}
where we have also suppressed the subscript $N_f$ for $\Gamma$.
In the above equation, the factor of $\kappa$ and the inverse powers of fermion masses can be read off from the correlator of Eq.~\eqref{eq:diagcorr} (identifying $\kappa=\kappa_{N_f}$). The above result and the relation $m^2_{\eta'}=8|\lambda| f_\pi^2$ from Section~\ref{sec:chiral_Lagrangian} imply
\begin{align}
\label{etaprime:mass}
  m^2_{\eta'}\propto\chi^{\rm gauge}_{\Omega_1} .
\end{align}
This extends the fully nonperturbative results of Refs.~\cite{Witten:1979vv,Veneziano:1979ec} for large $N_c$
to arbitrary $N_c$ in the dilute instanton gas approximation. By Eq.~\eqref{eq:Ocluster_fixed_sector}, the result~\eqref{eq:chi_omega_1} for the topological susceptibility in a subvolume holds for a fixed topological sector just as well. This implies that also in a finite volume---large enough to suppress artefacts---with periodic boundary conditions and fixed topology of the gauge field, there is a massive $\eta^\prime$ meson according to relation~\eqref{etaprime:mass}.

In the dilute instanton gas approximation, it can also make sense to calculate the instanton
number density $\langle n\rangle/\Omega$. Restricting the path integration for the full volume $\Omega$ to  a fixed topological sector with charge $\Delta m$, we find {(from inserting $n$ in Eq.~\eqref{eq:ZNf})}
\begin{align}\label{eq:deltanav}
\frac{\langle n \rangle_{\Delta m}}{\Omega}=\kappa\frac{I_{\Delta m -1}(2 \kappa\Omega)}{I_{\Delta m}(2 \kappa\Omega)}\sim\kappa,
\end{align}
while for the relative fluctuation one has
\begin{align}
\frac{\sqrt{\langle (n -\langle n\rangle)^2\rangle_{\Delta m}} }{\langle n\rangle_{\Delta m}}={\left(\frac{I_{\Delta m}(2 \kappa  \Omega ){}^2}{I_{\Delta m-1}(2 \kappa  \Omega ){}^2}+\frac{\Delta m I_{\Delta m}(2 \kappa  \Omega )}{\kappa  \Omega  \,I_{\Delta m-1}(2 \kappa  \Omega )}-1\right)^{1/2}}
{=\sqrt{\frac{\Delta m}{\kappa\Omega}}}\,,
\end{align}
where we have indicated the asymptotic behaviour for $\Omega\to \infty$. The value of $\langle  n\rangle/\Omega$ in Eq.~\eqref{eq:deltanav}  coincides with the result obtained when the path integral in $\Omega$ includes a sum over topological sectors after taking the infinite volume limit. When performing instead the sum over a finite volume, one finds
\begin{align}
 \frac{\langle n\rangle}{\Omega}=(-1)^{N_f}\kappa {\rm e}^{\ic\bar\theta}.
\end{align}

\begin{table}
\begin{center}
\begin{tabular}{l||l|l|l|l}
 &$\Delta n$ free, $\Omega_\infty$ first  & $\Delta n$ free, $\Omega_\infty$ last  & $\Delta n$ free, $\Omega_1\subset\Omega_\infty$, $\Omega_\infty$ first   &\\
 & or  $\Delta n$ fixed, $\Omega_\infty$ & or $\Delta n$ free, $\Omega_{
 \rm fin}$  & or $\Delta n$ fixed, $\Omega_1\subset\Omega_{\rm fin},\Omega_{\infty}$&  $\Delta n$ fixed, $\Omega_{\rm fin}$\\[5pt]
 \hline
 \hline
 \rule{0pt}{17pt}
$\chi$
& 0 & $2\kappa$ & $2\kappa$ &  $\Delta n^2/\Omega$\\[5pt]
\hline
\rule{0pt}{17pt}
${\langle n\rangle}/{\Omega}$ & $\kappa$ &  $(-1)^{N_f}\kappa \,{\rm e}^{\ic\bar\theta}$ & $\kappa$ & $\kappa I_{\Delta n-1}(2\kappa\Omega)/I_{\Delta n}(2\kappa\Omega)$\\[5pt]
\hline
\rule{0pt}{17pt}
${\langle \Delta n\rangle}/{\Omega}$ & 0 &  $2\ic(-1)^{N_f}\kappa \,{\sin\bar\theta}$  & 0& $\Delta n/\Omega$
\end{tabular}
\end{center}
\caption{\label{tab:summary}Values of the topological susceptibility $\chi$, average instanton density $\langle n\rangle/\Omega$ and average topological charge $\langle \Delta n\rangle/\Omega$ for different choices of free (i.e. summed over topological sectors) vs fixed topological charge, finite or infinite volume as well as for different orderings of the limit of infinite volume and the sum over the topological sectors. The symbols $\Omega_\infty$ and $\Omega_{\rm fin}$ denote an infinite and finite total volume $\Omega$, respectively, while $\Omega_1$ is assumed to be finite. ``$\Omega_\infty$ first/last'' refers to the infinite volume limit being taken before/after the sum over topological sectors.}
\end{table}

One can proceed along the previous lines to estimate  the topological susceptibility, the instanton number density $\langle n\rangle /\Omega$ and the average topological charge $\langle\Delta n\rangle /\Omega$ for different choices of infinite or finite volume and order of taking limits. We collect the results in Table~\ref{tab:summary}.
The topological susceptibility $\chi$ as well as the charge $\Delta n$ are quantities
that are well-defined independently of the dilute gas approximation because these are given in
terms of integrals over functions of the topological term. We  note that when summing over topological sectors for a finite volume or before taking the infinite volume limit, one finds the following result for $\langle \Delta n^2\rangle$ for arbitrary $\theta$:
\begin{align}
\frac{\langle \Delta n^2 \rangle}{\Omega}=2\kappa\left(\cos(N_f \pi+\bar\theta)-2\kappa\Omega \sin^2(N_f\pi+\bar\theta)\right)\,.
\end{align}
Due to the dependence on the spacetime volume it is again not clear how to interpret this result.
The term that depends on the spacetime volume vanishes when considering instead the susceptibility~(\ref{topological:susceptibility}) which however for $\bar\theta\not=\theta_0=1/2(1-(-1)^{N_f})\pi$ cannot be interpreted as
$\langle\Delta n^2\rangle/\Omega$.

For finite volume, the results in Table \ref{tab:summary} are compatible with the property
\begin{align}\label{eq:Deltan_chi}
\frac{ \langle \Delta n\rangle}{\Omega}=\ic \left(\theta-\theta_0\right) \left.\frac{\langle \Delta n^2\rangle}{\Omega}\right|_{\theta_0}+{\cal O}(\theta-\theta_0)^2
\end{align}
used in Ref.~\cite{Shifman:1979if}. This property can be derived by performing an expansion in $\theta$ within the path integral, which requires the $\theta$ dependence to remain analytic. While this is true for finite spacetimes (and thus provides a consistency check of the results in Table \ref{tab:summary}), as discussed in Section~\ref{sec:general:correlations} the analyticity is not retained when taking the infinite volume limit before the sum over topological sectors.  In Ref.~\cite{Shifman:1979if}, there is also a nonzero estimate for the topological susceptibility $\langle \Delta n^2\rangle/\Omega|_{\theta_0}$ based on current algebra theorems. As in Ref.~\cite{Witten:1979vv}, the topological susceptibility is defined there with an infrared regulator, and thus should correspond to a finite volume observable. It is in that sense in agreement with our result~(\ref{eq:chi_omega_1}) for $\chi_{\Omega_1}$ in finite subvolumes. Our argument deviates from the literature since in Ref.~\cite{Shifman:1979if}, the presence of $CP$ violation is concluded on the basis of relation~\eqref{eq:Deltan_chi}. As discussed before this is not valid for an infinite spacetime $\Omega$. When considering an infinite $\Omega$ and observables with support in a finite subvolume $\Omega_1\subset\Omega$, as in Section~\ref{sec:clustering:inf:vol}, one can obtain expectation values in terms of a path integral restricted to $\Omega_1$, according to Eq.~\eqref{exp:val:local}. From these observables, $\theta$ is absent, and thus there is no relation analogous to Eq.~\eqref{eq:Deltan_chi} and no $CP$ violation even though $\chi_{\Omega_1}\neq0$.

\subsection{Schr\"odinger picture}
\label{sec:schroedinger}

While the boundary conditions for the path integral are chosen to correspond to vanishing physical fields, even in the ground state there are quantum fluctuations that lead to nontrivial correlators. These are recovered when evaluating the path integral sufficiently far away from the boundaries. In the Schr\"odinger picture, the ground state corresponds to a wave functional $\Psi(\vec A^a)$, where the procedure of canonical quantization can be carried out in a gauge with $A^{a,0}=0$~\cite{Jackiw:1979ur}.

 Working in Minkowski spacetime and noting that ${g}{\vec{E}^a={-}\partial/\partial t\,\vec A^a}$ and ${g}{ \vec B^a=\nabla\times \vec A^a-1/2\,f^{abc} \vec{A}^a \times\vec{A}^b}$, the canonical momentum conjugate to $\vec A^a$ is given by
\begin{align}
{g}{ \vec{\Pi}^a={-} \vec E^a+\frac{g^2}{8\pi^2} \theta \vec B^a\,.}
\end{align}
The corresponding quantized operator must observe the commutation relations
\begin{align}
[A^{a,i}(\vec x),\Pi^{b,j}(\vec x^{\,\prime})]={\rm i} \delta^{ij}\delta^{ab}\delta^3(\vec x-\vec x^{\,\prime})
\,,
\quad
[\Pi^{a,i}(\vec x),\Pi^{b,j}(\vec x^{\,\prime})]=0\,.
\end{align}
Here, $i,j,\ldots$ are indices of three-dimensional space and $a,b,\ldots$ for the adjoint representation of { the gauge group}. These commutators hold for
\begin{align}
\label{E:canonicalmomentum}
\vec{\Pi}^a=\frac{\delta}{{\rm i}\delta \vec A^a}+\alpha\frac{g}{8\pi^2} \vec B^a\,,
\end{align}
where we are free to choose the parameter $\alpha$.

The Hamiltonian density is then given by
\begin{align}\label{eq:Hthetaalpha}
{\cal H}=\frac 12\left(({\vec E^a})^2+(\vec B^a)^2\right)
=\frac 12 \left(\left(g\frac{\delta}{{\rm i}\delta \vec A^a}-\frac{g^2}{8\pi^2} (\theta-\alpha)\vec B^a\right)^2+(\vec B^a)^2\right)\,.
\end{align}
Due to the freedom of choosing $\alpha$, the parameter $\theta$ turns out to be irrelevant for the form of the Schr\"odinger equation. For example, imposing that the operator ${\rm i}\delta/\delta\vec A^a$ corresponds to the field $\vec E^a$ is met by the choice $\alpha=\theta$.
 Note that changing $\alpha$ in Eq.~(\ref{E:canonicalmomentum}) does not have an impact on the extra constraint from the Gau{\ss} law that should be imposed---{$ \vec D_{ab} \cdot \vec E_b=0$}, where $\vec D$ is the covariant derivative---because $\vec D_{ab}\cdot \vec B_b=0$. Hence, we end up with a Hamiltonian and constraint that are $\theta$-independent, as will be the spectrum and the corresponding eigenstates. Nonetheless, one should be aware that the choice of $\alpha$ determines the coefficient of the topological term when constructing the action that appears in the path integral starting from the canonically quantized theory.

Now let ${\cal G}_n$ be a large gauge transformation that changes the Chern--Simons number by $n$ units. Since this operator commutes with the Hamiltonian, it is possible to find states that satisfy ${\cal H}\Psi = E \Psi$ and
\begin{align}
\label{theta:eigenvalue}
{\cal G}_n\Psi(\vec A^a)={\rm e}^{-{\rm i}n\theta^\prime }\Psi(\vec A^a)\,,
\end{align}
where the eigenvalue must be a pure phase in order to comply with gauge invariance. States with this property constitute subspaces invariant under the action of the Hamiltonian, i.e. $\theta^\prime$ is protected by a superselection rule. The question of whether the spectra of these subspaces are identical is crucial for the physical relevance of the vacuum angle. That is, even when starting with a Hamiltonian and constraint with no explicit dependence on a vacuum angle, if the spectrum of the Hamiltonian changed between sectors with different values of $\theta'$, one would conclude that  the latter can be in some way observable. The only viable analytical approach to this question appears to be given by the saddle point expansion, where we find that the angle is irrelevant and therefore all of these subspaces should have the same spectra. Given states with the property of Eq.~\eqref{theta:eigenvalue}  and evolving under the Hamiltonian of Eq.~\eqref{eq:Hthetaalpha} with $\alpha=\theta$ (so that there is no explicit dependence on the vacuum angle in $\cal H$) we consider new sates $\Psi^\prime$ defined as
\begin{align}
\label{theta:rephasing}
\Psi^\prime(\vec A^a)={\rm e}^{-{\rm i}\theta^\prime W(\vec A^a) }\Psi(\vec A^a)
\end{align}
with
\begin{align}
 W(\vec A^a)=\frac{1}{8\pi^2}\,\epsilon^{ijk}\int {\rm d}^3 x\left(\frac{1}{2}\,A^{a,i}\partial_j A^{a,k}-\frac{1}{6} \,f^{abc} A^{a,i} A^{b,j} A^{c,k}\right)\,.
\end{align}
This satisfies ${\cal G}_n\Psi^\prime=\Psi^\prime$ as well as
\begin{align}
\frac 12 \left(\left(g\frac{\delta}{{\rm i}\delta \vec A}+\frac{g^2}{8\pi^2}\theta^\prime\vec B\right)^2+\vec B^2\right)\Psi^\prime(\vec A)=:{\cal H}^\prime \Psi^\prime(\vec A)=E\,\Psi^\prime(\vec A)\,.
\end{align}
Hence,  for states that transform according to Eq.~(\ref{theta:eigenvalue}) the operator ${\cal H}$ has the same spectrum as ${\cal H}^\prime$ for states with ${\cal G}_n\Psi^\prime=\Psi^\prime$. Note that ${\cal H}^\prime$ corresponds to the Hamiltonian of Eq.~\eqref{eq:Hthetaalpha} with $\alpha=\theta-\theta'$. While ${\cal H}\not={\cal H}^\prime$ in general, they nonetheless lead to the same predictions for the observables.
 In order for ${\cal H}^\prime$ to correspond to the energy density, we
should identify ${\vec E}$ with  $-{\rm i}g\delta/\delta\vec A+g^2/(8\pi^2)\,\theta^\prime {\vec B}$ such that after all ${\cal H}^\prime(\vec E)={\cal H}(\vec E)$. Note that the latter identity holds also when adding some interaction terms to the Hamiltonian through which one can observe the field $\vec{E}$. Therefore, the predicted observables in the subspace of the states $\Psi^\prime$ are identical to those in the subspace of the states $\Psi$.
Now, while the phase has been removed from the relation ${\cal G}_n\Psi^\prime=\Psi^\prime$,
there is the parameter $\theta^\prime$ appearing in ${\cal H}^\prime$. Constructing the path integral from this Hamiltonian then leads to a topological term $1/(16\pi^2) \theta^\prime\, {\rm tr}\,F \widetilde F$ in the Lagrangian. We make use of this freedom of redefinition in Section~\ref{sec:path:int:fixed:top} when projecting on the vacuum states given by the wave functional. This is possible even though the wave functional is not known exactly because in the limit of infinite spacetime volumes only configurations that reduce to vanishing physical fields at infinity lead to saddle points, as discussed in Section~\ref{sec:bc}.

As a final remark, we note that the freedom of choosing $\alpha$ in Eq.~\eqref{E:canonicalmomentum}  can be understood in terms of quantum canonical transformations that preserve the commutation relations and do not affect the physics. Different choices of $\alpha$ lead to different sets of eigenfunctions which are related by unitary transformations that preserve the spectrum and the inner product. Equation~\eqref{theta:rephasing} for example corresponds to the mapping between eigenstates of Hamiltonians related by a canonical transformation corresponding to $\delta\alpha=-\theta'$.

\ifarXiv
\else{}
\fi

\section{\label{app:freeprop} Spectral decomposition of the free fermionic propagator in Min\-kow\-ski spacetime}

To illustrate the spectral decomposition and the $\vartheta$-adjoint defined in Eq.~(\ref{eq:modes}), we use here the techniques of Section~\ref{sec:cplx:mass:arbitrary} to derive the free Minkowski propagator in Eq.~\eqref{S:0inst}. Throughout this section, all objects are assumed to be defined in Minkowski spacetime. The free propagator is the inverse of the operator
\begin{align}\label{eq:MinkowskiDirac}
 \ic\slashed \partial-m{\rm e}^{\ic\alpha\gamma^5}\,,
\end{align}
and so we need to consider its eigenfunctions. One can construct a complete basis of continuum modes of the form
\begin{align}\label{eq:planewaves}
 \psi_{\{k\}}(x) = \frac{1}{(2\pi)^2}\,f(k){\rm e}^{-\ic kx}\,,
\end{align}
where $f(k)$ is a spinor depending on the four-momentum $k^\mu$.
Imposing
\begin{align}
  (\ic\slashed \partial-m{\rm e}^{\ic\alpha\gamma^5} )\psi_{\{k\}}(x)=\xi_{\{k\}} \psi_{\{k\}}(x)
\end{align}
gives
\begin{align}
 (\slashed k-m{\rm e}^{\ic\alpha\gamma^5})f(k)=\xi_{\{k\}}f(k)\,,
\end{align}
so that the spinors $f(k)$ are eigenvectors of the operator $\slashed k-m{\rm e}^{\ic\alpha\gamma^5}$. We can explicitly obtain these eigenvectors and their corresponding eigenvalues. The latter are:
\begin{align}
\label{eq:xiM}
 \xi_{\{k \},i}=\left\{-m_{\rm R}-\ic\sqrt{m_{\rm I}^2-k^2},-m_{\rm R}-\ic\sqrt{m_{\rm I}^2-k^2},-m_{\rm R}+\ic\sqrt{m_{\rm I}^2-k^2},-m_{\rm R}+\ic\sqrt{m_{\rm I}^2-k^2}\right\}\,,
\end{align}
which can be understood from the Euclidean results for the continuum spectrum given in Eq.~\eqref{eq:Euccontinuum} and the relation between rotated and Euclidean eigenvalues given in Eq.~\eqref{eq:continuum}.
The eigenvectors corresponding to the eigenvalues in Eq.~\eqref{eq:xiM} are

\begin{align}\label{eq:fs}
\begin{aligned}
 f_1(k)=&\,\left[
 \begin{array}{c}
\frac{k^2+\ic k^1}{\sqrt{2} \sqrt{\sqrt{m_{\rm I}^2-k^2} \left(\sqrt{m_{\rm I}^2-k^2}+m_{\rm I}\right)}}\\
\frac{\ic \left(k^0-k^3\right)}{\sqrt{2} \sqrt{\sqrt{m_{\rm I}^2-k^2} \left(\sqrt{m_{\rm I}^2-k^2}+m_{\rm I}\right)}}\\
0\\
\frac{1}{\sqrt{2}}\sqrt{\frac{\sqrt{m_{\rm I}^2-k^2}+m_{\rm I}}{\sqrt{m_{\rm I}^2-k^2}}}
 \end{array}
 \right]\,,
\\
 f_2(k)=&\,\left[
 \begin{array}{c}
\frac{\ic \left(k^0+k^3\right)}{\sqrt{2} \sqrt{\sqrt{m_{\rm I}^2-k^2} \left(\sqrt{m_{\rm I}^2-k^2}+m_{\rm I}\right)}}\\
\frac{-k^2+\ic k^1}{\sqrt{2} \sqrt{\sqrt{m_{\rm I}^2-k^2} \left(\sqrt{m_{\rm I}^2-k^2}+m_{\rm I}\right)}}\\
\frac{1}{\sqrt{2}}\sqrt{\frac{\sqrt{m_{\rm I}^2-k^2}+m_{\rm I}}{\sqrt{m_{\rm I}^2-k^2}}}\\
0
 \end{array}
 \right]\,,
\end{aligned}
 \end{align}
\begin{align}
\begin{aligned}
 f_3(k)=&\,\left[
 \begin{array}{c}
-\frac{1}{\sqrt{2}}
\frac{k^2+\ic k^1}{\sqrt{\sqrt{m_{\rm I}^2-k^2} \left(\sqrt{m_{\rm I}^2-k^2}-m_{\rm I}\right)}}\\
-
\frac{1}{\sqrt{2}}
\frac{\ic(k^0-k^3)}{\sqrt{\sqrt{m_{\rm I}^2-k^2} \left(\sqrt{m_{\rm I}^2-k^2}-m_{\rm I}\right)}}\\
0\\
\frac{1}{\sqrt{2}}
\sqrt{\frac{\sqrt{m_{\rm I}^2-k^2}-m_{\rm I}}{\sqrt{m_{\rm I}^2-k^2}}}
 \end{array}
 \right]\,,
\\
 f_4(k)=&\,\left[
 \begin{array}{c}
-\frac{1}{\sqrt{2}}
\frac{\ic \left(k^0+k^3\right)}{\sqrt{\sqrt{m_{\rm I}^2-k^2} \left(\sqrt{m_{\rm I}^2-k^2}-m_{\rm I}\right)}}\\
-\frac{1}{\sqrt{2}}
\frac{-k^2+\ic k^1}{\sqrt{\sqrt{m_{\rm I}^2-k^2} \left(\sqrt{m_{\rm I}^2-k^2}-m_{\rm I}\right)}}\\
\frac{1}{\sqrt{2}}
\sqrt{\frac{\sqrt{m_{\rm I}^2-k^2}-m_{\rm I}}{\sqrt{m_{\rm I}^2-k^2}}}\\
0
 \end{array}
 \right]\,.
\end{aligned}\end{align}

These eigenvectors satisfy orthogonality with respect to a $\vartheta$-adjoint inner product:
\begin{align}
 \tilde f_i(k) f_j(k)=\delta_{ij}\,,
\end{align}
with the tilde operation defined in accordance with our general arguments in Eq.~\eqref{eq:modes} applied to $\vartheta=0$:
\begin{align}
\label{eq:tildef}
 \tilde f_i(k)=f_i(k)^\dagger|_{k_0\rightarrow-k_0}\,.
\end{align}

Moreover, one can explicitly verify the completeness relation
\begin{align}
\label{eq:completenessk}
 \sum_{i=1}^4 f_i(k)\tilde f_i(k) =\mathbb{I}_4\,.
\end{align}

We can extend this property now to the eigenfunctions \eqref{eq:planewaves} in position space.
The $\vartheta$-adjoint defined in Eq.~\eqref{eq:modes} implies
\begin{align}
 \tilde\psi_{\{k\},j}(x)=\frac{1}{(2\pi)^2}\,\tilde f_j(k)\,{\rm e}^{\ic kx}\,,
\end{align}
and we have the $\vartheta$-adjoint inner products
\begin{align}
 (\psi_{\{k\},i}(x),\psi_{\{{k'}\},j}(x))=\int {\rm d}^4x \,\tilde\psi_{\{k\},i}(x)\psi_{\{k'\},j}(x)=\frac{1}{(2\pi)^4}\int {\rm d}^4x\, {\rm e}^{\ic(k-k')x} \tilde f_i(k)f_j(k)=\delta_{ij}\delta^4(k-k')\,,
\end{align}
which imply orthogonality of the eigenfunctions. Similarly, one can derive the completeness relation
\begin{align}
 \sum_i \int {\rm d}^4 k\,  \psi_{\{k\},i}(x)\tilde \psi_{\{k\},i}(x')=\frac{1}{(2\pi)^4} \int {\rm d}^4 k\, {\rm e}^{-\ic k(x-x')}\sum_i f_i(k)\tilde f_i(k)=\delta^4(x-x'\,)\mathbb{I}_4\,,
\end{align}
where we used Eq.~\eqref{eq:completenessk}. From the completeness and orthogonality it follows that we can write the Minkowskian free propagator as
\begin{align}
 S_0(x,x')=\sum_i\int {\rm d}^4 p \,\frac{1}{\xi_{\{p\},i}} \,\psi_{\{p\},i}(x)\tilde\psi_{\{p\},i}(x')=\frac{1}{(2\pi)^4}\int {\rm d}^4 p \,{\rm e}^{-{\rm i}p(x-x')}\,\sum_i\frac{1}{\xi_{\{p\},i}} \, f_{i}(p)\tilde f_{i}(p)\,.
\end{align}
Direct evaluation with the $\xi_i(p)$ in Eq.~\eqref{eq:xiM}, the eigenvectors $f_i(p)$ in Eqs.~\eqref{eq:fs} and the $\vartheta$-adjoint of Eq.~\eqref{eq:tildef} gives again the result of Eq.~\eqref{S:0inst}:
\begin{align}
  S_0(x,x')=\,\int\frac{{\rm d}^4p}{(2\pi)^4} \,{\rm e}^{-{\rm i}p(x^{\rm }-x^{{\rm }\prime})}\,\frac{(\slashed{p}+m{\rm e}^{-{\rm i}\gamma^5})}{p^2-m^2+{\rm i}\epsilon}\,.
\end{align}

To end, let us note that the operator in Eq.~\eqref{eq:MinkowskiDirac} is Hermitian under the Dirac adjoint inner product
\begin{align}
 \langle\psi_{\{k\},i},\psi_{\{k'\},j}\rangle \equiv\int {\rm d}^4x \,\bar\psi_{\{k\},i}(x)\psi_{\{k'\},j}\,.
\end{align}
However, this does not enter in conflict with the fact that the eigenvalues in Eq.~\eqref{eq:xiM} are complex, because
the eigenfunctions \eqref{eq:planewaves} have zero norm under the Dirac adjoint inner product, as can be verified by noting that the spinors $f_i(k)$ satisfy
\begin{align}
 \bar f_i(k)f_i(k)=0\,.
\end{align}
Moreover, the $f_i(k)$ are not orthogonal under the Dirac adjoint inner product, which again does not conflict with Hermiticity because the nonzero mixed products relate eigenvectors $f_i(k),f_j(k)$ with mutually conjugate eigenvalues. (Note that the eigenvalues in Eq.~\eqref{eq:xiM} come in conjugate pairs.) Hermiticity implies
\begin{align}
 \langle\psi_{\{k\}i},(\ic\slashed \partial-m{\rm e}^{\ic\alpha\gamma^5})\psi_{\{k\}j}\rangle=\xi_{\{k\},j}\langle\psi_{\{k\}i},\psi_{\{k\}j}\rangle=\langle(\ic\slashed \partial-m{\rm e}^{\ic\alpha\gamma^5})\psi_{\{k\}i},\psi_{\{k\}j}\rangle=\xi^*_{\{k\},i}\langle\psi_{\{k\}i},\psi_{\{k\}j}\rangle\,,
\end{align}
which allows nonzero mixed products $\langle\psi_{\{k\}i},\psi_{\{k\}j}\rangle$ as long as $\xi^*_{\{k\},i}=\xi_{\{k\},j}$.
The previous properties imply that the Dirac adjoint product does not allow to define orthogonal projectors that resolve the identity, in contrast to the $\vartheta$-adjoint inner product, from which one recovers the standard propagator.

\printbibliography

\end{document}